\documentclass[fleqn,usenatbib,useAMS]{mnras}
\usepackage{newtxtext,newtxmath}
\usepackage[T1]{fontenc}
\usepackage{graphicx}
\usepackage{amsmath}
\usepackage{caption}
\usepackage{multirow}
\usepackage{tabularx}
\usepackage{rotating}
\usepackage{adjustbox} 
\usepackage{xspace} 
\usepackage{xcolor}
\usepackage{xurl}
\usepackage{array}
\usepackage{orcidlink}

\newcommand{\kms}{km s$^{-1}$\xspace}

\newcommand{\so}{SO(2$_{3}-1_{2}$)\xspace}
\newcommand{\ha}{H40$\alpha$\xspace}
\newcommand{\hctn}{HC$_3$N($11-10$)\xspace}
\newcommand{\cs}{CS($2-1$)\xspace}
\newcommand{\chtcho}{CH$_3$CHO($5_{1,4}-4_{1,3}$)\xspace}

\newcommand{\arcdeg}{\mbox{$^\circ$}}

\defcitealias{Longmore2025}{Paper~I}
\defcitealias{Ginsburg2025}{Paper~II}
\defcitealias{Walker2025}{Paper~III}
\defcitealias{Lu2025}{Paper~IV}

\title[ACES Broadband Spectral Windows]{ALMA Central molecular zone Exploration Survey (ACES) V:\\ \cs, \so, \chtcho, \hctn and \ha lines data}

\newcounter{affcounter}

\newcommand{\defaffiliationlabel}[1]{%
  \refstepcounter{affcounter}%
  \expandafter\xdef\csname #1\endcsname{\theaffcounter}%
}

\newcommand{\affref}[1]{$^{\csname #1\endcsname}$}

\newcommand{\affrefs}[1]{%
  $^{%
    \@for\@ref:=#1\do{%
      \@ref\@ifnextchar\@nil{}{,}%
    }%
  }$%
}

\newcommand{\affrefTwo}[2]{$^{\csname #1\endcsname,\csname #2\endcsname}$}
\newcommand{\affrefThree}[3]{$^{\csname #1\endcsname,\csname #2\endcsname,\csname #3\endcsname}$}
\newcommand{\affrefFour}[4]{$^{\csname #1\endcsname,\csname #2\endcsname,\csname #3\endcsname,\csname #4\endcsname}$}

\newcommand{\printaffiliation}[2]{%
  $^{\csname #1\endcsname}$#2\\%
}
\defaffiliationlabel{naoj}
\defaffiliationlabel{ukarcnode}
\defaffiliationlabel{uflorida}
\defaffiliationlabel{eso}
\defaffiliationlabel{shao}
\defaffiliationlabel{naoc_key}
\defaffiliationlabel{ice_csic}
\defaffiliationlabel{ieec}
\defaffiliationlabel{umd}
\defaffiliationlabel{mpe}
\defaffiliationlabel{kansas}
\defaffiliationlabel{ljmu}
\defaffiliationlabel{mpia}
\defaffiliationlabel{ari_heidelberg}
\defaffiliationlabel{COOL}
\defaffiliationlabel{eso_chile}
\defaffiliationlabel{jao}
\defaffiliationlabel{nanjing}
\defaffiliationlabel{nanjing_key}
\defaffiliationlabel{cfa}
\defaffiliationlabel{mit}
\defaffiliationlabel{colorado}
\defaffiliationlabel{uconn}
\defaffiliationlabel{cab_csic}
\defaffiliationlabel{iaa_taipei}
\defaffiliationlabel{chalmers}
\defaffiliationlabel{clap}
\defaffiliationlabel{iop_epfl}
\defaffiliationlabel{oaq}
\defaffiliationlabel{inaf_arcetri}
\defaffiliationlabel{jbca}
\defaffiliationlabel{nrao}
\defaffiliationlabel{ita_heidelberg}
\defaffiliationlabel{anu}
\defaffiliationlabel{iaa_csic}
\defaffiliationlabel{ucn}
\defaffiliationlabel{cassaca}
\defaffiliationlabel{ias}
\defaffiliationlabel{ucl}
\defaffiliationlabel{izw_heidelberg}
\defaffiliationlabel{radcliffe}
\defaffiliationlabel{ipm}
\defaffiliationlabel{villanova}
\defaffiliationlabel{ulaserena}
\defaffiliationlabel{iff_csic}
\defaffiliationlabel{gbo}
\defaffiliationlabel{utokyo}
\defaffiliationlabel{umass}
\defaffiliationlabel{aberystwyth}
\defaffiliationlabel{kiaa_pku}
\defaffiliationlabel{pku_astro}
\author[Pei-Ying Hsieh \& ACES Team]{Pei-Ying Hsieh,\affref{naoj}\thanks{E-mail: pei-ying.hsieh@nao.ac.jp}\orcidlink{0000-0001-9155-39}
Daniel.~L.~Walker,\affref{ukarcnode}\orcidlink{0000-0001-7330-8856}
Adam Ginsburg,\affref{uflorida}\orcidlink{0000-0001-6431-9633}
Ashley~T.~Barnes,\affref{eso}\orcidlink{0000-0003-0410-4504}
Xing Lu,\affrefTwo{shao}{naoc_key}\orcidlink{0000-0003-2619-9305}
\newauthor
\'Alvaro S\'anchez-Monge,\affrefTwo{ice_csic}{ieec}\orcidlink{0000-0002-3078-9482}
Savannah R. Gramze,\affref{uflorida}\orcidlink{0000-0002-1313-429X}
Nazar Budaiev,\affref{uflorida}\orcidlink{0000-0002-0533-8575}
Marc W.~Pound,\affref{umd}\orcidlink{0000-0002-7269-342X}
\newauthor
Jaime E. Pineda,\affref{mpe}\orcidlink{0000-0002-3972-1978}
Claire Cook,\affref{kansas}
Jonathan D. Henshaw,\affrefTwo{ljmu}{mpia}\orcidlink{0000-0001-9656-7682}
\newauthor
Katharina Immer,\affref{eso}\orcidlink{0000-0003-4140-5138}
Namitha Issac,\affref{shao}\orcidlink{0000-0002-7881-689X}
Desmond Jeff,\affref{uflorida}\orcidlink{0000-0003-0416-4830}
Fu-Heng Liang,\affrefTwo{ari_heidelberg}{eso}\orcidlink{0000-0003-2496-1247}
Steven N. Longmore,\affrefTwo{ljmu}{COOL}\orcidlink{0000-0001-6353-0170}
\newauthor
Elisabeth A.C. Mills,\affref{kansas}\orcidlink{0000-0001-8782-1992}
Sergio Martín,\affrefTwo{eso_chile}{jao}\orcidlink{0000-0001-9281-2919}
Xing Pan,\affrefThree{nanjing}{nanjing_key}{cfa}\orcidlink{0000-0003-1337-9059}
Thushara G.S. Pillai,\affref{mit}\orcidlink{0000-0003-2133-4862}
\newauthor
Qizhou Zhang,\affref{cfa}\orcidlink{0000-0003-2384-6589}
John Bally,\affref{colorado}\orcidlink{0000-0001-8135-6612}
Cara Battersby,\affref{uconn}\orcidlink{0000-0002-6073-9320}
Laura Colzi,\affref{cab_csic}\orcidlink{0000-0001-8064-6394}
Paul T. P. Ho, \affref{iaa_taipei}\orcidlink{0000-0002-3412-4306}
\newauthor
Izaskun Jim\'enez-Serra,\affref{cab_csic}\orcidlink{0000-0003-4493-8714}
J.~M.~Diederik Kruijssen,\affref{COOL}\orcidlink{0000-0002-8804-0212}
Maya Petkova,\affref{chalmers}\orcidlink{0000-0002-6362-8159}
Mattia C. Sormani,\affref{clap}\orcidlink{0000-0001-6113-6241}
\newauthor
Robin G. Tress,\affref{iop_epfl}\orcidlink{0000-0002-9483-7164}
Jennifer Wallace,\affref{uconn}\orcidlink{0009-0002-7459-4174}
J. Armijos-Abenda\~no,\affref{oaq}\orcidlink{0000-0003-3341-6144}
Lucia Armillotta,\affref{inaf_arcetri}\orcidlink{0000-0002-5708-1927}
N. Bijas,\affref{jbca}\orcidlink{0000-0002-6398-7530}
\newauthor
Rojita Buddhacharya,\affref{ljmu}\orcidlink{0009-0004-0685-7678}
Laura A. Busch,\affref{mpe}
Natalie O. Butterfield,\affref{nrao}\orcidlink{0000-0002-4013-6469}
M\'elanie Chevance\affrefTwo{ita_heidelberg}{COOL}\orcidlink{0000-0002-5635-5180}
\newauthor
Ana Karla D\'iaz-Rodr\'iguez,\affref{ukarcnode}\orcidlink{0000-0001-9112-6474}
Christoph Federrath,\affref{anu}\orcidlink{0000-0002-0706-2306}
\newauthor
Rub\'{e}n Fedriani,\affref{iaa_csic}\orcidlink{0000-0003-4040-4934}
Pablo Garc{\'i}a,\affrefTwo{ucn}{cassaca}\orcidlink{0000-0002-8586-6721}
Qi-Lao Gu,\affref{shao}
\newauthor
Rebecca J. Houghton,\affref{ljmu}\orcidlink{0000-0002-9723-1088}
Yue Hu,\affref{ias}\orcidlink{0000-0002-8455-0805}
Janik Karoly,\affref{ucl}\orcidlink{0000-0001-5996-3600}
Ralf S.\ Klessen,\affrefFour{ita_heidelberg}{izw_heidelberg}{cfa}{radcliffe}\orcidlink{0000-0002-0560-3172}
Mark R. Krumholz,\affref{anu}\orcidlink{0000-0003-3893-854X}
\newauthor
Farideh Mazoochi,\affref{ipm}\orcidlink{0000-0003-3390-4893}
Francisco Nogueras-Lara,\affrefTwo{iaa_csic}{eso}\orcidlink{0000-0002-6379-7593}
Dylan Par\'e,\affref{villanova}\orcidlink{0000-0002-5811-0136}
Denise Riquelme-V\'asquez,\affref{ulaserena}\orcidlink{0000-0001-5389-0535}
\newauthor
V\'ictor M. Rivilla,\affref{cab_csic}\orcidlink{0000-0002-2887-5859}
Miriam G. Santa-Maria,\affrefTwo{uflorida}{iff_csic}\orcidlink{0000-0002-3941-0360}
Anika Schmiedeke,\affref{gbo}\orcidlink{0000-0002-1730-8832}
Yoshiaki Sofue,\affref{utokyo}\orcidlink{0000-0002-4268-6499}
\newauthor
Volker Tolls,\affref{cfa}\orcidlink{0000-0003-1841-2241}
Q. Daniel Wang,\affref{umass}\orcidlink{0000-0002-9279-4041}
Gwenllian M. Williams,\affref{aberystwyth}\orcidlink{0000-0001-5933-2147}
Fengwei Xu,\affrefTwo{kiaa_pku}{pku_astro}\orcidlink{0000-0001-5950-1932},
and Suinan Zhang\affref{shao}\orcidlink{0000-0002-8389-6695}
\\
$^{*}$Author affiliations are listed at the end of the paper
}

\begin{document}
\label{firstpage}
\pagerange{\pageref{firstpage}--\pageref{lastpage}}
\maketitle

\begin{abstract}

We present data from the ALMA Central Molecular Zone Exploration Survey (ACES) Large Program, which provides broad spectral-line and 3 mm continuum coverage of the Central Molecular Zone (CMZ) at a spatial resolution of 0.1 pc. The survey delivers homogeneous, wide-field mosaics that enable direct comparisons of the physical and chemical conditions across diverse environments in the Galactic center. In this data release paper, we present the \cs, \so, \chtcho, \hctn, and \ha lines observed simultaneously within two broad spectral windows.
These lines reveal pronounced spatial and chemical variations across the CMZ, tracing distinct components of molecular gas, shock-affected regions, and ionized structures. The high angular resolution and multi-line capability of the ACES dataset make it a powerful resource for future studies of gas dynamics, star formation activity, and the physical connection between the CMZ and Sgr A*.

\end{abstract}


\section{Introduction}
\label{sec:intro}

The central molecular zone (CMZ) is a dense reservoir of molecular gas within the innermost $\sim$300~pc of the Galactic center (GC), confined to a vertically thin disk with a scale height of 30--50~pc. The total gas mass is estimated at $\sim(5$--$10)\times10^7~M_\odot$, accounting for roughly 10\% of the Milky Way's molecular gas \citep[][and references therein]{Morris1996,Genzel2010}. Most of this mass is concentrated in compact giant molecular clouds (GMCs) with typical masses of $10^{5.5}$ to $10^{6.5}~M_\odot$. Hence, the large volume of molecular gas makes it an excellent environment for star formation.

Molecular gas in the CMZ exhibits physical conditions distinct from those in the Galactic disk. The CO-to-H$_2$ conversion factors and gas surface densities in the CMZ resemble those found in starburst galaxies \citep[see Figure 10 in][]{Tacconi2008}. The gas has high temperatures, turbulence, and pressure \citep{Ao2013, Mills2013, Ginsburg2016, Immer2016},  sharing characteristics with active star-forming regions in many different extragalactic environments \citep{Kruijssen2013}. However, despite the presence of dense gas, the current star formation rate (SFR) in the CMZ is only $\sim$10\% of what is expected from standard dense gas–SFR scaling relations \citep{Longmore2013b,Barnes2017,Lu2019a,Henshaw2023}. Understanding how these physical conditions are related to the suppressed star formation in the CMZ requires further observational and theoretical studies which link the global gas properties to the sub-parsec scales at which star formation occurs \citep{Kruijssen2014a, Kruijssen19, krumholz15, Krumholz16,Tress2020,Sormani2022}.

\begin{figure*}
\includegraphics[width=1.0\textwidth]{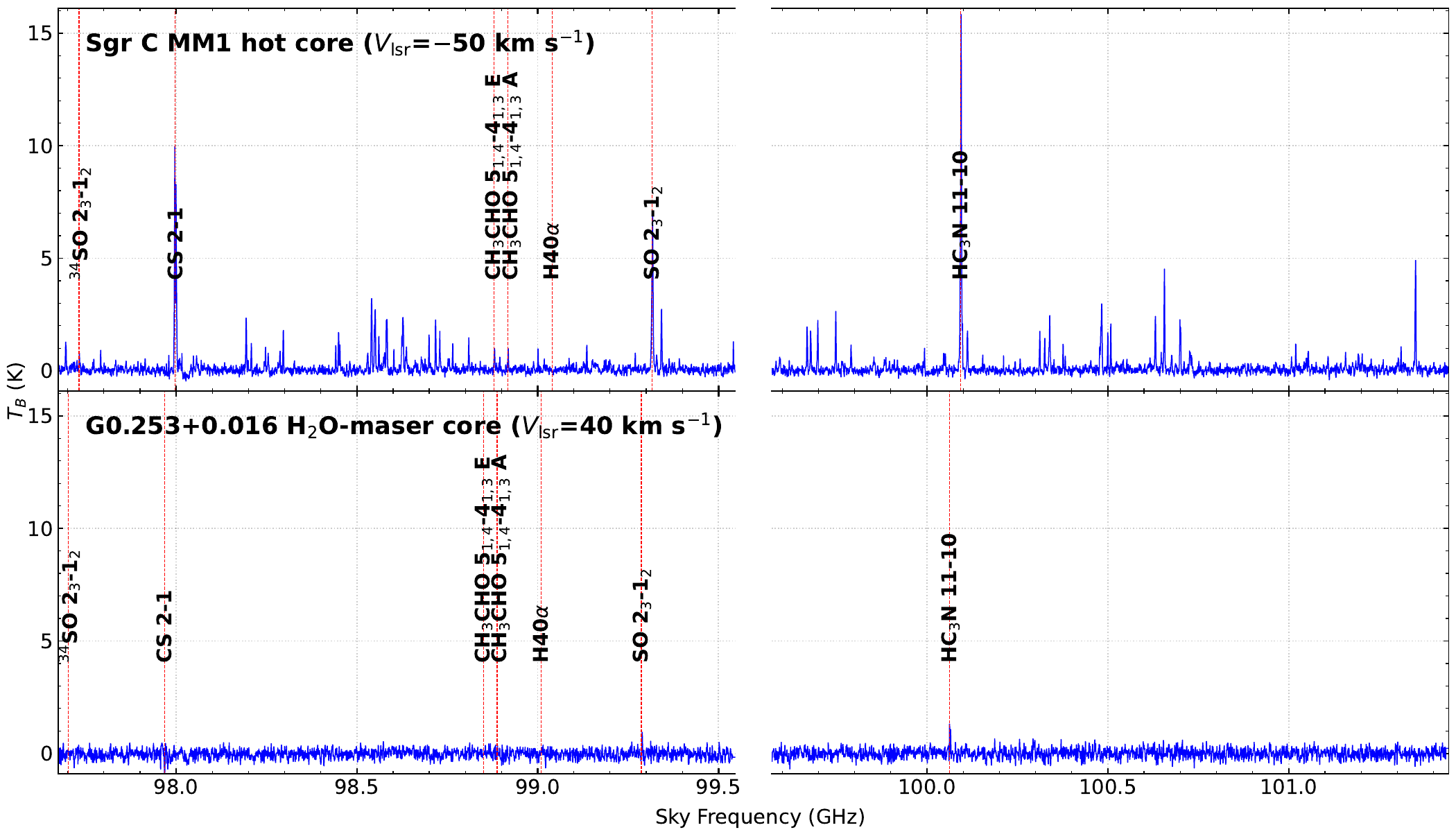}
\caption{The spectra of broad spectral windows (SPW 33, 35) towards the Sgr C MM1 hot core and water maser core of Brick cloud. The lines presented in this paper are labeled. The native channel width of broad spectral window is 1.48 km s$^{-1}$ with an effective bandwidth of $\sim1.86$ GHz ($\sim$ 5680 km s$^{-1}$ per spectral window). 
}
\label{fig:spectrum}
\end{figure*}

\begin{table*}
\centering
\caption{Key lines covered by the broadband spectral windows.\label{tab:lines}}
\begin{tabular}{ccccc}
\hline
Lines & Rest frequency  & $A_{\rm ul}$ & $E_{u}/k$ & $n_{\rm crit}^{\phantom{crit}a}$   \\
 & (GHz) & $s^{-1}$ & (K) & (cm$^{-3}$) \\
\hline
\cs & 97.98095 & 1.679E-5 & 7.1 & 2.8$\times10^5$ \\
\so & 99.29987 & 1.125E-5 & 9.2 & 2.9$\times10^5$\\
\chtcho E & 98.86331 & 3.104E-5 & 16.6 & 1$\times10^6$ \\
\chtcho A & 98.90095 &  3.108E-5 & 16.5  & 1$\times10^6$  \\
\hctn & 100.07639   & 0.777E-4 & 28.8  & 1.5$\times10^6$  \\
\ha   & 99.02295           &    -      &  -     & - \\
\hline
\end{tabular}
\begin{flushleft}
$^a$ Calculated assuming a gas temperature of 100~K. Critical densities are calculated as $n_{\rm crit} = A_{\rm ul}/\gamma_{\rm ul}$, where $A_{\rm ul}$ is the Einstein coefficient and $\gamma_{\rm ul}$ is the collisional rate. The values at 100~K are taken from the Leiden Atomic and Molecular Database \citep{Schoier2005} and the JPL spectroscopic database \citep{Pickett1998}. As no collisional rate is available for \chtcho, its critical density is estimated using the collisional rate of CH$_3$OH as a close approximation.\\
\end{flushleft}
\end{table*}

The Atacama Large Millimeter/submillimeter Array (ALMA) CMZ Exploration Survey (ACES) Large Program \citep[project code: 2021.1.00172.L; PI: S.~Longmore; hereafter  \citetalias{Longmore2025}]{Longmore2025} provides uniform Band~3 coverage of the inner $1.5^\circ \times 0.5^\circ$ of the CMZ, producing the largest mosaic yet made with ALMA and achieving an unprecedented angular resolution of $\sim$2\arcsec{} in both spectral-line and continuum emission. This new global, high-resolution and sensitivity view of the physical, kinematic and chemical structure facilitates ACES main science objective to build a global understanding of star formation, feedback, and the mass flows and energy cycles across the Central Molecular Zone.
ACES employed 6 spectral windows to span a wide range of frequency coverage. The continuum dataset, covering an effective bandwidth of 4.6 GHz, is described in detail by \citet{Ginsburg2025}. The high-resolution spectral-line data and details of the data combination will be presented by \citet{Walker2025}, while the intermediate-resolution spectral-line data will be presented by \citet{Lu2025}. In this paper, we present the data release of two broadband spectral windows that include the lines of \cs (carbon monosulfide), \so  (sulfur monoxide), \hctn (cyanoacetylene), \chtcho (acetaldehyde), and the radio recombination line \ha. The characteristics of these lines are summarized in Table~\ref{tab:lines}.


\begin{figure*}
\centering
\includegraphics[width=0.7\textwidth]{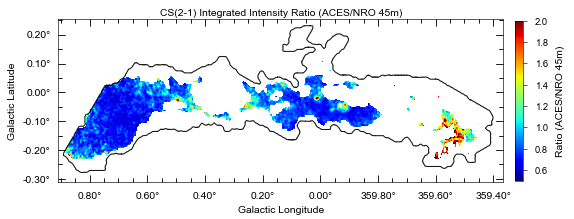}
\caption{The ratio of ACES and NRO 45m map. The ratio map was made with the integrated intensity maps.
}
\label{fig:ratio}
\end{figure*}

\begin{figure*}
\includegraphics[width=0.8\textwidth]{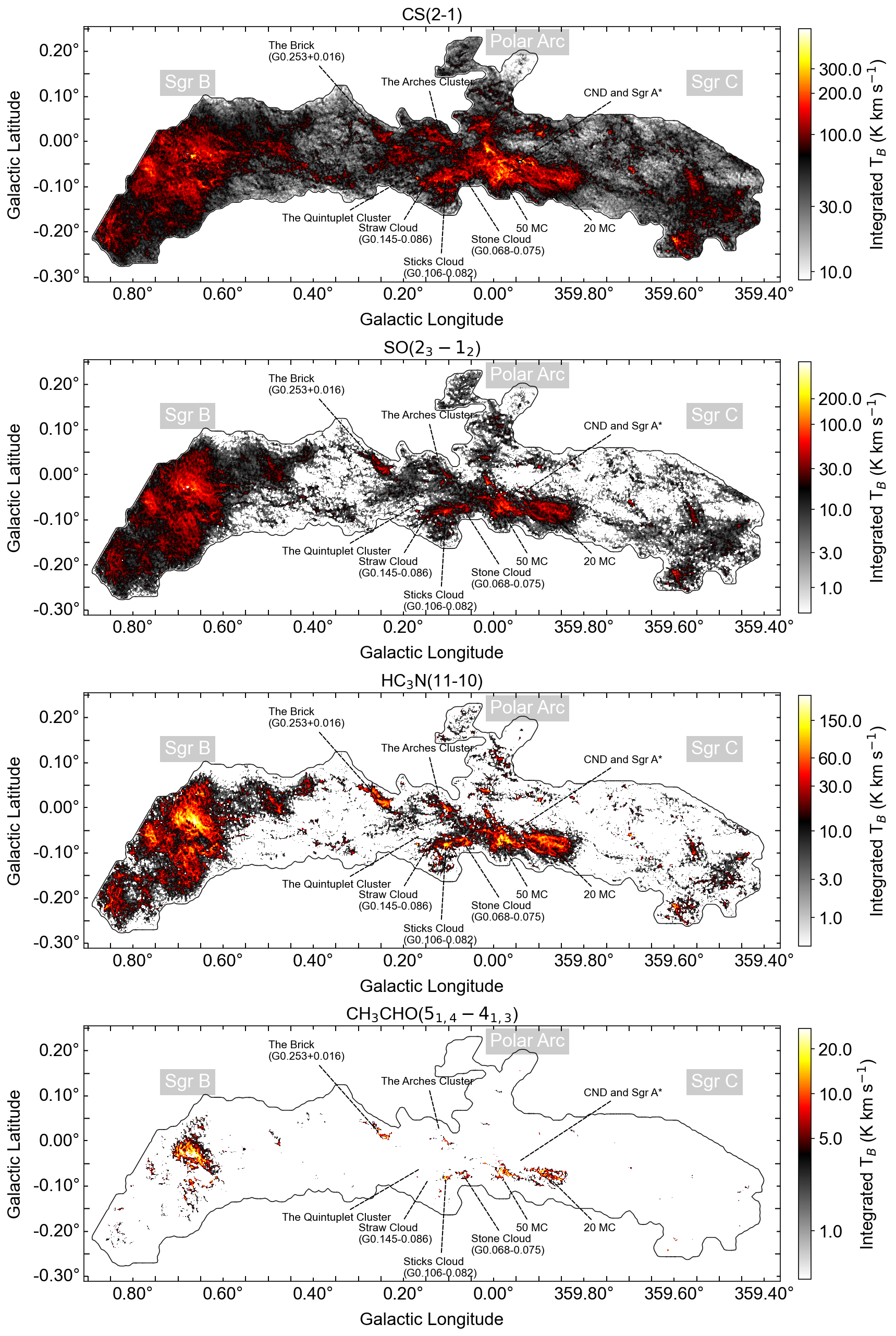}
\caption{
ALMA integrated intensity maps (\texttt{mom 0}) of \cs, \so, \hctn and \chtcho lines towards the CMZ. The nomenclature of previously identified features are labeled. The common beam size is 2.84\arcsec{} $\times$ 2.11\arcsec{} with a position angle of $-$89.46\arcdeg{} for the \cs, \so, and \chtcho lines. The common beam size is 2.65\arcsec{} $\times$ 1.97\arcsec{} with a position angle of $-$89.96\arcdeg{} for the \hctn line. Note that the acetaldehyde line  (\chtcho)  maps are integrated with \chtcho E and \chtcho A.
}
\label{fig:m0_mosaic}
\end{figure*}

\begin{figure*}
\includegraphics[width=0.8\textwidth]{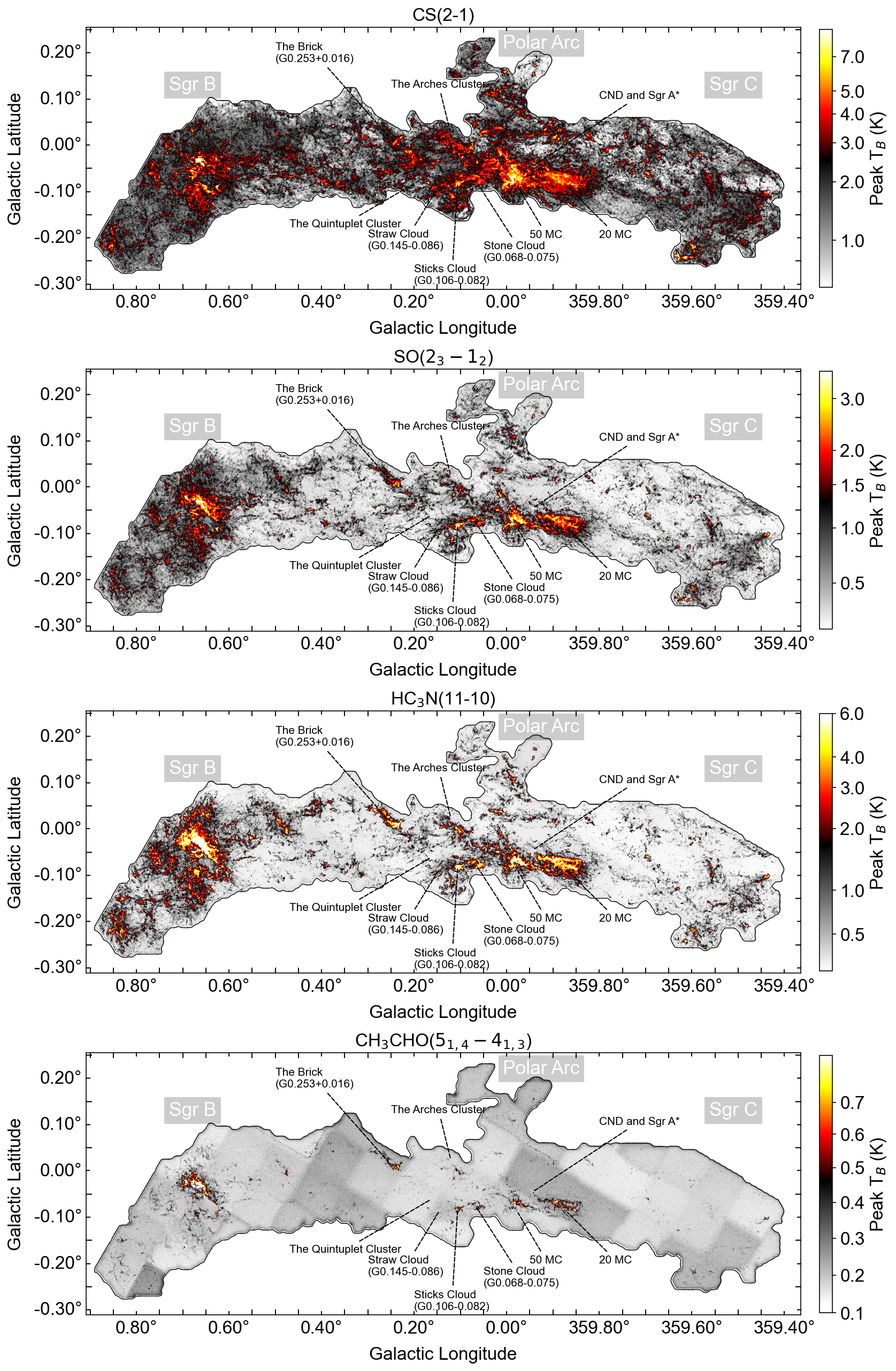}
\caption{
ALMA peak intensity maps (\texttt{mom 8}) of  \cs, \so, \hctn and \chtcho  lines towards the CMZ. Note that the acetaldehyde line  (\chtcho)  maps are stacked with \chtcho E and \chtcho A (also see Figure~\ref{fig:spectrum}).
}
\label{fig:m8_mosaic}
\end{figure*}

\begin{figure*}
\includegraphics[width=1.0\textwidth]{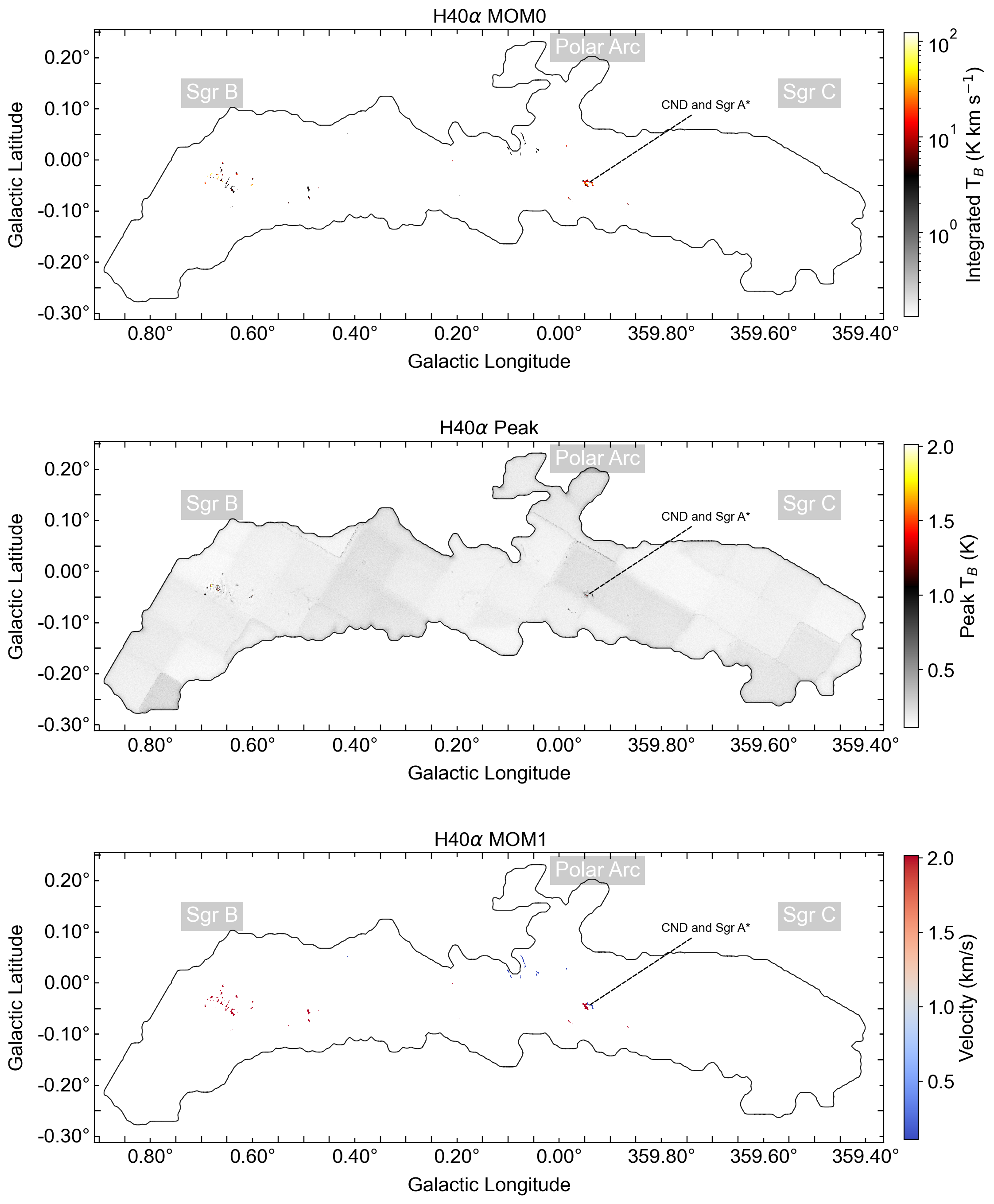}
\caption{
ALMA mosaic map of \ha line towards the CMZ. From the top to the bottom we present the \texttt{mom 0}, peak intensity, and \texttt{mom 1} images. The common beam size is 2.84\arcsec{} $\times$ 2.11\arcsec{} with a position angle of $-$89.46\arcdeg{}.
}
\label{fig:h40a_mosaic}
\end{figure*}

\begin{figure*}
\includegraphics[width=0.85\textwidth]{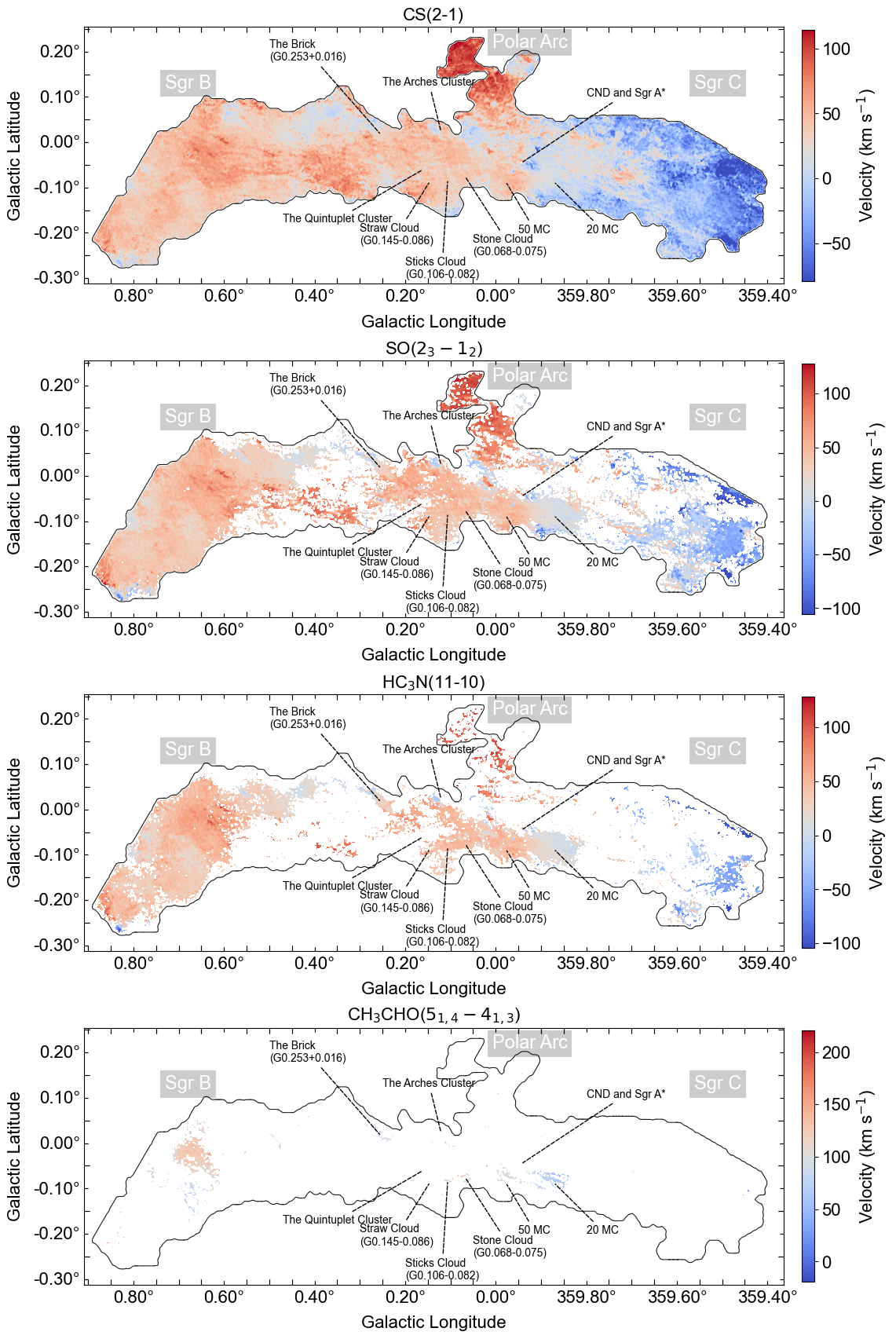}
\caption{
The intensity weighted velocity map (\texttt{mom 1}) of \cs, \so, \hctn and \chtcho lines towards the CMZ.
}
\label{fig:m1_cmz}
\end{figure*}

\section{Observations and Data Processing}
\label{sec:obs}
Forty-five individual mosaics were used to image the central $1.5^\circ \times 0.5^\circ$ of the CMZ with the ALMA Band~3 receiver and the 12-m array. The 7-m and TP arrays were also employed to recover the short-spacing information. The overall survey design and observational setup are described in \citet[][hereafter \citetalias{Longmore2025}]{Longmore2025}, while details of the data calibration and imaging procedures are given in \citet[][hereafter \citetalias{Ginsburg2025}; \citetalias{Walker2025}]{Ginsburg2025,Walker2025}. The data product releases of the continuum and spectral-line cubes are presented in \citetalias{Ginsburg2025}, \citetalias{Walker2025}, \citet[][hereafter \citetalias{Lu2025}]{Lu2025}, and in this work.

We use the CASA package \citep{CASATeam2022} for calibration and imaging.
The majority of the data were processed with CASA version 6.2.1.7 and version 6.4.1.12 for the later dataset, as the project extended across different observing cycles and pipeline version releases. The full details of the CASA versions used can be found in \citetalias{Ginsburg2025}.  In this paper we present the spectral cubes of the broad spectral windows (SPW 33 and SPW 35 in Table 1 of \citetalias{Walker2025}).
The spectrum of broad spectral window towards the water maser core of the Brick cloud is shown in Figure~\ref{fig:spectrum}. The conversion factors from Jy beam$^{-1}$ to K are 21.2, 20.7, 23.4, and 20.8 for \cs, \so, \hctn and \chtcho, respectively. Detailed descriptions of the data issues, continuum subtraction, deconvolution (image reconstruction), array combination, and mosaic procedures are provided in \citetalias{Walker2025} and in the  \texttt{merge\_tclean\_commands.py} script in the ACES GitHub repository\footnote{\url{https://github.com/ACES-CMZ/reduction_ACES}}.

The CASA task  \texttt{feather} was used to combine the images from the 12 m, 7 m, and TP arrays for the data release. Section 3.5 of \citetalias{Walker2025} provides a detailed discussion of various array combination methods using CASA. In this work, we adopted the “feather-only” approach as described in Section 3.5.3 of \citetalias{Walker2025}. Additionally, \citetalias{Lu2025} discussed alternative combination methods using Miriad \citep{Sault1995}.
The native channel width of the broad spectral window is 1.48 km s$^{-1}$ with an effective bandwidth of $\sim1.86$ GHz ($\sim$ 5680 km s$^{-1}$). The image rms noise per 1.48~\kms{} channel is not uniform across the mosaics, but varies between regions typically in the range of 3--5 mJy beam$^{-1}$. The cube statistics are shown in Table~\ref{tab:cubestats_spw33-35}. The id of each field can also be found in \citetalias{Walker2025} and \citetalias{Ginsburg2025}. For the purpose of mosaic, image cubes of individual fields were  smoothed to a common beam to produce the final, full-CMZ mosaics. The common beam size is applied to the mosaic maps, while most of the individual fields have high angular resolution, the final beam is compromised with the lowest resolution of field am. The common beam sizes of the SPW 33 and SPW 35 are 2.84\arcsec{} $\times$ 2.11\arcsec{} (position angle=$-$89.46\arcdeg{}) and 2.65\arcsec{} $\times$ 1.97\arcsec{} (position angle=$-$89.96\arcdeg{}), respectively.


We compare the ACES \cs data with the NRO~45 m \cs data \citep{Hsieh2016}. The spatial and spectral resolutions of the 45 m data are 38$''$ and 2.5~km~s$^{-1}$, respectively. The total bandwidth of the 45,m data is 400~km~s$^{-1}$, which is narrower than that of the ACES data. The 45\,m image was obtained in units of $T_{\rm A}^*$, and we converted it to $T_{\rm B}$ by dividing by the main-beam efficiency of 0.43~\citep{Hsieh2016}.
To compare with the ACES map, we generated \texttt{mom0} maps for both datasets using the same method described in Section~\ref{subsec:moment}, integrating over the same velocity range and smoothing to a common beam size of 38$''$. The ratio image is shown in Figure~\ref{fig:ratio}. The mean ratio across the masked region is approximately 0.9, indicating that our flux calibration is consistent with the 45 m data. Elevated ratios near Sgr~C may be affected by edge effects in the maps, while the high values near Sgr~A* are likely due to residual continuum emission in the ACES data.

\section{ACES Data Products and Structural Highlights}
\label{sec:products}
\subsection{Moment Maps of CMZ Mosaics}
\label{subsec:moment}
We present the integrated intensity (\texttt{mom0}) and peak intensity (\texttt{mom8} mosaic images (a nomenclature within the immoments task of CASA) of these four molecular lines in Figure~\ref{fig:m0_mosaic},  Figure~\ref{fig:m8_mosaic} and  Figure~\ref{fig:h40a_mosaic}. The zoomed-in images of \ha are presented in Figure~\ref{fig:h40a_sgra}.

Note that the \chtcho \texttt{mom0} map includes both the \chtcho~E and \chtcho~A transitions to improve the signal-to-noise ratio. These two lines are close in frequency, separated by approximately $\pm$150~km~s$^{-1}$. If high-velocity components (e.g., $\pm$150~km~s$^{-1}$), such as those from the circumnuclear disk (CND) are present, the lines may become spectrally blended in our data. However, since the CND is not clearly detected in either transition, the \chtcho \texttt{mom0} map shown in Figure~\ref{fig:m0_mosaic} is not affected by blending from high-velocity components. Users can generate individual \texttt{mom0} maps for each transition directly from the mosaic cube. Details on generating moment maps can be found in \citetalias{Walker2025}.

The moment maps provide a detailed view of the spatial distribution of emission across the CMZ. In general the ALMA maps show a combination of extended emission and compact features. 
The \so and \cs maps reveal a smooth and diffuse emission structure, highlighting extended regions of molecular gas across the CMZ. In contrast, the  \chtcho map is more concentrated towards the Brick, the 20  km s$^{-1}$ cloud, the 50 km s$^{-1}$ cloud, and the Sgr B region. The \hctn\ emission exhibits an intermediate morphology, sharing characteristics of both the extended \so\ and \cs\ emission and the compact \chtcho\ structures.
The \ha map displays distinct, compact emission structures that are characteristic of ionized gas. A zoomed-in view of the \ha map with significant detection is shown in Figure~\ref{fig:h40a_sgra}. Together, these maps provide a multi-line view of the gas properties, distinguishing between ionized and molecular components, and reveal the physical and chemical conditions of the observed region. However, since the molecular lines (\cs, \chtcho, \so, and \hctn) primarily trace molecular gas that is spatially anti-correlated with the ionized gas traced by \ha (cf. Figure~\ref{fig:2d-matrix}), a comparison between the \ha emission and the continuum is more physically relevant. Therefore, we do not discuss the \ha properties in detail in this paper. 
Readers are referred to \citet{Ginsburg2025} for a detailed description of the continuum emission. Although that work does not discuss the H40$\alpha$ line, a direct comparison with the maps presented here may help to examine the spatial relation between ionized and molecular components.

The characteristic kinetic temperature in the CMZ has been reported as 65$\pm$10~K by \citet{Ao2013} and within the range of 50–100~K \citep{Ginsburg2016, Krieger2017}. As shown in Figure~\ref{fig:m8_mosaic}, \cs exhibits a peak brightness temperature of approximately 10~K, particularly toward the Sgr~A and Sgr~B regions. This relatively high brightness temperature, comparable to the typical kinetic temperature in the CMZ, suggests that \cs emission is generally optically thick or moderately thick, tracing gas with significant optical depth across most regions.

In contrast, \so  and \hctn  show peak brightness temperatures of about 4–6~K. Although lower than that of \cs, these values indicate that the lines are moderately optically thin to moderately thick, likely reflecting somewhat lower column densities or smaller beam filling factors compared to \cs.

Meanwhile, \chtcho shows a peak brightness temperature of $\leq$1~K, indicating that its emission might be optically thin across most of the mapped region, similar to the HC$^{15}$N line reported in \citetalias{Lu2025}. With additional SO transitions available in the ACES data \citepalias{Longmore2025}, including the isotopologue line $^{34}$SO ($2_{3}$–$1_{2}$), which is expected to be optically thin, as well as SO ($2_{2}$–$1_{1}$) and SO ($5_{4}$–$4_{4}$), future studies can derive detailed temperature and density distributions to better characterize the excitation conditions in various regions of the CMZ.


\subsection{\texttt{mom 1} map and longitude-velocity maps}
In Figure~\ref{fig:m1_cmz}, we present the \texttt{mom 1} images of four molecular lines of the CMZ. The \texttt{mom 1} map of \ha is shown separately in Figure~\ref{fig:h40a_mosaic}. Despite being derived from high-resolution ALMA data, the velocity fields reveal large-scale coherent rotation, with redshifted emission at positive longitudes and blueshifted emission at negative longitudes, consistent with previous single-dish observations \citep[e.g.,][]{Jones2012}. On smaller scales, several regions show overlapping red- and blueshifted components, indicative of multiple gas streamers or distinct kinematic features along the line of sight \citep[e.g.,][]{Henshaw2016b, Lipman2025, Walker2025b}.

\begin{figure*}
\includegraphics[width=0.7\textwidth]{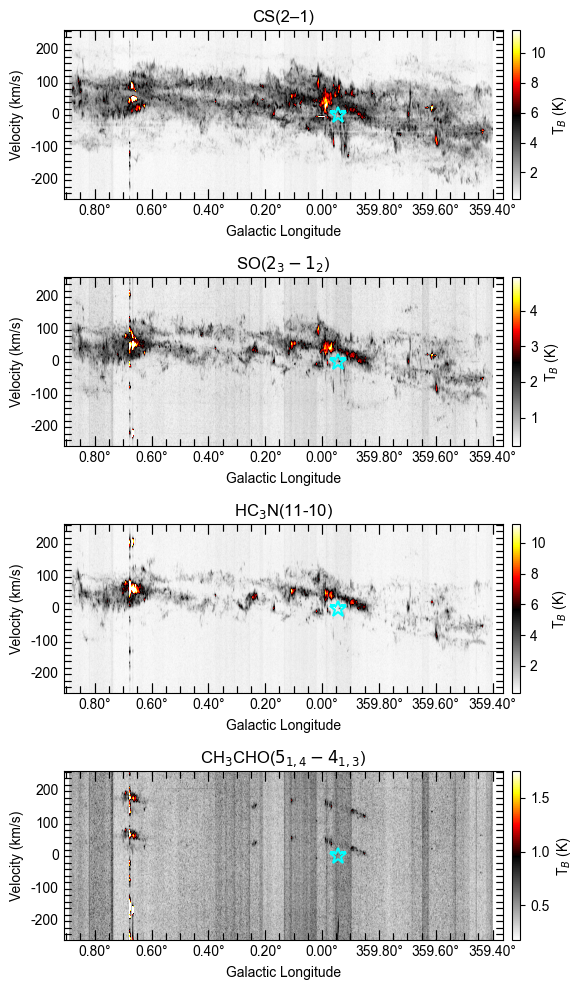}
\caption{
The Longitude-velocity diagram of \cs, \so, \hctn and \chtcho lines towards the CMZ. The location of Sgr A* is labeled with the cyan star. 
The \chtcho E  ($f_{\rm rest}=98.86331$ GHz). and \chtcho A ($f_{\rm rest}=98.9005$ GHz) can be seen together in the diagram, where we set the rest frequency of \chtcho A.
}
\label{fig:pv_cmz}
\end{figure*}

In Figure~\ref{fig:pv_cmz}, we present the longitude–velocity ($l$–$v$) diagrams of the four molecular lines showing intensity peaks. In the \cs line, the horizontal absorption features at $-52$, $-28$, and $-3$~km~s$^{-1}$ are due to foreground absorption, which is attributed to gas in spiral arms along the line of sight at these velocities \citep{Jones2012}. The same features can be seen in other abundant molecules such as HCO$^{+}$ \citepalias{Walker2025}. However, these foreground absorption features are not found in the other lines presented in this paper.

\begin{figure*}
\centering
\raisebox{0mm}{\includegraphics[width=0.45\textwidth]{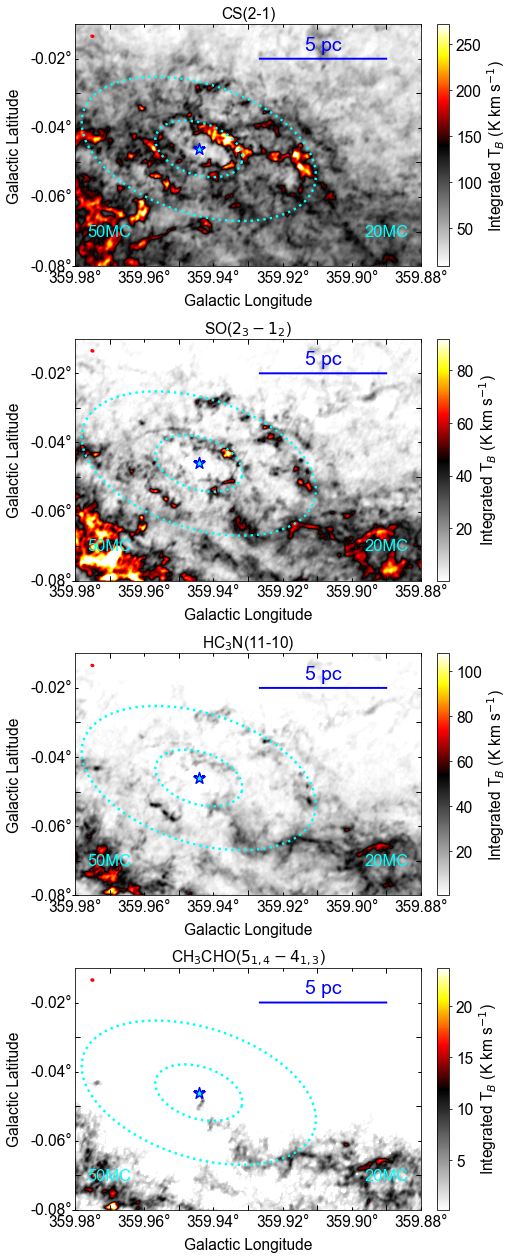}}
\includegraphics[width=0.42\textwidth]{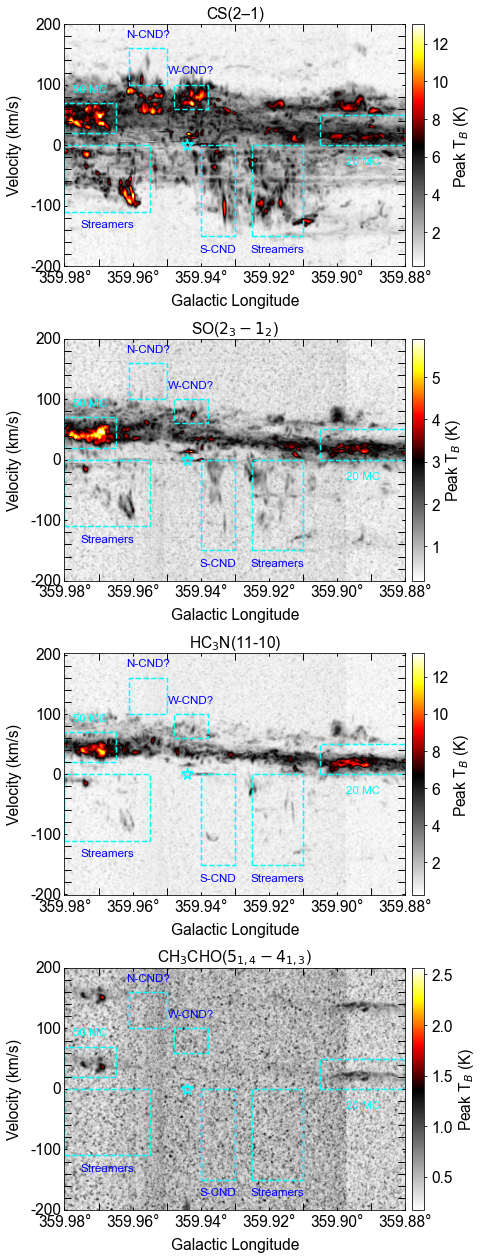}
\caption{
Left: the \texttt{mom 0} image of the CND. Sgr A* is marked with a cyan star, and the CND is outlined with a small cyan dotted ellipse. The streamers are outlined with the big ellipse \citep{Liu2013,Hsieh2017}.  The \cs, \so, \hctn and \chtcho lines  images are presented. The beam is shown in the top left corner as a red filled eliipse. Right: the longitude-velocity diagrams of the four molecular lines zoomed-in to the region of the CND.
}
\label{fig:m0_mosaic_cnd}
\end{figure*}

These complex motions associated with sub-pc scale filaments and broad-linewidth clumps observed in the velocity map are more clearly revealed in the $l$–$v$ diagrams.
Towards the Sgr A complex region, in Figure~\ref{fig:m0_mosaic_cnd}, we present the $l$–$v$ diagram of the CND. The streamers are shown as in the previous work by \citet{Hsieh2017}. The high-velocity gas standing out from the CMZ ring (associated with the 50 and 20 km s$^{-1}$ clouds in this figure) represents the CND and the streamers, which are kinematically distinct in both velocity structure and spatial location. These features also exhibit significantly broader linewidths ($\ge30$ km s$^{-1}$) compared to the surrounding CMZ gas, further highlighting their dynamic nature associated with tidal disruption in the deep potential well. The high spectral and spatial resolution mosaic map will enable study the infall of gas further out.

In comparison to the $l$–$v$ diagrams in \citet{Jones2012} at 39$''$ resolution, while the Mopra data successfully traces the large-scale spiral arm structure and provides a comprehensive view of the galactic-scale kinematics across the CMZ at sub-pc scale. The low spatial resolution limits the study of motion of gas connected the CMZ and the CND. The broad linewidths clumps present in several places of the CMZ are also marginally resolved in the Mopra data. The broad linewidth clumps characteristic of the turbulent and shocked gas in this environment, leaving the physical origin and role of these velocity components unclear. ALMA data represent a crucial intermediate scale that bridges the gap between the large-scale dynamics captured by single-dish surveys and the sub-pc clumps, showing the necessity of high angular resolution observations to fully characterize the complex interplay between inflow, turbulence, and star formation in Galactic center environments.

\begin{figure*}
\centering
\includegraphics[width=1\textwidth]{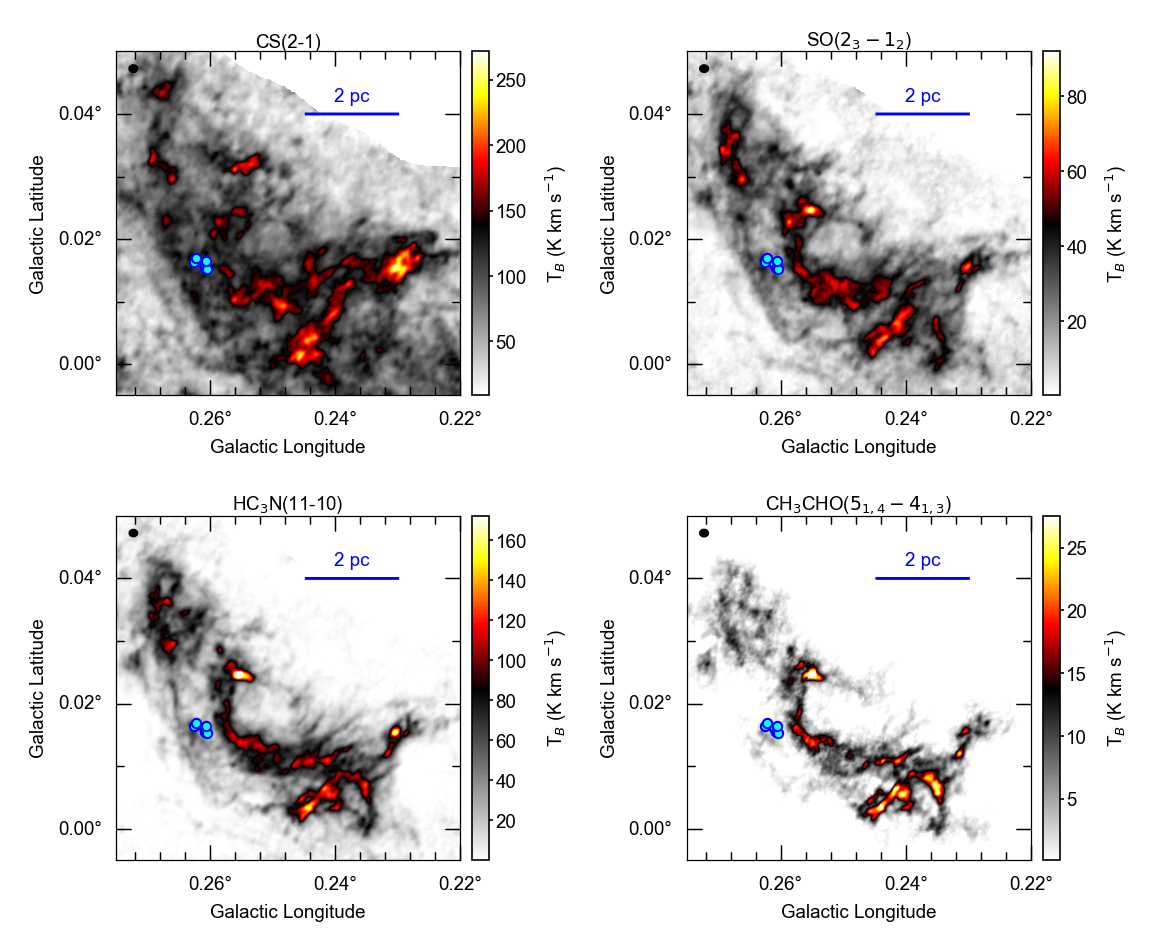}
\caption{
The \texttt{mom 0} images of the Brick cloud (G0.253+0.016).  The \cs, \so, \hctn and \chtcho line images are presented. The outflow positions featuring ongoing star formation are labeled with the cyan circles \citep{Walker2021}. The synthesized beam is shown on the top left corner with a red ellipse.
}
\label{fig:m0_mosaic_brick}
\end{figure*}

\begin{figure*}
\centering
\includegraphics[width=1\textwidth]{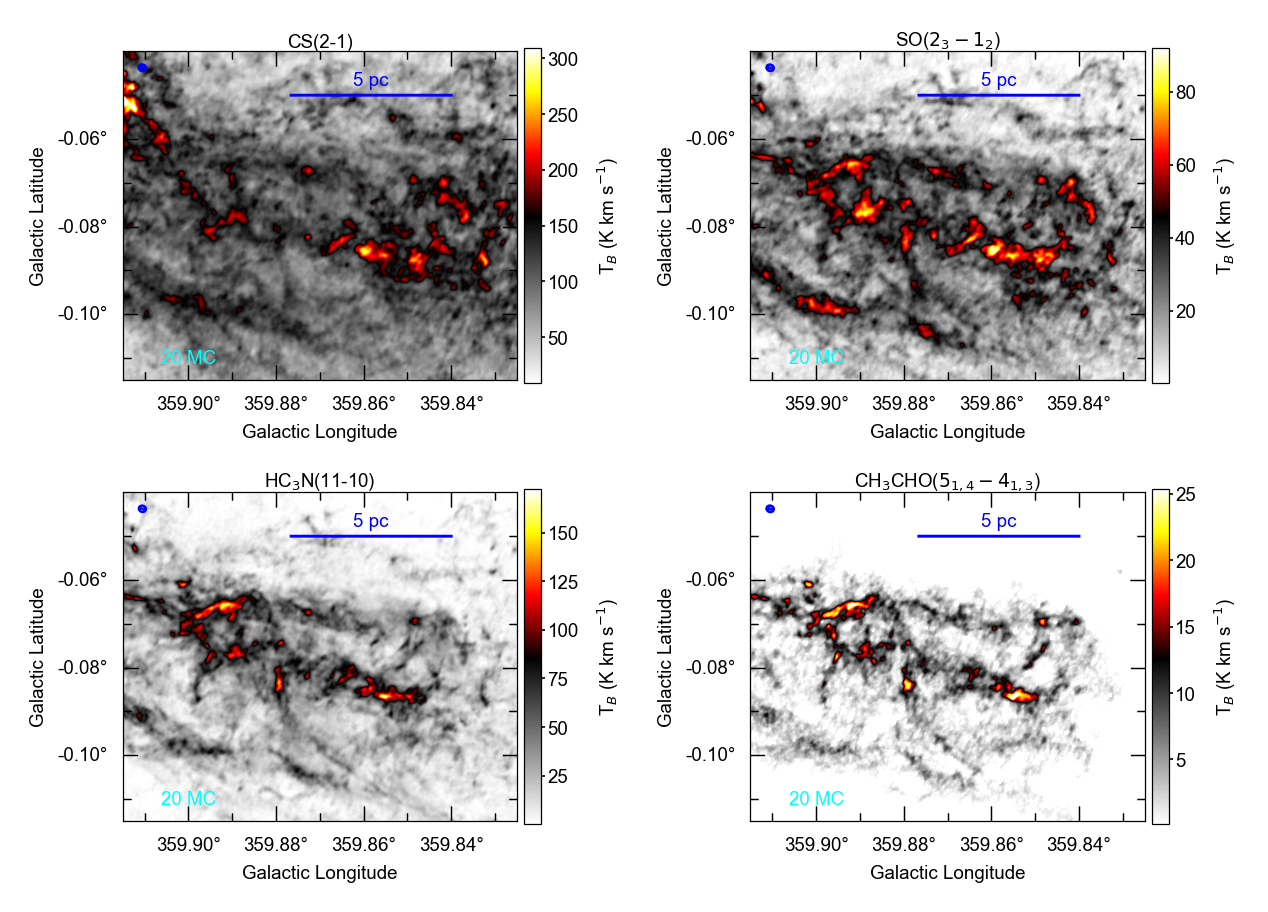}
\caption{
The \texttt{mom 0} images of the 20 km s$^{-1}$ cloud. The \cs, \so, \hctn and \chtcho line images are presented. The synthesized beam is shown on the top left corner with a blue ellipse.
}
\label{fig:m0_mosaic_20mc}
\end{figure*}

\begin{figure*}
\centering
\includegraphics[width=1\textwidth]{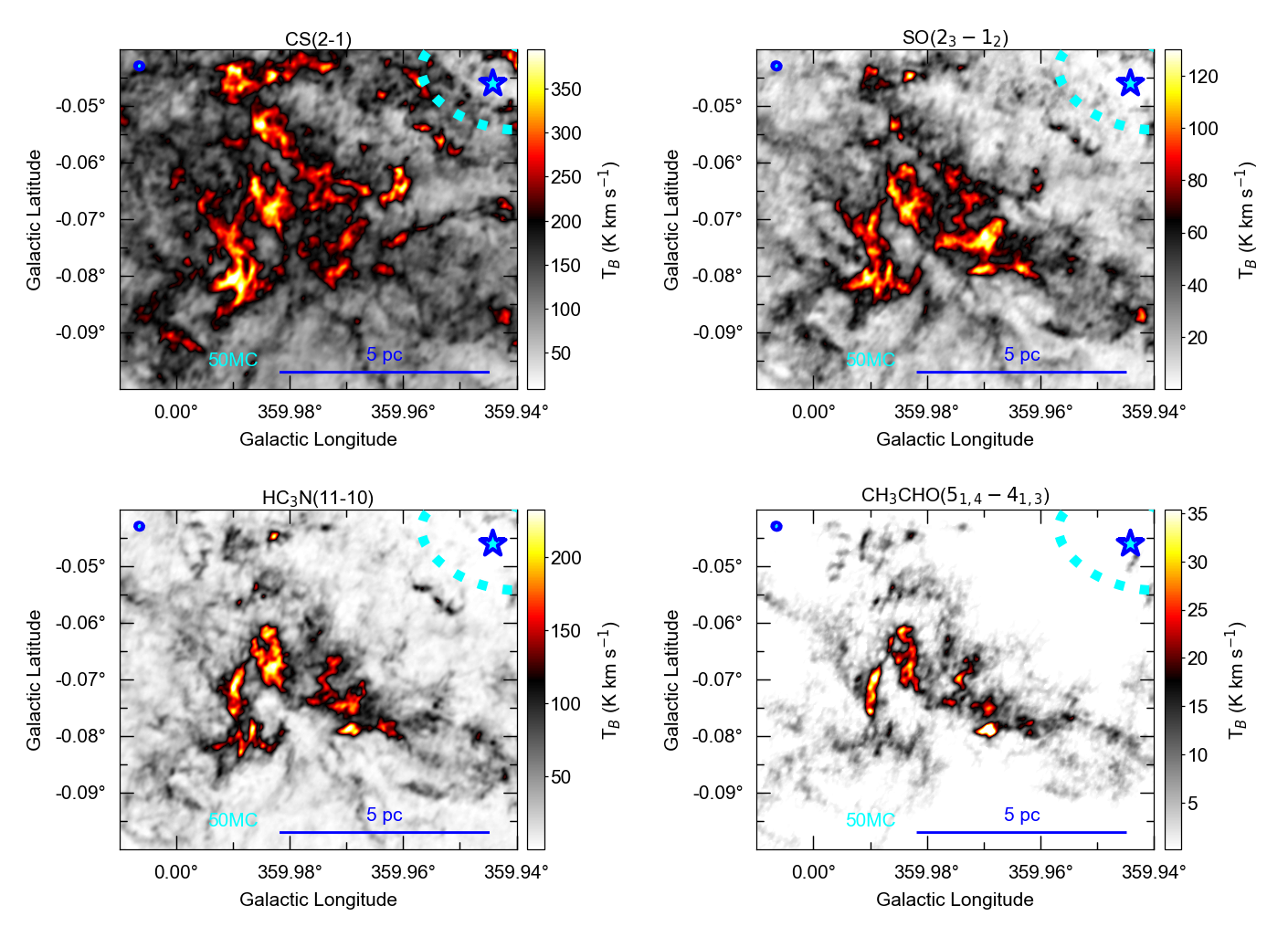}
\caption{
The \texttt{mom 0} images of the 50 km s$^{-1}$ cloud. The Sgr A* is labeled with the blue star. The CND is outlined with a cyan dotted line. The \cs, \so, \hctn and \chtcho line images are presented. The synthesized beam is shown on the top left corner with a blue ellipse.
}
\label{fig:m0_mosaic_50mc}
\end{figure*}

\begin{figure*}
\includegraphics[width=0.8\textwidth]{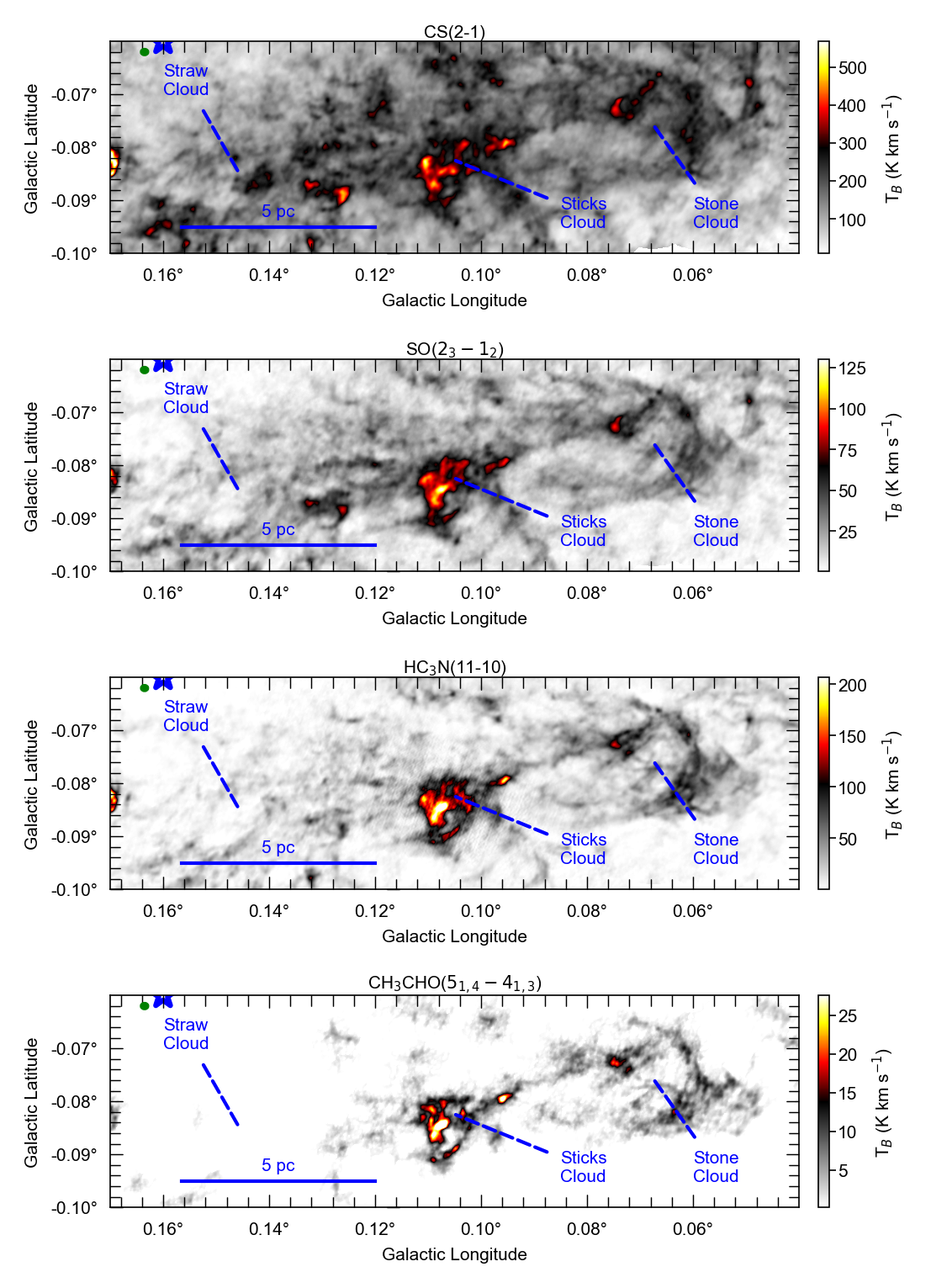}
\caption{
The \texttt{mom 0} images of the three little pigs cloud (stone, sticks, and straw cloud). The \cs, \so, \hctn and \chtcho line images are presented. The synthesized beam is shown on the top left corner with a blue ellipse.
}
\label{fig:m0_mosaic_pigs}
\end{figure*}

\subsection{Key Structures}

In Figures~\ref{fig:m0_mosaic_cnd}, \ref{fig:m0_mosaic_brick}, \ref{fig:m0_mosaic_20mc}, \ref{fig:m0_mosaic_50mc}, and \ref{fig:m0_mosaic_pigs}, we show zoomed-in images of the four molecular lines toward the well-known features interesting for star formation in the CMZ, namely, the CND, the Brick cloud, the 20 and 50 km s$^{-1}$ clouds, and the Three Little Pigs (Sticks, Straw, and Stone clouds).

\subsubsection{The circumnuclear disk}
The circumnuclear disk (CND) is a molecular ring rotating around Sgr A*. It contains a molecular gas mass of approximately $10^{4}$ M$_{\odot}$ \citep[e.g.][]{RequenaTorres2012,Mills2013,Hsieh2021} and with a radius of $\sim$2 pc. Within this region lie the ionized streamers known as the mini-spiral \citep[or Sgr A West;][]{Zhao2009} (Figure~\ref{fig:h40a_sgra} in this paper) and the nuclear star cluster \citep{Figer1999b}. The streamers surrounding the CND are thought to be tidally stretched gas originating from nearby clouds (the 50 and 20 km s$^{-1}$ clouds) that is accreting onto the CND. As the most massive and closest molecular reservoir to Sgr A*, the CND is of great interest for understanding both the fueling of the central black hole and nuclear star formation. 
The ACES high-resolution mosaic of the nuclear region enables studies of gas inflow kinematics and chemical conditions from the CMZ to the CND \citep[e.g.][]{Sofue2025}.


In Figure~\ref{fig:m0_mosaic_cnd}, the CND and the streamers \citep{Liu2013,Hsieh2017,Hsieh2021} are outlined with small and large cyan dotted ellipses, respectively. Both the CND and the streamers are detected in the \cs, \so, and faint emission with \hctn lines; however, only the streamers visually connecting to the 20 km s$^{-1}$ cloud are detected in the \chtcho line.

In the vicinity of the nuclear star cluster and Sgr A*, UV fields can photodissociate and deplete HC$_3$N, although the molecule may still survive in dense, well-shielded clumps \citep{Rodriguez-Fernandez1998, Meier2005, Martin2012, Harada2015, Lindberg2011}. The weak \hctn emission in the ACES map could therefore arise from a combination of UV photodissociation and excitation effects, as the warm gas in the CND may preferentially populate higher-$J$ transitions. To quantify the relative effect of UV shielding and excitation, future observations of high-excitation HC$_3$N lines and detailed abundance analyses will be required, which we leave for community follow-up.

\subsubsection{The Brick} G0.253+0.016, an infrared dark cloud known as ''the Brick'', is one of the most massive and dense molecular clouds in the CMZ. It has a low dust temperature of 20~K, a $n_{\rm H_2}$ of $\ge10^{4}$~cm$^{-3}$, and a large gas mass of $\ge10^{5}$~M$_\odot$ within size of 2--3~pc. It is considered an ideal site for forming high-mass stars \citep{Lis1994a, Lis2001}. Despite having the right conditions for star formation, it shows little to no star formation activity \citep[e.g.][]{Immer2012a, Kauffmann2013, Rathborne2014a}. Although the Brick does not yet harbor fully formed massive stars or clusters, previous ALMA observations have shown evidence of ongoing intermediate- to high-mass star formation, including the detection of seven outflow features \citep{Walker2021}.
    
\chtcho and \so were previously imaged towards the Brick cloud by \citet{Rathborne2015} with ALMA. Although the sensitivity and resolution are slightly higher in their data, we have imaged additional \cs and \hctn lines with a larger mosaic. In Figure~\ref{fig:m0_mosaic_brick}, the Brick cloud exhibits arm-like internal structures \citep{Johnston2014}, and \chtcho emission is relatively brighter in the southern part of the cloud. While both \cs and \so exhibit widespread emission tracing the filaments and seem to be immersed in the diffuse arm-like structures in the CMZ, the emission from \chtcho line is much more compact and spatially confined.

The \chtcho emission appears to be well separated from the diffuse large-scale arms and instead traces distinct clumps concentrated within the central ridges of the Brick. The \chtcho line exhibits morphologies similar to those of \hctn, tracing a coherent network of compact clumps and ridges. These features are especially prominent along the southern ridge and central arc. This visual similarity is quantified by a high 2D morphological correlation coefficient of 0.9 (Figure~\ref{fig:2d-matrix}, more description in \citetalias{Lu2025}). The \so line also shows significant overlap with \hctn, capturing many of the same structures, though with slightly more diffuse emission. The \so--\hctn correlation is also high, at 0.94, indicating strong morphological similarity. In contrast, the \cs emission displays a more extended and diffuse distribution, particularly toward the northern regions. It shows weaker overlap with both \chtcho  (correlation coefficient=0.35) and \hctn  (correlation coefficient=0.66), but aligns relatively well with \so (correlation coefficient=0.8). The variation in internal structures among the four lines suggests that the compact features within the Brick vary chemically and physically.


\subsubsection{The 20 and 50 km s$^{-1}$ cloud}
The 20 km s$^{-1}$ cloud and the 50 km s$^{-1}$ cloud are prominent giant molecular clouds in the CMZ \citep[e.g.,][]{Uehara2019,Lu2021}, specifically, near the Sgr A*. The 20 km s$^{-1}$ cloud shows signs of early-stage star formation, including the detection of protostellar outflows in the SiO(5--4) line \citep{Lu2021}. In contrast, the 50~km~s$^{-1}$ cloud does not exhibit clear protostellar outflow signatures and currently has low star formation rate \citep[$\le0.3\times10^{-3}$ M$_\odot$, ][]{Lu2021}, but it is likely in a later stage of star formation \citep{Mills2011}, as it lies adjacent to the compact H II region complex G$-0.02-0.07$ (Figure~\ref{fig:h40a_sgra} in this paper) \citep{Tsuboi2019, Mills2011}, which shares a similar velocity structure. While the 50 km s$^{-1}$ cloud is currently inactive in star formation, \citet{Miyawaki2021} reported the detection of 28 hot molecular core candidates, which are in quasi-static equilibrium.

Figures~\ref{fig:m0_mosaic_20mc} and \ref{fig:m0_mosaic_50mc} show the detection of \chtcho\ emission toward the 20 and 50 km s$^{-1}$ clouds, respectively. As in the Brick cloud, CH$_3$CHO shows a high correlation with HC$_3$N (0.9), a moderate correlation with SO (0.69), and a low correlation with CS (0.34).

\subsubsection{The three little pigs (TLP)}
The Three Little Pigs (TLP) cloud complex consists of four molecular clouds known as the Straw (M0.145-0.086), the Sticks (M0.106-0.082) , and the Stone (M0.068-0.075) \citep{Battersby2020} from the left to the right side in Figure~\ref{fig:m0_mosaic_pigs}. Consistent with the previous SMA 1 mm image \citep{Battersby2020}, the morphology of the TLP clouds appears to become progressively more complex from the straw cloud to the stone cloud. The straw cloud is very faint in lines, possibly quiescent or less evolved. The sticks cloud appears to be the most prominent in these four molecular lines, showing compact clumps in all three tracers. Similar to the aforementioned clouds, we find a consistent trend in the morphological correlations among these four lines: CH$_3$CHO shows the highest correlation with HC$_3$N (0.89), an intermediate correlation with SO (0.77), and the lowest correlation with CS (0.59).

\section{Morphological Correlations Between the \cs, \so, \hctn, \chtcho and SiO(2-1)}

In Figure~\ref{fig:2d-matrix}, we find that \chtcho shows the highest correlation with \hctn (0.84–0.9 across the four highlighted regions), an intermediate correlation with \so (0.64–0.69), and the weakest correlation with \cs (0.25–0.59). This consistent pattern across all regions suggests that \chtcho emission is more spatially similar to \hctn than with \cs.
Although the CS molecule is widely used dense-gas tracer because of the large dipole moment and high critical densities, its high abundance often leads to significant optical depth, limiting its reliability for tracing the most compact, high-column structures. \citet{Pety2017} showed that CS emission mainly traces gas at intermediate column densities and saturates in the densest regions, while CS(1–0) in the CMZ shows moderate optical depths of $\tau\lesssim2$–3 \citep{Tsuboi1999}, consistent with its widespread distribution across the CMZ clouds.

The critical densities of \cs and \so are roughly an order of magnitude lower than those of \chtcho and \hctn. Although we have not directly derived the gas densities here, at first glance, the correlation coefficients suggest that \cs and \so trace more extended dense gas, whereas \chtcho and \hctn highlight more compact, denser substructures. However, \chtcho emission does not necessarily trace dense cores exclusively \citep{Ikeda2001,Jorgensen2016,Codella2017,Oya2017}; given its relatively low abundance and the subthermal excitation of its low-$J$ transitions at moderate densities ($\sim10^{4-5}$~cm$^{-3}$), its detection may be biased toward high-column-density regions that exceed the sensitivity threshold rather than reflecting a true absence elsewhere \citep[e.g.,][]{RequenaTorres2006, JimenezSerra2025a}. Observations of optically thin isotopologues such as HC$^{15}$N \citepalias{Walker2025} could therefore help disentangle optical-depth effects and confirm the effectiveness of these molecules as dense-gas tracers.


In Figure~\ref{fig:2d-matrix}, \so, \hctn, and \chtcho show stronger correlations with SiO than with CS across the four regions and in the CMZ average \citepalias{Walker2025}. SiO is a well-established shock tracer \citep{Schilke1997}, and its widespread CMZ emission likely reflects large-scale shocks driven by, for example, expanding H,\textsc{ii} regions or supernova remnants \citep{Inutsuka2021,Henshaw2022}, cloud–cloud collisions \citep{Tsuboi2015a,Zeng2020}, or Galactic shear \citep{Kruijssen2014a,Petkova2021}. Additional processes such as cosmic-ray heating \citep{Yusef-Zadeh2013d} and X-ray irradiation \citep{Martin-Pintado2000} may also influence Si-bearing chemistry.

Sulphur-bearing species such as SO can be enhanced in shock and post-shock gas \citep{Pineau1993,Lis2001,Lu2017,Callanan2023}, although SO is not unique to shocks and also appears in relatively quiescent gas in both low- and high-mass star-forming regions \citep[e.g.][]{Hacar2013}. \hctn is a well-known tracer of dense, early-phase cores, forming mainly through reactions involving CN and C$_2$H$_2$ radicals \citep[e.g.][]{Walmsley1980,Suzuki1992}, but its abundance can also increase in low-velocity shocks associated with stellar outflows \citep[e.g.,][]{Bachiller1997,Mendoza2018,Hoque2025}. \chtcho is typically linked to hot cores and outflows in the Galactic disk \citep{Ikeda2001,Jorgensen2016,Codella2017,Oya2017}, yet in the CMZ, COMs including \chtcho are widespread and associated with low-velocity shocks largely independent of local star formation activity \citep{RequenaTorres2006,JimenezSerra2025b,Li2025}.

Across these species, the correlation strengths with SiO—SO–SiO (0.90) $>$ HC$_3$N–SiO (0.83) $>$ CS–SiO (0.79) $>$ CH$_3$CHO–SiO (0.67) suggest different physical and chemical conditions. A detailed determination of SiO abundances and dedicated chemical modeling, along with further examination of the associations among \so, \hctn, and \chtcho, will be essential to better understand the extreme environment of the CMZ. These analyses are beyond the scope of this paper and the in-depth studies are left for future studies.

\section{Summary}

In this paper, we present an overview of the ACES broadband spectral-line data and the public release of the \cs, \so, \chtcho, \ha, and \hctn\ mosaic images. We provide the corresponding moment maps, longitude–velocity ($l$–$v$) diagrams, and a morphological correlation analysis comparing 14 lines with the 3 mm continuum \citepalias{Ginsburg2025, Walker2025, Lu2025}.
The emissions from these molecules are widely distributed across the CMZ. The \cs\ and \so\ lines exhibit more extended spatial distributions, whereas \hctn\ and \chtcho\ are more concentrated toward the denser regions. Comparisons with other lines, such as SO transitions and isotopologues \citepalias{Longmore2025} will also be useful for constraining the temperature and density.

The morphological correlation analysis reveals systematic differences in how molecular species trace the physical and chemical structures within the CMZ. The stronger correlations of \so, \hctn, and \chtcho\ with SiO indicate that these species are more closely linked to shocked gas. These results highlight the potential of the CMZ as a laboratory for studying the properties and history of large-scale shocks and their impact on molecular chemistry. Although we have not determined SiO or other shock-related abundances here, the observed correlations provide valuable constraints for future investigations at high spatial resolution. In addition to shocks, other energetic drivers such as cosmic rays, X rays, and photodissociation regions (PDRs), as indicated by the H II regions traced by H40$\alpha$, are also likely to influence the chemical and thermal state of the gas.  A comprehensive study of these processes, combining chemical modeling and multi-line diagnostics, will be essential to deepen our understanding of the interplay between dynamics and chemistry in the CMZ.

\section*{Acknowledgments}

We thank the anonymous referee for the careful reading and constructive comments that helped improve this paper.
This paper makes use of the following ALMA data: ADS/JAO.ALMA\#2021.1.00172.L. ALMA is a partnership of ESO (representing its member states), NSF (USA) and NINS (Japan), together with NRC (Canada), NSTC and ASIAA (Taiwan), and KASI (Republic of Korea), in cooperation with the Republic of Chile. The Joint ALMA Observatory is operated by ESO, AUI/NRAO and NAOJ. The authors are grateful to the staff throughout the ALMA organisation, particularly those at the European ALMA Regional Centre, the Joint ALMA Observatory, and the UK ALMA Regional Centre Node, for their extensive support, which was essential to the success of this challenging Large Program.

Data analysis was in part carried out on the Multi-wavelength Data Analysis System operated by the Astronomy Data Center (ADC), National Astronomical Observatory of Japan.
D.L.W gratefully acknowledges support from the UK ALMA Regional Centre (ARC) Node, which is supported by the Science and Technology Facilities Council [grant numbers ST/Y004108/1 and ST/T001488/1].
R.F. acknowledges support from the grants PID2023-146295NB-I00, and from the Severo Ochoa grant CEX2021-001131-S funded by MCIN/AEI/ 10.13039/501100011033 and by ``European Union NextGenerationEU/PRTR''. 
I.J-.S., L.C., and V.M.R. acknowledge support from the grant PID2022-136814NB-I00 by the Spanish Ministry of Science, Innovation and Universities/State Agency of Research MICIU/AEI/10.13039/501100011033 and by ERDF, UE. I.J-.S. also acknowledges the ERC Consolidator grant OPENS (project number 101125858) funded by the European Union. V.M.R. also acknowledges the grant RYC2020-029387-I funded by MICIU/AEI/10.13039/501100011033 and by "ESF, Investing in your future", and from the Consejo Superior de Investigaciones Cient{\'i}ficas (CSIC) and the Centro de Astrobiolog{\'i}a (CAB) through the project 20225AT015 (Proyectos intramurales especiales del CSIC); and from the grant CNS2023-144464 funded by MICIU/AEI/10.13039/501100011033 and by “European Union NextGenerationEU/PRTR.
The authors acknowledge UFIT Research Computing for providing computational resources and support that have contributed to the research results reported in this publication. 
AG acknowledges support from the NSF under grants CAREER 2142300, AAG 2008101, and particularly AAG 2206511 that supports the ACES large program.
M.G.S.-M.\ acknowledges support from the NSF under grant CAREER 2142300. M.G.S.-M.\ also thanks the Spanish MICINN for funding support under grant PID2023-146667NB-I00.
A.S.-M.\ acknowledges support from the RyC2021-032892-I grant funded by MCIN/AEI/10.13039/501100011033 and by the European Union `Next GenerationEU’/PRTR, as well as the program Unidad de Excelencia María de Maeztu CEX2020-001058-M, and support from the PID2023-146675NB-I00 (MCI-AEI-FEDER, UE).
KMD acknowledges support from the European Research Council (ERC) Advanced Grant MOPPEX 833460.vii.
X.L.\ acknowledges support from the Strategic Priority Research Program of the Chinese Academy of Sciences (CAS) Grant No.\ XDB0800300, the National Key R\&D Program of China (No.\ 2022YFA1603101), State Key Laboratory of Radio Astronomy and Technology (CAS), the National Natural Science Foundation of China (NSFC) through grant Nos.\ 12273090 and 12322305, the Natural Science Foundation of Shanghai (No.\ 23ZR1482100), and the CAS ``Light of West China'' Program No.\ xbzg-zdsys-202212.
Q. Z. gratefully acknowledges the support from the National Science Foundation under Award No. AST-2206512, and the Smithsonian Institute FY2024 Scholarly Studies Program.
FHL acknowledges support from the ESO Studentship Programme, the Scatcherd European Scholarship of the University of Oxford, and the European Research Council’s starting grant ERC StG-101077573 (`ISM-METALS').
C.~F.~acknowledges funding provided by the Australian Research Council (Discovery Projects DP230102280 and DP250101526), and the Australia-Germany Joint Research Cooperation Scheme (UA-DAAD).
F. M acknowledges financial support from the School of Astronomy at the Institute for Research in Fundamental Sciences-IPM.
J.K. is supported by the Royal Society under grant number RF\textbackslash ERE\textbackslash231132, as part of project URF\textbackslash R1\textbackslash211322.
MCS acknowledges financial support from the European Research Council under the ERC Starting Grant ``GalFlow'' (grant 101116226) and from Fondazione Cariplo under the grant ERC attrattivit\`{a} n. 2023-3014.
M.C.\ gratefully acknowledges funding from the DFG through an Emmy Noether Research Group (grant number CH2137/1-1).
COOL Research DAO \citep{Chevance2025} is a Decentralized Autonomous Organization supporting research in astrophysics aimed at uncovering our cosmic origins.
J.Wallace gratefully acknowledges funding from National Science Foundation under Award Nos. 2108938 and 2206510.
E.A.C.\ Mills  gratefully  acknowledges  funding  from the National  Science  Foundation  under  Award  Nos. 1813765, 2115428, 2206509, and CAREER 2339670.
F.N.-L. gratefully acknowledges financial support from grant PID2024-162148NA-I00, funded by MCIN/AEI/10.13039/501100011033 and the European Regional Development Fund (ERDF) “A way of making Europe”, from the Ramón y Cajal programme (RYC2023-044924-I) funded by MCIN/AEI/10.13039/501100011033 and FSE+, and from the Severo Ochoa grant CEX2021-001131-S, funded by MCIN/AEI/10.13039/501100011033.

\section*{Software}

This paper made use of the following software packages: \textbf{CASA} \citep{CASATeam2022} to carry out calibration and imaging of the ALMA data, \textbf{CARTA} \citep{carta} to visualise and analyse the data, \textbf{astropy} \citep{astropy:2013, astropy:2018, astropy:2022} to analyse the images and make plots, \textbf{spectral-cube} \citep{Ginsburg2019spectralcube} to process the image cubes and produce the moment maps, \textbf{radio-beam} \citep{Koch2025radiobeam} to determine (1) a common beam among multiple channels in a cube, (2) to smooth the individual channel beams to this common resolution, (3) to determine the Jy-K conversion, and \textbf{APLpy} \citep{aplpy} and \textbf{seaborn} \citep{Waskom2021} to make plots.

\section*{Data Availability}

All data products and associated documentation can be found at \url{https://almascience.org/alma-data/lp/aces}. To maximise usability and provide a single, comprehensive resource for users, we also provide an online, machine-readable table, which contains the full, detailed list of all data products released for the entire survey, including hyperlinks to the full files from the ALMA archive.

All code and data processing issues are available at the public GitHub repository here: \url{https://github.com/ACES-CMZ/reduction_ACES}.

The ACES data reduction was a monolithic work that was written up in 5 papers.  If you use the ACES data, please cite the appropriate works, which includes \citet{Longmore2025} and the data papers: continuum \citep{Ginsburg2025} and cubes in high-resolution (this work), medium-resolution \citep{Lu2025}, and low-resolution \citep{Hsieh2025} spectral windows. Note that all papers should be cited for use of any data; the continuum imaging relied on the line papers, and vice-versa.

\bibliographystyle{mnras}
\bibliography{refs}

@ARTICLE{Pety2017,
       author = {{Pety}, J{\'e}r{\^o}me and {Guzm{\'a}n}, Viviana V. and {Orkisz}, Jan H. and {Liszt}, Harvey S. and {Gerin}, Maryvonne and {Bron}, Emeric and {Bardeau}, S{\'e}bastien and {Goicoechea}, Javier R. and {Gratier}, Pierre and {Le Petit}, Franck and {Levrier}, Fran{\c{c}}ois and {{\"O}berg}, Karin I. and {Roueff}, Evelyne and {Sievers}, Albrecht},
        title = "{The anatomy of the Orion B giant molecular cloud: A local template for studies of nearby galaxies}",
      journal = {\aap},
         year = 2017,
        month = jan,
       volume = {599},
          eid = {A98},
        pages = {A98},
          doi = {10.1051/0004-6361/201629862},
archivePrefix = {arXiv},
       eprint = {1611.04037},
 primaryClass = {astro-ph.GA},
       adsurl = {https://ui.adsabs.harvard.edu/abs/2017A&A...599A..98P}
}

@ARTICLE{Pineau1993,
       author = {{Pineau des Forets}, G. and {Roueff}, E. and {Schilke}, P. and {Flower}, D.~R.},
        title = "{Sulphur-bearing molecules as tracers of shocks in interstellar clouds}",
      journal = {\mnras},
         year = 1993,
        month = jun,
       volume = {262},
       number = {4},
        pages = {915-928},
          doi = {10.1093/mnras/262.4.915},
       adsurl = {https://ui.adsabs.harvard.edu/abs/1993MNRAS.262..915P}
}

@ARTICLE{Chevance2025,
       author = {{Chevance}, M\'elanie and {Kruijssen}, J.~M. Diederik and {Longmore}, Steven~N.},
        title = "{COOL Research DAO Whitepaper - Towards community-owned astrophysics for everyone}",
      journal = {arXiv e-prints},
         year = 2025,
        month = dec,
          eid = {arXiv:2501.13160},
        pages = {arXiv:2501.13160},
          doi = {10.48550/arXiv.2501.13160},
archivePrefix = {arXiv},
       eprint = {2501.13160},
 primaryClass = {astro-ph.IM},
       adsurl = {https://ui.adsabs.harvard.edu/abs/2025arXiv250113160C}
}

@ARTICLE{Liu2013,
       author = {{Liu}, Hauyu Baobab and {Ho}, Paul T.~P. and {Wright}, Melvyn C.~H. and {Su}, Yu-Nung and {Hsieh}, Pei-Ying and {Sun}, Ai-Lei and {Kim}, Sungsoo S. and {Minh}, Young Chol},
        title = "{Interstellar Medium Processing in the Inner 20 pc in Galactic Center}",
      journal = {\apj},
         year = 2013,
        month = jun,
       volume = {770},
       number = {1},
          eid = {44},
        pages = {44},
          doi = {10.1088/0004-637X/770/1/44},
archivePrefix = {arXiv},
       eprint = {1304.7573},
 primaryClass = {astro-ph.GA},
       adsurl = {https://ui.adsabs.harvard.edu/abs/2013ApJ...770...44L}
}

@ARTICLE{Sofue2025,
       author = {{Sofue}, Yoshiaki and {Oka}, Tomoharu and {Longmore}, Steven N. and {Walker}, Daniel and {Ginsburg}, Adam and {Henshaw}, Jonathan D. and {Bally}, John and {Barnes}, Ashley T. and {Battersby}, Cara and {Colzi}, Laura and {Ho}, Paul and {Jimenez-Serra}, Izaskun and {Kruijssen}, J.~M. Diederik and {Mills}, Elizabeth and {Petkova}, Maya A. and {Sormani}, Mattia C. and {Wallace}, Jennifer and {Armijos-Abenda{\~n}o}, Jairo and {Dutkowska}, Katarzyna M. and {Enokiya}, Rei and {Garc{\'\i}a}, Pablo and {Gramze}, Savannah and {Henkel}, Christian and {Hsieh}, Pei-Ying and {Hu}, Yue and {Immer}, Katharina and {Iwata}, Yuhei and {Karoly}, Janik and {Klessen}, Ralf S. and {Kohno}, Kotaro and {Krumholz}, Mark R. and {Lipman}, Dani and {Morris}, Mark R. and {Nogueras-Lara}, Francisco and {Pineda}, Jaime E. and {Mart{\'\i}n}, Sergio and {Requenatorres}, Miguel Angel and {Rivilla}, V{\'\i}ctor M. and {Riquelme-V{\'a}squez}, Denise and {S{\'a}nchez-Monge}, {\'A}lvaro and {Santa-Maria}, Miriam G. and {Smith}, Howard A. and {Tolls}, Volker and {Wang}, Q. Daniel},
        title = "{Circum-nuclear eccentric gas flow in the Galactic Center revealed by ALMA CMZ Exploration Survey (ACES)}",
      journal = {arXiv e-prints},
         year = 2025,
        month = jun,
          eid = {arXiv:2506.11553},
        pages = {arXiv:2506.11553},
          doi = {10.48550/arXiv.2506.11553},
archivePrefix = {arXiv},
       eprint = {2506.11553},
 primaryClass = {astro-ph.GA},
       adsurl = {https://ui.adsabs.harvard.edu/abs/2025arXiv250611553S}
}

@ARTICLE{Tacconi2008,
       author = {{Tacconi}, L.~J. and {Genzel}, R. and {Smail}, I. and {Neri}, R. and {Chapman}, S.~C. and {Ivison}, R.~J. and {Blain}, A. and {Cox}, P. and {Omont}, A. and {Bertoldi}, F. and {Greve}, T. and {F{\"o}rster Schreiber}, N.~M. and {Genel}, S. and {Lutz}, D. and {Swinbank}, A.~M. and {Shapley}, A.~E. and {Erb}, D.~K. and {Cimatti}, A. and {Daddi}, E. and {Baker}, A.~J.},
        title = "{Submillimeter Galaxies at z \raisebox{-0.5ex}\textasciitilde 2: Evidence for Major Mergers and Constraints on Lifetimes, IMF, and CO-H$_{2}$ Conversion Factor}",
      journal = {\apj},
         year = 2008,
        month = jun,
       volume = {680},
       number = {1},
        pages = {246-262},
          doi = {10.1086/587168},
archivePrefix = {arXiv},
       eprint = {0801.3650},
 primaryClass = {astro-ph},
       adsurl = {https://ui.adsabs.harvard.edu/abs/2008ApJ...680..246T}
}

@ARTICLE{Figer1999b,
   author = {{Figer}, D.~F. and {Kim}, S.~S. and {Morris}, M. and {Serabyn}, E. and 
	{Rich}, R.~M. and {McLean}, I.~S.},
    title = "{Hubble Space Telescope/NICMOS Observations of Massive Stellar Clusters near the Galactic Center}",
  journal = {\apj},
   eprint = {astro-ph/9906299},
     year = 1999,
    month = nov,
   volume = 525,
    pages = {750-758},
      doi = {10.1086/307937},
   adsurl = {http://adsabs.harvard.edu/abs/1999ApJ...525..750F}
}

@ARTICLE{Zhao2009,
   author = {{Zhao}, J.-H. and {Morris}, M.~R. and {Goss}, W.~M. and {An}, T.
	},
    title = "{Dynamics of Ionized Gas at the Galactic Center: Very Large Array Observations of the Three-dimensional Velocity Field and Location of the Ionized Streams in Sagittarius A West}",
  journal = {\apj},
archivePrefix = "arXiv",
   eprint = {0904.3133},
 primaryClass = "astro-ph.GA",
     year = 2009,
    month = jul,
   volume = 699,
    pages = {186-214},
      doi = {10.1088/0004-637X/699/1/186},
   adsurl = {http://adsabs.harvard.edu/abs/2009ApJ...699..186Z}
}

@ARTICLE{RequenaTorres2012,
   author = {{Requena-Torres}, M.~A. and {G{\"u}sten}, R. and {Wei{\ss}}, A. and 
	{Harris}, A.~I. and {Mart{\'{\i}}n-Pintado}, J. and {Stutzki}, J. and 
	{Klein}, B. and {Heyminck}, S. and {Risacher}, C.},
    title = "{GREAT confirms transient nature of the circum-nuclear disk}",
  journal = {\aap},
archivePrefix = "arXiv",
   eprint = {1203.6687},
 primaryClass = "astro-ph.GA",
     year = 2012,
    month = jun,
   volume = 542,
      eid = {L21},
    pages = {L21},
      doi = {10.1051/0004-6361/201219068},
   adsurl = {http://adsabs.harvard.edu/abs/2012A%26A...542L..21R}
}

@ARTICLE{Harada2015,
       author = {{Harada}, N. and {Riquelme}, D. and {Viti}, S. and {Jim{\'e}nez-Serra}, I. and {Requena-Torres}, M.~A. and {Menten}, K.~M. and {Mart{\'\i}n}, S. and {Aladro}, R. and {Martin-Pintado}, J. and {Hochg{\"u}rtel}, S.},
        title = "{Chemical features in the circumnuclear disk of the Galactic center}",
      journal = {\aap},
         year = 2015,
        month = dec,
       volume = {584},
          eid = {A102},
        pages = {A102},
          doi = {10.1051/0004-6361/201526994},
archivePrefix = {arXiv},
       eprint = {1510.02904},
 primaryClass = {astro-ph.GA},
       adsurl = {https://ui.adsabs.harvard.edu/abs/2015A&A...584A.102H}
}

@ARTICLE{JimenezSerra2025a,
       author = {{Jimenez-Serra}, Izaskun and {Codella}, Claudio and {Belloche}, Arnaud},
        title = "{Observations of complex organic molecules in the gas phase of the interstellar medium}",
      journal = {arXiv e-prints},
         year = 2025,
        month = mar,
          eid = {arXiv:2503.17104},
        pages = {arXiv:2503.17104},
          doi = {10.48550/arXiv.2503.17104},
archivePrefix = {arXiv},
       eprint = {2503.17104},
 primaryClass = {astro-ph.GA},
       adsurl = {https://ui.adsabs.harvard.edu/abs/2025arXiv250317104J}
}

@ARTICLE{JimenezSerra2025b,
       author = {{Jimenez-Serra}, Izaskun},
        title = "{Chemistry in the Galactic Center}",
      journal = {arXiv e-prints},
         year = 2025,
        month = jan,
          eid = {arXiv:2501.01782},
        pages = {arXiv:2501.01782},
          doi = {10.48550/arXiv.2501.01782},
archivePrefix = {arXiv},
       eprint = {2501.01782},
 primaryClass = {astro-ph.GA},
       adsurl = {https://ui.adsabs.harvard.edu/abs/2025arXiv250101782J}
}

@ARTICLE{Yusef-Zadeh2013d,
       author = {{Yusef-Zadeh}, F. and {Wardle}, M. and {Lis}, D. and {Viti}, S. and {Brogan}, C. and {Chambers}, E. and {Pound}, M. and {Rickert}, M.},
        title = "{74 MHz Nonthermal Emission from Molecular Clouds: Evidence for a Cosmic Ray Dominated Region at the Galactic Center}",
      journal = {Journal of Physical Chemistry A},
         year = 2013,
        month = oct,
       volume = {117},
       number = {39},
        pages = {9404-9419},
          doi = {10.1021/jp311240h},
archivePrefix = {arXiv},
       eprint = {1305.1047},
 primaryClass = {astro-ph.GA},
       adsurl = {https://ui.adsabs.harvard.edu/abs/2013JPCA..117.9404Y}
}

@ARTICLE{Martin-Pintado2000,
       author = {{Mart{\'\i}n-Pintado}, J. and {de Vicente}, P. and {Rodr{\'\i}guez-Fern{\'a}ndez}, N.~J. and {Fuente}, A. and {Planesas}, P.},
        title = "{A correlation between the SiO and the Fe 6.4 keV line emission from the Galactic center}",
      journal = {\aap},
         year = 2000,
        month = apr,
       volume = {356},
        pages = {L5-L8},
          doi = {10.48550/arXiv.astro-ph/0003063},
archivePrefix = {arXiv},
       eprint = {astro-ph/0003063},
 primaryClass = {astro-ph},
       adsurl = {https://ui.adsabs.harvard.edu/abs/2000A&A...356L...5M}
}

@ARTICLE{Ikeda2001,
       author = {{Ikeda}, M. and {Ohishi}, M. and {Nummelin}, A. and {Dickens}, J.~E. and {Bergman}, P. and {Hjalmarson}, {\r{A}}. and {Irvine}, W.~M.},
        title = "{Survey Observations of c-C$_{2}$H$_{4}$O and CH$_{3}$CHO toward Massive Star-forming Regions}",
      journal = {\apj},
         year = 2001,
        month = oct,
       volume = {560},
       number = {2},
        pages = {792-805},
          doi = {10.1086/322957},
       adsurl = {https://ui.adsabs.harvard.edu/abs/2001ApJ...560..792I}
}

@ARTICLE{RequenaTorres2006,
       author = {{Requena-Torres}, M.~A. and {Mart{\'\i}n-Pintado}, J. and {Rodr{\'\i}guez-Franco}, A. and {Mart{\'\i}n}, S. and {Rodr{\'\i}guez-Fern{\'a}ndez}, N.~J. and {de Vicente}, P.},
        title = "{Organic molecules in the Galactic center. Hot core chemistry without hot cores}",
      journal = {\aap},
         year = 2006,
        month = sep,
       volume = {455},
       number = {3},
        pages = {971-985},
          doi = {10.1051/0004-6361:20065190},
archivePrefix = {arXiv},
       eprint = {astro-ph/0605031},
 primaryClass = {astro-ph},
       adsurl = {https://ui.adsabs.harvard.edu/abs/2006A&A...455..971R}
}

@ARTICLE{Oya2017,
       author = {{Oya}, Yoko and {Sakai}, Nami and {Watanabe}, Yoshimasa and {Higuchi}, Aya E. and {Hirota}, Tomoya and {L{\'o}pez-Sepulcre}, Ana and {Sakai}, Takeshi and {Aikawa}, Yuri and {Ceccarelli}, Cecilia and {Lefloch}, Bertrand and {Caux}, Emmanuel and {Vastel}, Charlotte and {Kahane}, Claudine and {Yamamoto}, Satoshi},
        title = "{L483: Warm Carbon-chain Chemistry Source Harboring Hot Corino Activity}",
      journal = {\apj},
         year = 2017,
        month = mar,
       volume = {837},
       number = {2},
          eid = {174},
        pages = {174},
          doi = {10.3847/1538-4357/aa6300},
archivePrefix = {arXiv},
       eprint = {1703.03653},
 primaryClass = {astro-ph.SR},
       adsurl = {https://ui.adsabs.harvard.edu/abs/2017ApJ...837..174O}
}

@ARTICLE{Codella2017,
       author = {{Codella}, C. and {Ceccarelli}, C. and {Caselli}, P. and {Balucani}, N. and {Barone}, V. and {Fontani}, F. and {Lefloch}, B. and {Podio}, L. and {Viti}, S. and {Feng}, S. and {Bachiller}, R. and {Bianchi}, E. and {Dulieu}, F. and {Jim{\'e}nez-Serra}, I. and {Holdship}, J. and {Neri}, R. and {Pineda}, J.~E. and {Pon}, A. and {Sims}, I. and {Spezzano}, S. and {Vasyunin}, A.~I. and {Alves}, F. and {Bizzocchi}, L. and {Bottinelli}, S. and {Caux}, E. and {Chac{\'o}n-Tanarro}, A. and {Choudhury}, R. and {Coutens}, A. and {Favre}, C. and {Hily-Blant}, P. and {Kahane}, C. and {Jaber Al-Edhari}, A. and {Laas}, J. and {L{\'o}pez-Sepulcre}, A. and {Ospina}, J. and {Oya}, Y. and {Punanova}, A. and {Puzzarini}, C. and {Quenard}, D. and {Rimola}, A. and {Sakai}, N. and {Skouteris}, D. and {Taquet}, V. and {Testi}, L. and {Theul{\'e}}, P. and {Ugliengo}, P. and {Vastel}, C. and {Vazart}, F. and {Wiesenfeld}, L. and {Yamamoto}, S.},
        title = "{Seeds of Life in Space (SOLIS). II. Formamide in protostellar shocks: Evidence for gas-phase formation}",
      journal = {\aap},
         year = 2017,
        month = sep,
       volume = {605},
          eid = {L3},
        pages = {L3},
          doi = {10.1051/0004-6361/201731249},
archivePrefix = {arXiv},
       eprint = {1708.04663},
 primaryClass = {astro-ph.EP},
       adsurl = {https://ui.adsabs.harvard.edu/abs/2017A&A...605L...3C}
}

@ARTICLE{Jorgensen2016,
       author = {{J{\o}rgensen}, J.~K. and {van der Wiel}, M.~H.~D. and {Coutens}, A. and {Lykke}, J.~M. and {M{\"u}ller}, H.~S.~P. and {van Dishoeck}, E.~F. and {Calcutt}, H. and {Bjerkeli}, P. and {Bourke}, T.~L. and {Drozdovskaya}, M.~N. and {Favre}, C. and {Fayolle}, E.~C. and {Garrod}, R.~T. and {Jacobsen}, S.~K. and {{\"O}berg}, K.~I. and {Persson}, M.~V. and {Wampfler}, S.~F.},
        title = "{The ALMA Protostellar Interferometric Line Survey (PILS). First results from an unbiased submillimeter wavelength line survey of the Class 0 protostellar binary IRAS 16293-2422 with ALMA}",
      journal = {\aap},
         year = 2016,
        month = nov,
       volume = {595},
          eid = {A117},
        pages = {A117},
          doi = {10.1051/0004-6361/201628648},
archivePrefix = {arXiv},
       eprint = {1607.08733},
 primaryClass = {astro-ph.SR},
       adsurl = {https://ui.adsabs.harvard.edu/abs/2016A&A...595A.117J}
}

@ARTICLE{Miyawaki2021,
       author = {{Miyawaki}, Ryosuke and {Tsuboi}, Masato and {Uehara}, Kenta and {Miyazaki}, Atsushi},
        title = "{Hot molecular core candidates in the Galactic center 50 km s$^{-1}$ molecular cloud}",
      journal = {\pasj},
         year = 2021,
        month = aug,
       volume = {73},
       number = {4},
        pages = {943-969},
          doi = {10.1093/pasj/psab056},
archivePrefix = {arXiv},
       eprint = {2105.09449},
 primaryClass = {astro-ph.GA},
       adsurl = {https://ui.adsabs.harvard.edu/abs/2021PASJ...73..943M}
}

@ARTICLE{Lipman2025,
       author = {{Lipman}, Dani and {Battersby}, Cara and {Walker}, Daniel L. and {Sormani}, Mattia C. and {Bally}, John and {Barnes}, Ashley and {Ginsburg}, Adam and {Glover}, Simon C.~O. and {Henshaw}, Jonathan D. and {Hatchfield}, H. Perry and {Immer}, Katharina and {Klessen}, Ralf S. and {Longmore}, Steven N. and {Mills}, Elisabeth A.~C. and {Smith}, Rowan and {Tress}, R.~G. and {Alboslani}, Danya and {Zhang}, Qizhou},
        title = "{3D CMZ. IV. Distinguishing Near versus Far Distances in the Galactic Center Using Spitzer and Herschel}",
      journal = {\apj},
         year = 2025,
        month = may,
       volume = {984},
       number = {2},
          eid = {159},
        pages = {159},
          doi = {10.3847/1538-4357/adb5ee},
archivePrefix = {arXiv},
       eprint = {2410.17321},
 primaryClass = {astro-ph.GA},
       adsurl = {https://ui.adsabs.harvard.edu/abs/2025ApJ...984..159L}
}

@ARTICLE{Hsieh2025,
       author = {{Hsieh}, P.},
        title = "{}",
      journal = {ArXiv e-prints},
         year = 2025,
        month = jan,
     note    = {In preparation}
}

@ARTICLE{Longmore2025,
       author = {{Longmore}, Steven and {ACES Team}},
        title = "{ACES overview paper}",
      journal = {\apj},
         year = 2025,
        month = jun,
       volume = {1},
       number = {1},
          eid = {1},
        pages = {1},
          doi = {10.1088/0000-0000}
}

@ARTICLE{Ginsburg2025,
       author = {{Ginsburg}, Adam and {ACES Team}},
        title = "{ACES data release paper I}",
      journal = {\apj},
         year = 2025,
        month = jun,
       volume = {1},
       number = {1},
          eid = {1},
        pages = {1},
          doi = {10.1088/0000-0000}
}

@ARTICLE{Walker2025,
       author = {{Walker}, Dan and {ACES Team}},
        title = "{ACES data release paper II}",
      journal = {\apj},
         year = 2025,
        month = jun,
       volume = {1},
       number = {1},
          eid = {1},
        pages = {1},
          doi = {10.1088/0000-0000}
}

@ARTICLE{Lu2025,
       author = {{Lu}, X.},
        title = "{}",
      journal = {in prep.},
         year = 2025,
        month = jan,
     note    = {In preparation}
}

@ARTICLE{Walker2025b,
       author = {{Walker}, Daniel L. and {Battersby}, Cara and {Lipman}, Dani and {Sormani}, Mattia C. and {Ginsburg}, Adam and {Glover}, Simon C.~O. and {Henshaw}, Jonathan D. and {Longmore}, Steven N. and {Klessen}, Ralf S. and {Immer}, Katharina and {Alboslani}, Danya and {Bally}, John and {Barnes}, Ashley and {Hatchfield}, H. Perry and {Mills}, Elisabeth A.~C. and {Smith}, Rowan and {Tress}, Robin G. and {Zhang}, Qizhou},
        title = "{3D CMZ. III. Constraining the 3D Structure of the Central Molecular Zone via Molecular Line Emission and Absorption}",
      journal = {\apj},
         year = 2025,
        month = may,
       volume = {984},
       number = {2},
          eid = {158},
        pages = {158},
          doi = {10.3847/1538-4357/adb5ef},
archivePrefix = {arXiv},
       eprint = {2410.17320},
 primaryClass = {astro-ph.GA},
       adsurl = {https://ui.adsabs.harvard.edu/abs/2025ApJ...984..158W}
}

@ARTICLE{CASATeam2022,
       author = {{CASA Team} and {Bean}, Ben and {Bhatnagar}, Sanjay and {Castro}, Sandra and {Donovan Meyer}, Jennifer and {Emonts}, Bjorn and {Garcia}, Enrique and {Garwood}, Robert and {Golap}, Kumar and {Gonzalez Villalba}, Justo and {Harris}, Pamela and {Hayashi}, Yohei and {Hoskins}, Josh and {Hsieh}, Mingyu and {Jagannathan}, Preshanth and {Kawasaki}, Wataru and {Keimpema}, Aard and {Kettenis}, Mark and {Lopez}, Jorge and {Marvil}, Joshua and {Masters}, Joseph and {McNichols}, Andrew and {Mehringer}, David and {Miel}, Renaud and {Moellenbrock}, George and {Montesino}, Federico and {Nakazato}, Takeshi and {Ott}, Juergen and {Petry}, Dirk and {Pokorny}, Martin and {Raba}, Ryan and {Rau}, Urvashi and {Schiebel}, Darrell and {Schweighart}, Neal and {Sekhar}, Srikrishna and {Shimada}, Kazuhiko and {Small}, Des and {Steeb}, Jan-Willem and {Sugimoto}, Kanako and {Suoranta}, Ville and {Tsutsumi}, Takahiro and {van Bemmel}, Ilse M. and {Verkouter}, Marjolein and {Wells}, Akeem and {Xiong}, Wei and {Szomoru}, Arpad and {Griffith}, Morgan and {Glendenning}, Brian and {Kern}, Jeff},
        title = "{CASA, the Common Astronomy Software Applications for Radio Astronomy}",
      journal = {\pasp},
         year = 2022,
        month = nov,
       volume = {134},
       number = {1041},
          eid = {114501},
        pages = {114501},
          doi = {10.1088/1538-3873/ac9642},
archivePrefix = {arXiv},
       eprint = {2210.02276},
 primaryClass = {astro-ph.IM},
       adsurl = {https://ui.adsabs.harvard.edu/abs/2022PASP..134k4501C}
}

@ARTICLE{Sormani2022,
       author = {{Sormani}, Mattia C. and {Sanders}, Jason L. and {Fritz}, Tobias K. and {Smith}, Leigh C. and {Gerhard}, Ortwin and {Sch{\"o}del}, Rainer and {Magorrian}, John and {Neumayer}, Nadine and {Nogueras-Lara}, Francisco and {Feldmeier-Krause}, Anja and {Mastrobuono-Battisti}, Alessandra and {Schultheis}, Mathias and {Shahzamanian}, Banafsheh and {Vasiliev}, Eugene and {Klessen}, Ralf S. and {Lucas}, Philip and {Minniti}, Dante},
        title = "{Self-consistent modelling of the Milky Way's nuclear stellar disc}",
      journal = {\mnras},
         year = 2022,
        month = may,
       volume = {512},
       number = {2},
        pages = {1857-1884},
          doi = {10.1093/mnras/stac639},
archivePrefix = {arXiv},
       eprint = {2111.12713},
 primaryClass = {astro-ph.GA},
       adsurl = {https://ui.adsabs.harvard.edu/abs/2022MNRAS.512.1857S}
}

@INPROCEEDINGS{Henshaw2023,
       author = {{Henshaw}, J.~D. and {Barnes}, A.~T. and {Battersby}, C. and {Ginsburg}, A. and {Sormani}, M.~C. and {Walker}, D.~L.},
        title = "{Star Formation in the Central Molecular Zone of the Milky Way}",
    booktitle = {Protostars and Planets VII},
         year = 2023,
       editor = {{Inutsuka}, S. and {Aikawa}, Y. and {Muto}, T. and {Tomida}, K. and {Tamura}, M.},
       series = {Astronomical Society of the Pacific Conference Series},
       volume = {534},
        month = jul,
        pages = {83},
          doi = {10.48550/arXiv.2203.11223},
archivePrefix = {arXiv},
       eprint = {2203.11223},
 primaryClass = {astro-ph.GA},
       adsurl = {https://ui.adsabs.harvard.edu/abs/2023ASPC..534...83H}
}

@INPROCEEDINGS{Inutsuka2021,
       author = {{Inutsuka}, S.},
        title = "{Star Formation in The Galactic Disk and The Galactic Center}",
    booktitle = {New Horizons in Galactic Center Astronomy and Beyond},
         year = 2021,
       editor = {{Tsuboi}, M. and {Oka}, T.},
       series = {Astronomical Society of the Pacific Conference Series},
       volume = {528},
        month = jul,
        pages = {271},
       adsurl = {https://ui.adsabs.harvard.edu/abs/2021ASPC..528..271I}
}

@ARTICLE{Henshaw2022,
       author = {{Henshaw}, Jonathan D. and {Krumholz}, Mark R. and {Butterfield}, Natalie O. and {Mackey}, Jonathan and {Ginsburg}, Adam and {Haworth}, Thomas J. and {Nogueras-Lara}, Francisco and {Barnes}, Ashley T. and {Longmore}, Steven N. and {Bally}, John and {Kruijssen}, J.~M. Diederik and {Mills}, Elisabeth A.~C. and {Beuther}, Henrik and {Walker}, Daniel L. and {Battersby}, Cara and {Bulatek}, Alyssa and {Henning}, Thomas and {Ott}, Juergen and {Soler}, Juan D.},
        title = "{A wind-blown bubble in the Central Molecular Zone cloud G0.253+0.016}",
      journal = {\mnras},
         year = 2022,
        month = feb,
       volume = {509},
       number = {4},
        pages = {4758-4774},
          doi = {10.1093/mnras/stab3039},
archivePrefix = {arXiv},
       eprint = {2110.11367},
 primaryClass = {astro-ph.GA},
       adsurl = {https://ui.adsabs.harvard.edu/abs/2022MNRAS.509.4758H}
}

@ARTICLE{Schoier2005,
       author = {{Sch{\"o}ier}, F.~L. and {van der Tak}, F.~F.~S. and {van Dishoeck}, E.~F. and {Black}, J.~H.},
        title = "{An atomic and molecular database for analysis of submillimetre line observations}",
      journal = {\aap},
         year = 2005,
        month = mar,
       volume = {432},
       number = {1},
        pages = {369-379},
          doi = {10.1051/0004-6361:20041729},
archivePrefix = {arXiv},
       eprint = {astro-ph/0411110},
 primaryClass = {astro-ph},
       adsurl = {https://ui.adsabs.harvard.edu/abs/2005A&A...432..369S}
}

@ARTICLE{Bachiller1997,
       author = {{Bachiller}, R. and {P{\'e}rez Guti{\'e}rrez}, M.},
        title = "{Shock Chemistry in the Young Bipolar Outflow L1157}",
      journal = {\apjl},
         year = 1997,
        month = sep,
       volume = {487},
       number = {1},
        pages = {L93-L96},
          doi = {10.1086/310877},
       adsurl = {https://ui.adsabs.harvard.edu/abs/1997ApJ...487L..93B}
}

@ARTICLE{Mendoza2018,
       author = {{Mendoza}, Edgar and {Lefloch}, B. and {Ceccarelli}, C. and {Kahane}, C. and {Jaber}, A.~A. and {Podio}, L. and {Benedettini}, M. and {Codella}, C. and {Viti}, S. and {Jimenez-Serra}, I. and {Lepine}, J.~R.~D. and {Boechat-Roberty}, H.~M. and {Bachiller}, R.},
        title = "{A search for cyanopolyynes in L1157-B1}",
      journal = {\mnras},
         year = 2018,
        month = apr,
       volume = {475},
       number = {4},
        pages = {5501-5512},
          doi = {10.1093/mnras/sty180},
archivePrefix = {arXiv},
       eprint = {1801.03461},
 primaryClass = {astro-ph.GA},
       adsurl = {https://ui.adsabs.harvard.edu/abs/2018MNRAS.475.5501M}
}

@ARTICLE{Suzuki1992,
       author = {{Suzuki}, Hiroko and {Yamamoto}, Satoshi and {Ohishi}, Masatoshi and {Kaifu}, Norio and {Ishikawa}, Shin-Ichi and {Hirahara}, Yasuhiro and {Takano}, Shuro},
        title = "{A Survey of CCS, HC 3N, HC 5N, and NH 3 toward Dark Cloud Cores and Their Production Chemistry}",
      journal = {\apj},
         year = 1992,
        month = jun,
       volume = {392},
        pages = {551},
          doi = {10.1086/171456},
       adsurl = {https://ui.adsabs.harvard.edu/abs/1992ApJ...392..551S}
}

@ARTICLE{Callanan2023,
       author = {{Callanan}, Daniel and {Longmore}, Steven N. and {Battersby}, Cara and {Hatchfield}, H. Perry and {Walker}, Daniel L. and {Henshaw}, Jonathan and {Keto}, Eric and {Barnes}, Ashley and {Ginsburg}, Adam and {Kauffmann}, Jens and {Kruijssen}, J.~M. Diederik and {Lu}, Xing and {Mills}, Elisabeth A.~C. and {Pillai}, Thushara and {Zhang}, Qizhou and {Bally}, John and {Butterfield}, Natalie and {Contreras}, Yanett A. and {Ho}, Luis C. and {Immer}, Katharina and {Johnston}, Katharine G. and {Ott}, Juergen and {Patel}, Nimesh and {Tolls}, Volker},
        title = "{CMZoom III: Spectral line data release}",
      journal = {\mnras},
         year = 2023,
        month = apr,
       volume = {520},
       number = {3},
        pages = {4760-4778},
          doi = {10.1093/mnras/stad388},
archivePrefix = {arXiv},
       eprint = {2301.04699},
 primaryClass = {astro-ph.GA},
       adsurl = {https://ui.adsabs.harvard.edu/abs/2023MNRAS.520.4760C}
}

@ARTICLE{Hsieh2021,
       author = {{Hsieh}, Pei-Ying and {Koch}, Patrick M. and {Kim}, Woong-Tae and {Mart{\'\i}n}, Sergio and {Yen}, Hsi-Wei and {Carpenter}, John M. and {Harada}, Nanase and {Turner}, Jean L. and {Ho}, Paul T.~P. and {Tang}, Ya-Wen and {Beck}, Sara},
        title = "{The Circumnuclear Disk Revealed by ALMA. I. Dense Clouds and Tides in the Galactic Center}",
      journal = {\apj},
         year = 2021,
        month = jun,
       volume = {913},
       number = {2},
          eid = {94},
        pages = {94},
          doi = {10.3847/1538-4357/abf4cd},
archivePrefix = {arXiv},
       eprint = {2104.02078},
 primaryClass = {astro-ph.GA},
       adsurl = {https://ui.adsabs.harvard.edu/abs/2021ApJ...913...94H}
}

@ARTICLE{Walker2021,
       author = {{Walker}, Daniel L. and {Longmore}, Steven N. and {Bally}, John and {Ginsburg}, Adam and {Kruijssen}, J.~M. Diederik and {Zhang}, Qizhou and {Henshaw}, Jonathan D. and {Lu}, Xing and {Alves}, Jo{\~a}o and {Barnes}, Ashley T. and {Battersby}, Cara and {Beuther}, Henrik and {Contreras}, Yanett A. and {G{\'o}mez}, Laura and {Ho}, Luis C. and {Jackson}, James M. and {Kauffmann}, Jens and {Mills}, Elisabeth A.~C. and {Pillai}, Thushara},
        title = "{Star formation in 'the Brick': ALMA reveals an active protocluster in the Galactic centre cloud G0.253+0.016}",
      journal = {\mnras},
         year = 2021,
        month = may,
       volume = {503},
       number = {1},
        pages = {77-95},
          doi = {10.1093/mnras/stab415},
archivePrefix = {arXiv},
       eprint = {2102.03560},
 primaryClass = {astro-ph.GA},
       adsurl = {https://ui.adsabs.harvard.edu/abs/2021MNRAS.503...77W}
}

@ARTICLE{Petkova2021,
       author = {{Petkova}, Maya A. and {Kruijssen}, J.~M. Diederik and {Kluge}, A. Louise and {Glover}, Simon C.~O. and {Walker}, Daniel L. and {Longmore}, Steven N. and {Henshaw}, Jonathan D. and {Reissl}, Stefan and {Dale}, James E.},
        title = "{The complex multi-scale structure in simulated and observed emission maps of the proto-cluster cloud G0.253+0.016 (`the Brick')}",
      journal = {arXiv e-prints},
         year = 2021,
        month = apr,
          eid = {arXiv:2104.09558},
        pages = {arXiv:2104.09558},
archivePrefix = {arXiv},
       eprint = {2104.09558},
 primaryClass = {astro-ph.GA},
       adsurl = {https://ui.adsabs.harvard.edu/abs/2021arXiv210409558P}
}

@ARTICLE{Li2025,
       author = {{Li}, Zi-Yang and {Liu}, Xunchuan and {Liu}, Tie and {Qin}, Sheng-Li and {Goldsmith}, Paul F. and {Garc{\'\i}a}, Pablo and {Peng}, Yaping and {Chen}, Li and {Jiao}, Yunfan and {Kou}, Zhiping and {Li}, Chuanshou and {Zou}, Jiahang and {Tang}, Mengyao and {Li}, Shanghuo and {Liu}, Meizhu and {Garay}, Guido and {Xu}, Fengwei and {Jiao}, Wenyu and {Luo}, Qiu-Yi and {Zhang}, Suinan and {Gu}, Qi-Lao and {Mai}, Xiaofeng and {Zhang}, Yan-Kun and {Weng}, Jixiang and {Lee}, Chang Won and {Sanhueza}, Patricio and {Dib}, Sami and {Das}, Swagat R. and {Tang}, Xindi and {Bronfman}, Leonardo and {Gorai}, Prasanta and {Tatematsu}, Ken'ichi and {Liu}, Hong-Li and {Yang}, Dongting and {Zhang}, Zhenying and {Shen}, Xianjin},
        title = "{The ALMA-ATOMS survey: A sample of weak hot core candidates identified through line stacking}",
      journal = {\aap},
         year = 2025,
        month = may,
       volume = {697},
          eid = {A190},
        pages = {A190},
          doi = {10.1051/0004-6361/202452762},
archivePrefix = {arXiv},
       eprint = {2504.06802},
 primaryClass = {astro-ph.GA},
       adsurl = {https://ui.adsabs.harvard.edu/abs/2025A&A...697A.190L}
}

@ARTICLE{Simpson2021,
       author = {{Simpson}, Janet P. and {Colgan}, Sean W.~J. and {Cotera}, Angela S. and {Kaufman}, Michael J. and {Stolovy}, Susan R.},
        title = "{Sagittarius B1: A Patchwork of H II Regions and Photodissociation Regions}",
      journal = {\apj},
         year = 2021,
        month = mar,
       volume = {910},
       number = {1},
          eid = {59},
        pages = {59},
          doi = {10.3847/1538-4357/abe636},
archivePrefix = {arXiv},
       eprint = {2102.07031},
 primaryClass = {astro-ph.GA},
       adsurl = {https://ui.adsabs.harvard.edu/abs/2021ApJ...910...59S}
}

@ARTICLE{Lu2021,
       author = {{Lu}, Xing and {Li}, Shanghuo and {Ginsburg}, Adam and {Longmore}, Steven N. and {Kruijssen}, J.~M. Diederik and {Walker}, Daniel L. and {Feng}, Siyi and {Zhang}, Qizhou and {Battersby}, Cara and {Pillai}, Thushara and {Mills}, Elisabeth A.~C. and {Kauffmann}, Jens and {Cheng}, Yu and {Inutsuka}, Shu-ichiro},
        title = "{ALMA Observations of Massive Clouds in the Central Molecular Zone: Ubiquitous Protostellar Outflows}",
      journal = {\apj},
         year = 2021,
        month = mar,
       volume = {909},
       number = {2},
          eid = {177},
        pages = {177},
          doi = {10.3847/1538-4357/abde3c},
archivePrefix = {arXiv},
       eprint = {2101.07925},
 primaryClass = {astro-ph.GA},
       adsurl = {https://ui.adsabs.harvard.edu/abs/2021ApJ...909..177L}
}

@ARTICLE{Tress2020,
       author = {{Tress}, Robin G. and {Sormani}, Mattia C. and {Glover}, Simon C.~O. and {Klessen}, Ralf S. and {Battersby}, Cara D. and {Clark}, Paul C. and {Hatchfield}, H. Perry and {Smith}, Rowan J.},
        title = "{Simulations of the Milky Way's central molecular zone - I. Gas dynamics}",
      journal = {\mnras},
         year = 2020,
        month = dec,
       volume = {499},
       number = {3},
        pages = {4455-4478},
          doi = {10.1093/mnras/staa3120},
archivePrefix = {arXiv},
       eprint = {2004.06724},
 primaryClass = {astro-ph.GA},
       adsurl = {https://ui.adsabs.harvard.edu/abs/2020MNRAS.499.4455T}
}

@ARTICLE{Zeng2020,
       author = {{Zeng}, S. and {Zhang}, Q. and {Jim{\'e}nez-Serra}, I. and {Tercero}, B. and {Lu}, X. and {Mart{\'\i}n-Pintado}, J. and {de Vicente}, P. and {Rivilla}, V.~M. and {Li}, S.},
        title = "{Cloud-cloud collision as drivers of the chemical complexity in Galactic Centre molecular clouds}",
      journal = {\mnras},
         year = 2020,
        month = oct,
       volume = {497},
       number = {4},
        pages = {4896-4909},
          doi = {10.1093/mnras/staa2187},
archivePrefix = {arXiv},
       eprint = {2007.14362},
 primaryClass = {astro-ph.GA},
       adsurl = {https://ui.adsabs.harvard.edu/abs/2020MNRAS.497.4896Z}
}

@ARTICLE{Battersby2020,
       author = {{Battersby}, Cara and {Keto}, Eric and {Walker}, Daniel and {Barnes}, Ashley and {Callanan}, Daniel and {Ginsburg}, Adam and {Hatchfield}, H. Perry and {Henshaw}, Jonathan and {Kauffmann}, Jens and {Kruijssen}, J.~M. Diederik and {Longmore}, Steven N. and {Lu}, Xing and {Mills}, Elisabeth A.~C. and {Pillai}, Thushara and {Zhang}, Qizhou and {Bally}, John and {Butterfield}, Natalie and {Contreras}, Yanett A. and {Ho}, Luis C. and {Ott}, J{\"u}rgen and {Patel}, Nimesh and {Tolls}, Volker},
        title = "{CMZoom: Survey Overview and First Data Release}",
      journal = {\apjs},
         year = 2020,
        month = aug,
       volume = {249},
       number = {2},
          eid = {35},
        pages = {35},
          doi = {10.3847/1538-4365/aba18e},
archivePrefix = {arXiv},
       eprint = {2007.05023},
 primaryClass = {astro-ph.GA},
       adsurl = {https://ui.adsabs.harvard.edu/abs/2020ApJS..249...35B}
}

@ARTICLE{Pickett1998,
       author = {{Pickett}, H.~M. and {Poynter}, R.~L. and {Cohen}, E.~A. and {Delitsky}, M.~L. and {Pearson}, J.~C. and {M{\"u}ller}, H.~S.~P.},
        title = "{Submillimeter, millimeter and microwave spectral line catalog.}",
      journal = {\jqsrt},
         year = 1998,
        month = nov,
       volume = {60},
       number = {5},
        pages = {883-890},
          doi = {10.1016/S0022-4073(98)00091-0},
       adsurl = {https://ui.adsabs.harvard.edu/abs/1998JQSRT..60..883P}
}

@ARTICLE{Tsuboi2019,
       author = {{Tsuboi}, Masato and {Kitamura}, Yoshimi and {Uehara}, Kenta and {Miyazaki}, Atsushi and {Miyawaki}, Ryosuke and {Tsutsumi}, Takahiro and {Miyoshi}, Makoto},
        title = "{G-0.02-0.07, the compact H II region complex nearest to the galactic center with ALMA}",
      journal = {\pasj},
         year = 2019,
        month = dec,
       volume = {71},
       number = {6},
          eid = {128},
        pages = {128},
          doi = {10.1093/pasj/psz116},
archivePrefix = {arXiv},
       eprint = {1909.13405},
 primaryClass = {astro-ph.GA},
       adsurl = {https://ui.adsabs.harvard.edu/abs/2019PASJ...71..128T}
}

@ARTICLE{Lu2019a,
       author = {{Lu}, Xing and {Mills}, Elisabeth A.~C. and {Ginsburg}, Adam and {Walker}, Daniel L. and {Barnes}, Ashley T. and {Butterfield}, Natalie and {Henshaw}, Jonathan D. and {Battersby}, Cara and {Kruijssen}, J.~M. Diederik and {Longmore}, Steven N. and {Zhang}, Qizhou and {Bally}, John and {Kauffmann}, Jens and {Ott}, J{\"u}rgen and {Rickert}, Matthew and {Wang}, Ke},
        title = "{A Census of Early-phase High-mass Star Formation in the Central Molecular Zone}",
      journal = {\apjs},
         year = 2019,
        month = oct,
       volume = {244},
       number = {2},
          eid = {35},
        pages = {35},
          doi = {10.3847/1538-4365/ab4258},
archivePrefix = {arXiv},
       eprint = {1909.02338},
 primaryClass = {astro-ph.GA},
       adsurl = {https://ui.adsabs.harvard.edu/abs/2019ApJS..244...35L}
}

@ARTICLE{Lu2019b,
       author = {{Lu}, Xing and {Zhang}, Qizhou and {Kauffmann}, Jens and {Pillai}, Thushara and {Ginsburg}, Adam and {Mills}, Elisabeth A.~C. and {Kruijssen}, J.~M. Diederik and {Longmore}, Steven N. and {Battersby}, Cara and {Liu}, Hauyu Baobab and {Gu}, Qiusheng},
        title = "{Star Formation Rates of Massive Molecular Clouds in the Central Molecular Zone}",
      journal = {\apj},
         year = 2019,
        month = feb,
       volume = {872},
       number = {2},
          eid = {171},
        pages = {171},
          doi = {10.3847/1538-4357/ab017d},
archivePrefix = {arXiv},
       eprint = {1901.07779},
 primaryClass = {astro-ph.GA},
       adsurl = {https://ui.adsabs.harvard.edu/abs/2019ApJ...872..171L}
}

@ARTICLE{Uehara2019,
       author = {{Uehara}, Kenta and {Tsuboi}, Masato and {Kitamura}, Yoshimi and {Miyawaki}, Ryosuke and {Miyazaki}, Atsushi},
        title = "{Molecular Cloud Cores in the Galactic Center 50 km s$^{-1}$ Molecular Cloud}",
      journal = {\apj},
         year = 2019,
        month = feb,
       volume = {872},
       number = {2},
          eid = {121},
        pages = {121},
          doi = {10.3847/1538-4357/aafee7},
archivePrefix = {arXiv},
       eprint = {1903.01759},
 primaryClass = {astro-ph.GA},
       adsurl = {https://ui.adsabs.harvard.edu/abs/2019ApJ...872..121U}
}

@INPROCEEDINGS{Sault1995,
       author = {{Sault}, R.~J. and {Teuben}, P.~J. and {Wright}, M.~C.~H.},
        title = "{A Retrospective View of MIRIAD}",
    booktitle = {Astronomical Data Analysis Software and Systems IV},
         year = 1995,
       editor = {{Shaw}, R.~A. and {Payne}, H.~E. and {Hayes}, J.~J.~E.},
       series = {Astronomical Society of the Pacific Conference Series},
       volume = {77},
        month = jan,
        pages = {433},
          doi = {10.48550/arXiv.astro-ph/0612759},
archivePrefix = {arXiv},
       eprint = {astro-ph/0612759},
 primaryClass = {astro-ph},
       adsurl = {https://ui.adsabs.harvard.edu/abs/1995ASPC...77..433S}
}

@ARTICLE{Simpson2018a,
       author = {{Simpson}, Janet P. and {Colgan}, Sean W.~J. and {Cotera}, Angela S. and {Kaufman}, Michael J. and {Stolovy}, Susan R.},
        title = "{SOFIA FIFI-LS Observations of Sgr B1: Ionization Structure and Sources of Excitation}",
      journal = {\apjl},
         year = 2018,
        month = nov,
       volume = {867},
       number = {1},
          eid = {L13},
        pages = {L13},
          doi = {10.3847/2041-8213/aae8e4},
archivePrefix = {arXiv},
       eprint = {1810.08301},
 primaryClass = {astro-ph.GA},
       adsurl = {https://ui.adsabs.harvard.edu/abs/2018ApJ...867L..13S}
}

@ARTICLE{Krieger2017,
       author = {{Krieger}, Nico and {Ott}, J{\"u}rgen and {Beuther}, Henrik and {Walter}, Fabian and {Kruijssen}, J.~M. Diederik and {Meier}, David S. and {Mills}, Elisabeth A.~C. and {Contreras}, Yanett and {Edwards}, Phil and {Ginsburg}, Adam and {Henkel}, Christian and {Henshaw}, Jonathan and {Jackson}, James and {Kauffmann}, Jens and {Longmore}, Steven and {Mart{\'\i}n}, Sergio and {Morris}, Mark R. and {Pillai}, Thushara and {Rickert}, Matthew and {Rosolowsky}, Erik and {Shinnaga}, Hiroko and {Walsh}, Andrew and {Yusef-Zadeh}, Farhad and {Zhang}, Qizhou},
        title = "{The Survey of Water and Ammonia in the Galactic Center (SWAG): Molecular Cloud Evolution in the Central Molecular Zone}",
      journal = {\apj},
         year = 2017,
        month = nov,
       volume = {850},
       number = {1},
          eid = {77},
        pages = {77},
          doi = {10.3847/1538-4357/aa951c},
archivePrefix = {arXiv},
       eprint = {1710.06902},
 primaryClass = {astro-ph.GA},
       adsurl = {https://ui.adsabs.harvard.edu/abs/2017ApJ...850...77K}
}

@ARTICLE{Hsieh2017,
       author = {{Hsieh}, Pei-Ying and {Koch}, Patrick M. and {Ho}, Paul T.~P. and {Kim}, Woong-Tae and {Tang}, Ya-Wen and {Wang}, Hsiang-Hsu and {Yen}, Hsi-Wei and {Hwang}, Chorng-Yuan},
        title = "{Molecular Gas Feeding the Circumnuclear Disk of the Galactic Center}",
      journal = {\apj},
         year = 2017,
        month = sep,
       volume = {847},
       number = {1},
          eid = {3},
        pages = {3},
          doi = {10.3847/1538-4357/aa8329},
archivePrefix = {arXiv},
       eprint = {1708.08579},
 primaryClass = {astro-ph.GA},
       adsurl = {https://ui.adsabs.harvard.edu/abs/2017ApJ...847....3H}
}

@ARTICLE{Hsieh2016,
       author = {{Hsieh}, Pei-Ying and {Ho}, Paul T.~P. and {Hwang}, Chorng-Yuan and {Shimajiri}, Yoshito and {Matsushita}, Satoki and {Koch}, Patrick M. and {Iono}, Daisuke},
        title = "{The Fossil Nuclear Outflow in the Central 30 pc of the Galactic Center}",
      journal = {\apj},
         year = 2016,
        month = nov,
       volume = {831},
       number = {1},
          eid = {72},
        pages = {72},
          doi = {10.3847/0004-637X/831/1/72},
archivePrefix = {arXiv},
       eprint = {1607.03673},
 primaryClass = {astro-ph.GA},
       adsurl = {https://ui.adsabs.harvard.edu/abs/2016ApJ...831...72H}
}

@ARTICLE{Barnes2017,
       author = {{Barnes}, A.~T. and {Longmore}, S.~N. and {Battersby}, C. and {Bally}, J. and {Kruijssen}, J.~M.~D. and {Henshaw}, J.~D. and {Walker}, D.~L.},
        title = "{Star formation rates and efficiencies in the Galactic Centre}",
      journal = {\mnras},
         year = 2017,
        month = aug,
       volume = {469},
       number = {2},
        pages = {2263-2285},
          doi = {10.1093/mnras/stx941},
archivePrefix = {arXiv},
       eprint = {1704.03572},
 primaryClass = {astro-ph.GA},
       adsurl = {https://ui.adsabs.harvard.edu/abs/2017MNRAS.469.2263B}
}

@ARTICLE{Lu2017,
       author = {{Lu}, Xing and {Zhang}, Qizhou and {Kauffmann}, Jens and {Pillai}, Thushara and {Longmore}, Steven N. and {Kruijssen}, J.~M. Diederik and {Battersby}, Cara and {Liu}, Hauyu Baobab and {Ginsburg}, Adam and {Mills}, Elisabeth A.~C. and {Zhang}, Zhi-Yu and {Gu}, Qiusheng},
        title = "{The Molecular Gas Environment in the 20 km s$^{-1}$ Cloud in the Central Molecular Zone}",
      journal = {\apj},
         year = 2017,
        month = apr,
       volume = {839},
       number = {1},
          eid = {1},
        pages = {1},
          doi = {10.3847/1538-4357/aa67f7},
archivePrefix = {arXiv},
       eprint = {1703.06551},
 primaryClass = {astro-ph.GA},
       adsurl = {https://ui.adsabs.harvard.edu/abs/2017ApJ...839....1L}
}

@ARTICLE{Immer2016,
       author = {{Immer}, K. and {Kauffmann}, J. and {Pillai}, T. and {Ginsburg}, A. and {Menten}, K.~M.},
        title = "{Temperature structures in Galactic center clouds. Direct evidence for gas heating via turbulence}",
      journal = {\aap},
         year = 2016,
        month = nov,
       volume = {595},
          eid = {A94},
        pages = {A94},
          doi = {10.1051/0004-6361/201628777},
archivePrefix = {arXiv},
       eprint = {1607.03535},
 primaryClass = {astro-ph.GA},
       adsurl = {https://ui.adsabs.harvard.edu/abs/2016A&A...595A..94I}
}

@ARTICLE{Steinke2016,
       author = {{Steinke}, M. and {Oskinova}, L.~M. and {Hamann}, W. -R. and {Sander}, A. and {Liermann}, A. and {Todt}, H.},
        title = "{Analysis of the WN star WR 102c, its WR nebula, and the associated cluster of massive stars in the Sickle Nebula}",
      journal = {\aap},
         year = 2016,
        month = apr,
       volume = {588},
          eid = {A9},
        pages = {A9},
          doi = {10.1051/0004-6361/201527692},
archivePrefix = {arXiv},
       eprint = {1601.03395},
 primaryClass = {astro-ph.SR},
       adsurl = {https://ui.adsabs.harvard.edu/abs/2016A&A...588A...9S}
}

@ARTICLE{Henshaw2016b,
       author = {{Henshaw}, J.~D. and {Longmore}, S.~N. and {Kruijssen}, J.~M.~D. and {Davies}, B. and {Bally}, J. and {Barnes}, A. and {Battersby}, C. and {Burton}, M. and {Cunningham}, M.~R. and {Dale}, J.~E. and {Ginsburg}, A. and {Immer}, K. and {Jones}, P.~A. and {Kendrew}, S. and {Mills}, E.~A.~C. and {Molinari}, S. and {Moore}, T.~J.~T. and {Ott}, J. and {Pillai}, T. and {Rathborne}, J. and {Schilke}, P. and {Schmiedeke}, A. and {Testi}, L. and {Walker}, D. and {Walsh}, A. and {Zhang}, Q.},
        title = "{Molecular gas kinematics within the central 250 pc of the Milky Way}",
      journal = {\mnras},
         year = 2016,
        month = apr,
       volume = {457},
       number = {3},
        pages = {2675-2702},
          doi = {10.1093/mnras/stw121},
archivePrefix = {arXiv},
       eprint = {1601.03732},
 primaryClass = {astro-ph.GA},
       adsurl = {https://ui.adsabs.harvard.edu/abs/2016MNRAS.457.2675H}
}

@ARTICLE{Ginsburg2016,
       author = {{Ginsburg}, Adam and {Henkel}, Christian and {Ao}, Yiping and {Riquelme}, Denise and {Kauffmann}, Jens and {Pillai}, Thushara and {Mills}, Elisabeth A.~C. and {Requena-Torres}, Miguel A. and {Immer}, Katharina and {Testi}, Leonardo and {Ott}, Juergen and {Bally}, John and {Battersby}, Cara and {Darling}, Jeremy and {Aalto}, Susanne and {Stanke}, Thomas and {Kendrew}, Sarah and {Kruijssen}, J.~M. Diederik and {Longmore}, Steven and {Dale}, James and {Guesten}, Rolf and {Menten}, Karl M.},
        title = "{Dense gas in the Galactic central molecular zone is warm and heated by turbulence}",
      journal = {\aap},
         year = 2016,
        month = feb,
       volume = {586},
          eid = {A50},
        pages = {A50},
          doi = {10.1051/0004-6361/201526100},
archivePrefix = {arXiv},
       eprint = {1509.01583},
 primaryClass = {astro-ph.GA},
       adsurl = {https://ui.adsabs.harvard.edu/abs/2016A&A...586A..50G}
}

@ARTICLE{Tsuboi2015a,
       author = {{Tsuboi}, Masato and {Miyazaki}, Atsushi and {Uehara}, Kenta},
        title = "{Cloud-cloud collision in the Galactic center 50 km s$^{-1}$ molecular cloud}",
      journal = {\pasj},
         year = 2015,
        month = dec,
       volume = {67},
       number = {6},
          eid = {109},
        pages = {109},
          doi = {10.1093/pasj/psv076},
archivePrefix = {arXiv},
       eprint = {1507.08351},
 primaryClass = {astro-ph.GA},
       adsurl = {https://ui.adsabs.harvard.edu/abs/2015PASJ...67..109T}
}

@ARTICLE{Rathborne2015,
       author = {{Rathborne}, J.~M. and {Longmore}, S.~N. and {Jackson}, J.~M. and {Alves}, J.~F. and {Bally}, J. and {Bastian}, N. and {Contreras}, Y. and {Foster}, J.~B. and {Garay}, G. and {Kruijssen}, J.~M.~D. and {Testi}, L. and {Walsh}, A.~J.},
        title = "{A Cluster in the Making: ALMA Reveals the Initial Conditions for High-mass Cluster Formation}",
      journal = {\apj},
         year = 2015,
        month = apr,
       volume = {802},
       number = {2},
          eid = {125},
        pages = {125},
          doi = {10.1088/0004-637X/802/2/125},
archivePrefix = {arXiv},
       eprint = {1501.07368},
 primaryClass = {astro-ph.GA},
       adsurl = {https://ui.adsabs.harvard.edu/abs/2015ApJ...802..125R}
}

@ARTICLE{Rathborne2014a,
       author = {{Rathborne}, J.~M. and {Longmore}, S.~N. and {Jackson}, J.~M. and {Kruijssen}, J.~M.~D. and {Alves}, J.~F. and {Bally}, J. and {Bastian}, N. and {Contreras}, Y. and {Foster}, J.~B. and {Garay}, G. and {Testi}, L. and {Walsh}, A.~J.},
        title = "{Turbulence Sets the Initial Conditions for Star Formation in High-pressure Environments}",
      journal = {\apjl},
         year = 2014,
        month = nov,
       volume = {795},
       number = {2},
          eid = {L25},
        pages = {L25},
          doi = {10.1088/2041-8205/795/2/L25},
archivePrefix = {arXiv},
       eprint = {1409.0935},
 primaryClass = {astro-ph.GA},
       adsurl = {https://ui.adsabs.harvard.edu/abs/2014ApJ...795L..25R}
}

@ARTICLE{Hacar2013,
       author = {{Hacar}, A. and {Tafalla}, M. and {Kauffmann}, J. and {Kov{\'a}cs}, A.},
        title = "{Cores, filaments, and bundles: hierarchical core formation in the L1495/B213 Taurus region}",
      journal = {\aap},
         year = 2013,
        month = jun,
       volume = {554},
          eid = {A55},
        pages = {A55},
          doi = {10.1051/0004-6361/201220090},
archivePrefix = {arXiv},
       eprint = {1303.2118},
 primaryClass = {astro-ph.GA},
       adsurl = {https://ui.adsabs.harvard.edu/abs/2013A&A...554A..55H}
}

@ARTICLE{Hoque2025,
       author = {{Hoque}, Ariful and {Baug}, Tapas and {Dewangan}, Lokesh K. and {Juvela}, Mika and {Tej}, Anandmayee and {Goldsmith}, Paul F. and {Garc{\'\i}a}, Pablo and {Stutz}, Amelia M. and {Liu}, Tie and {Lee}, Chang Won and {Xu}, Fengwei and {Sanhueza}, Patricio and {Bhadari}, N.~K. and {Tatematsu}, K. and {Liu}, Xunchuan and {Liu}, Hong-Li and {Zhang}, Yong and {Tang}, Xindi and {Garay}, Guido and {Wang}, Ke and {Zhang}, Siju and {T{\'o}th}, L. Viktor and {Nazeer}, Hafiz and {Hwang}, Jihye and {Gorai}, Prasanta and {Bronfman}, Leonardo and {Das}, Swagat Ranjan and {Sinha}, Tirthendu},
        title = "{The ALMA-ATOMS Survey: Exploring Protostellar Outflows in HC$_{3}$N}",
      journal = {\apj},
         year = 2025,
        month = jul,
       volume = {987},
       number = {2},
          eid = {197},
        pages = {197},
          doi = {10.3847/1538-4357/add928},
archivePrefix = {arXiv},
       eprint = {2505.04164},
 primaryClass = {astro-ph.GA},
       adsurl = {https://ui.adsabs.harvard.edu/abs/2025ApJ...987..197H}
}

@ARTICLE{Johnston2014,
       author = {{Johnston}, K.~G. and {Beuther}, H. and {Linz}, H. and {Schmiedeke}, A. and {Ragan}, S.~E. and {Henning}, Th.},
        title = "{The dynamics and star-forming potential of the massive Galactic centre cloud G0.253+0.016}",
      journal = {\aap},
         year = 2014,
        month = aug,
       volume = {568},
          eid = {A56},
        pages = {A56},
          doi = {10.1051/0004-6361/201423943},
archivePrefix = {arXiv},
       eprint = {1404.1372},
 primaryClass = {astro-ph.GA},
       adsurl = {https://ui.adsabs.harvard.edu/abs/2014A&A...568A..56J}
}

@ARTICLE{Kruijssen2014a,
       author = {{Kruijssen}, J.~M. Diederik and {Longmore}, Steven N. and {Elmegreen}, Bruce G. and {Murray}, Norman and {Bally}, John and {Testi}, Leonardo and {Kennicutt}, Robert C.},
        title = "{What controls star formation in the central 500 pc of the Galaxy?}",
      journal = {\mnras},
         year = 2014,
        month = jun,
       volume = {440},
       number = {4},
        pages = {3370-3391},
          doi = {10.1093/mnras/stu494},
archivePrefix = {arXiv},
       eprint = {1303.6286},
 primaryClass = {astro-ph.GA},
       adsurl = {https://ui.adsabs.harvard.edu/abs/2014MNRAS.440.3370K}
}

@ARTICLE{Mills2013,
       author = {{Mills}, E.~A.~C. and {G{\"u}sten}, R. and {Requena-Torres}, M.~A. and {Morris}, M.~R.},
        title = "{The Excitation of HCN and HCO$^{+}$ in the Galactic Center Circumnuclear Disk}",
      journal = {\apj},
         year = 2013,
        month = dec,
       volume = {779},
       number = {1},
          eid = {47},
        pages = {47},
          doi = {10.1088/0004-637X/779/1/47},
archivePrefix = {arXiv},
       eprint = {1309.7412},
 primaryClass = {astro-ph.GA},
       adsurl = {https://ui.adsabs.harvard.edu/abs/2013ApJ...779...47M}
}

@ARTICLE{Ao2013,
       author = {{Ao}, Y. and {Henkel}, C. and {Menten}, K.~M. and {Requena-Torres}, M.~A. and {Stanke}, T. and {Mauersberger}, R. and {Aalto}, S. and {M{\"u}hle}, S. and {Mangum}, J.},
        title = "{The thermal state of molecular clouds in the Galactic center: evidence for non-photon-driven heating}",
      journal = {\aap},
         year = 2013,
        month = feb,
       volume = {550},
          eid = {A135},
        pages = {A135},
          doi = {10.1051/0004-6361/201220096},
archivePrefix = {arXiv},
       eprint = {1211.7142},
 primaryClass = {astro-ph.GA},
       adsurl = {https://ui.adsabs.harvard.edu/abs/2013A&A...550A.135A}
}

@ARTICLE{Kruijssen2013,
       author = {{Kruijssen}, J.~M. Diederik and {Longmore}, Steven N.},
        title = "{Comparing molecular gas across cosmic time-scales: the Milky Way as both a typical spiral galaxy and a high-redshift galaxy analogue}",
      journal = {\mnras},
         year = 2013,
        month = nov,
       volume = {435},
       number = {3},
        pages = {2598-2603},
          doi = {10.1093/mnras/stt1634},
archivePrefix = {arXiv},
       eprint = {1309.0505},
 primaryClass = {astro-ph.CO},
       adsurl = {https://ui.adsabs.harvard.edu/abs/2013MNRAS.435.2598K}
}

@ARTICLE{Jones2011,
       author = {{Jones}, P.~A. and {Burton}, M.~G. and {Tothill}, N.~F.~H. and {Cunningham}, M.~R.},
        title = "{Spectral imaging of the Sagittarius B2 region in multiple 7-mm molecular lines}",
      journal = {\mnras},
         year = 2011,
        month = mar,
       volume = {411},
       number = {4},
        pages = {2293-2310},
          doi = {10.1111/j.1365-2966.2010.17849.x},
archivePrefix = {arXiv},
       eprint = {1010.2449},
 primaryClass = {astro-ph.GA},
       adsurl = {https://ui.adsabs.harvard.edu/abs/2011MNRAS.411.2293J}
}

@ARTICLE{Kauffmann2013,
       author = {{Kauffmann}, Jens and {Pillai}, Thushara and {Zhang}, Qizhou},
        title = "{The Galactic Center Cloud G0.253+0.016: A Massive Dense Cloud with low Star Formation Potential}",
      journal = {\apjl},
         year = 2013,
        month = mar,
       volume = {765},
       number = {2},
          eid = {L35},
        pages = {L35},
          doi = {10.1088/2041-8205/765/2/L35},
archivePrefix = {arXiv},
       eprint = {1301.1338},
 primaryClass = {astro-ph.GA},
       adsurl = {https://ui.adsabs.harvard.edu/abs/2013ApJ...765L..35K}
}

@ARTICLE{Longmore2013b,
       author = {{Longmore}, S.~N. and {Bally}, J. and {Testi}, L. and {Purcell}, C.~R. and {Walsh}, A.~J. and {Bressert}, E. and {Pestalozzi}, M. and {Molinari}, S. and {Ott}, J. and {Cortese}, L. and {Battersby}, C. and {Murray}, N. and {Lee}, E. and {Kruijssen}, J.~M.~D. and {Schisano}, E. and {Elia}, D.},
        title = "{Variations in the Galactic star formation rate and density thresholds for star formation}",
      journal = {\mnras},
         year = 2013,
        month = feb,
       volume = {429},
       number = {2},
        pages = {987-1000},
          doi = {10.1093/mnras/sts376},
archivePrefix = {arXiv},
       eprint = {1208.4256},
 primaryClass = {astro-ph.GA},
       adsurl = {https://ui.adsabs.harvard.edu/abs/2013MNRAS.429..987L}
}

@ARTICLE{Immer2012a,
       author = {{Immer}, K. and {Menten}, K.~M. and {Schuller}, F. and {Lis}, D.~C.},
        title = "{A multi-wavelength view of the Galactic center dust ridge reveals little star formation}",
      journal = {\aap},
         year = 2012,
        month = dec,
       volume = {548},
          eid = {A120},
        pages = {A120},
          doi = {10.1051/0004-6361/201219182},
archivePrefix = {arXiv},
       eprint = {1211.0455},
 primaryClass = {astro-ph.GA},
       adsurl = {https://ui.adsabs.harvard.edu/abs/2012A&A...548A.120I}
}

@ARTICLE{Martin2012,
       author = {{Mart{\'\i}n}, S. and {Mart{\'\i}n-Pintado}, J. and {Montero-Casta{\~n}o}, M. and {Ho}, P.~T.~P. and {Blundell}, R.},
        title = "{Surviving the hole. I. Spatially resolved chemistry around Sagittarius A$^{{\ensuremath{*}}}$}",
      journal = {\aap},
         year = 2012,
        month = mar,
       volume = {539},
          eid = {A29},
        pages = {A29},
          doi = {10.1051/0004-6361/201117268},
archivePrefix = {arXiv},
       eprint = {1112.0566},
 primaryClass = {astro-ph.GA},
       adsurl = {https://ui.adsabs.harvard.edu/abs/2012A&A...539A..29M}
}

@ARTICLE{Jones2012,
       author = {{Jones}, P.~A. and {Burton}, M.~G. and {Cunningham}, M.~R. and {Requena-Torres}, M.~A. and {Menten}, K.~M. and {Schilke}, P. and {Belloche}, A. and {Leurini}, S. and {Mart{\'\i}n-Pintado}, J. and {Ott}, J. and {Walsh}, A.~J.},
        title = "{Spectral imaging of the Central Molecular Zone in multiple 3-mm molecular lines}",
      journal = {\mnras},
         year = 2012,
        month = feb,
       volume = {419},
       number = {4},
        pages = {2961-2986},
          doi = {10.1111/j.1365-2966.2011.19941.x},
archivePrefix = {arXiv},
       eprint = {1110.1421},
 primaryClass = {astro-ph.GA},
       adsurl = {https://ui.adsabs.harvard.edu/abs/2012MNRAS.419.2961J}
}

@ARTICLE{DePree2011,
       author = {{De Pree}, C.~G. and {Wilner}, D.~J. and {Goss}, W.~M.},
        title = "{Ionized Gas Kinematics and Morphology in Sgr B2 Main on 1000 AU Scales}",
      journal = {\aj},
         year = 2011,
        month = nov,
       volume = {142},
       number = {5},
          eid = {177},
        pages = {177},
          doi = {10.1088/0004-6256/142/5/177},
archivePrefix = {arXiv},
       eprint = {1109.4198},
 primaryClass = {astro-ph.GA},
       adsurl = {https://ui.adsabs.harvard.edu/abs/2011AJ....142..177D}
}

@ARTICLE{Mills2011,
       author = {{Mills}, E. and {Morris}, M.~R. and {Lang}, C.~C. and {Dong}, H. and {Wang}, Q.~D. and {Cotera}, A. and {Stolovy}, S.~R.},
        title = "{Properties of the Compact H II Region Complex G-0.02-0.07}",
      journal = {\apj},
         year = 2011,
        month = jul,
       volume = {735},
       number = {2},
          eid = {84},
        pages = {84},
          doi = {10.1088/0004-637X/735/2/84},
archivePrefix = {arXiv},
       eprint = {1102.2533},
 primaryClass = {astro-ph.GA},
       adsurl = {https://ui.adsabs.harvard.edu/abs/2011ApJ...735...84M}
}

@ARTICLE{Genzel2010,
       author = {{Genzel}, Reinhard and {Eisenhauer}, Frank and {Gillessen}, Stefan},
        title = "{The Galactic Center massive black hole and nuclear star cluster}",
      journal = {Reviews of Modern Physics},
         year = 2010,
        month = oct,
       volume = {82},
       number = {4},
        pages = {3121-3195},
          doi = {10.1103/RevModPhys.82.3121},
archivePrefix = {arXiv},
       eprint = {1006.0064},
 primaryClass = {astro-ph.GA},
       adsurl = {https://ui.adsabs.harvard.edu/abs/2010RvMP...82.3121G}
}

@ARTICLE{Rodriguez-Fernandez1998,
       author = {{Rodriguez-Franco}, A. and {Martin-Pintado}, J. and {Fuente}, A.},
        title = "{CN emission in Orion. The high density interface between the H II region and the molecular cloud}",
      journal = {\aap},
         year = 1998,
        month = jan,
       volume = {329},
        pages = {1097-1110},
       adsurl = {https://ui.adsabs.harvard.edu/abs/1998A&A...329.1097R}
}

@ARTICLE{Meier2005,
       author = {{Meier}, David S. and {Turner}, Jean L.},
        title = "{Spatially Resolved Chemistry in Nearby Galaxies. I. The Center of IC 342}",
      journal = {\apj},
         year = 2005,
        month = jan,
       volume = {618},
       number = {1},
        pages = {259-280},
          doi = {10.1086/426499},
archivePrefix = {arXiv},
       eprint = {astro-ph/0410039},
 primaryClass = {astro-ph},
       adsurl = {https://ui.adsabs.harvard.edu/abs/2005ApJ...618..259M}
}

@ARTICLE{Lang2002,
       author = {{Lang}, Cornelia C. and {Goss}, W.~M. and {Morris}, Mark},
        title = "{The Molecular Component of the Galactic Center Arched Filaments H II Complex: OVRO Observations of the CS J=2-1 Line}",
      journal = {\aj},
         year = 2002,
        month = nov,
       volume = {124},
       number = {5},
        pages = {2677-2692},
          doi = {10.1086/344159},
archivePrefix = {arXiv},
       eprint = {astro-ph/0208324},
 primaryClass = {astro-ph},
       adsurl = {https://ui.adsabs.harvard.edu/abs/2002AJ....124.2677L}
}

@ARTICLE{Lindberg2011,
       author = {{Lindberg}, J.~E. and {Aalto}, S. and {Costagliola}, F. and {P{\'e}rez-Beaupuits}, J.-P. and {Monje}, R. and {Muller}, S.},
        title = "{A survey of HC$_{3}$N in extragalactic sources. Is HC$_{3}$N a tracer of activity in ULIRGs?}",
      journal = {\aap},
         year = 2011,
        month = mar,
       volume = {527},
          eid = {A150},
        pages = {A150},
          doi = {10.1051/0004-6361/201015565},
archivePrefix = {arXiv},
       eprint = {1101.1751},
 primaryClass = {astro-ph.CO},
       adsurl = {https://ui.adsabs.harvard.edu/abs/2011A&A...527A.150L}
}

@ARTICLE{Lis2001,
       author = {{Lis}, D.~C. and {Serabyn}, E. and {Zylka}, R. and {Li}, Y.},
        title = "{Quiescent Giant Molecular Cloud Cores in the Galactic Center}",
      journal = {\apj},
         year = 2001,
        month = apr,
       volume = {550},
       number = {2},
        pages = {761-777},
          doi = {10.1086/319815},
       adsurl = {https://ui.adsabs.harvard.edu/abs/2001ApJ...550..761L}
}

@ARTICLE{Tsuboi1999,
       author = {{Tsuboi}, Masato and {Handa}, Toshihiro and {Ukita}, Nobuharu},
        title = "{Dense Molecular Clouds in the Galactic Center Region. I. Observations and Data}",
      journal = {\apjs},
         year = 1999,
        month = jan,
       volume = {120},
       number = {1},
        pages = {1-39},
          doi = {10.1086/313165},
       adsurl = {https://ui.adsabs.harvard.edu/abs/1999ApJS..120....1T}
}

@ARTICLE{Schilke1997,
       author = {{Schilke}, P. and {Walmsley}, C.~M. and {Pineau des Forets}, G. and {Flower}, D.~R.},
        title = "{SiO production in interstellar shocks.}",
      journal = {\aap},
         year = 1997,
        month = may,
       volume = {321},
        pages = {293-304},
       adsurl = {https://ui.adsabs.harvard.edu/abs/1997A&A...321..293S}
}

@ARTICLE{Morris1996,
       author = {{Morris}, Mark and {Serabyn}, Eugene},
        title = "{The Galactic Center Environment}",
      journal = {\araa},
         year = 1996,
        month = jan,
       volume = {34},
        pages = {645-702},
          doi = {10.1146/annurev.astro.34.1.645},
       adsurl = {https://ui.adsabs.harvard.edu/abs/1996ARA&A..34..645M}
}

@ARTICLE{Lis1994a,
       author = {{Lis}, D.~C. and {Carlstrom}, J.~E.},
        title = "{Submillimeter Continuum Survey of the Galactic Center}",
      journal = {\apj},
         year = 1994,
        month = mar,
       volume = {424},
        pages = {189},
          doi = {10.1086/173882},
       adsurl = {https://ui.adsabs.harvard.edu/abs/1994ApJ...424..189L}
}

@ARTICLE{Mehringer1993a,
       author = {{Mehringer}, David M. and {Palmer}, Patrick and {Goss}, W.~M. and {Yusef-Zadeh}, F.},
        title = "{Radio Continuum and Radio Recombination Line Observations of Sagittarius B2}",
      journal = {\apj},
         year = 1993,
        month = aug,
       volume = {412},
        pages = {684},
          doi = {10.1086/172954},
       adsurl = {https://ui.adsabs.harvard.edu/abs/1993ApJ...412..684M}
}

@ARTICLE{Mehringer1992,
       author = {{Mehringer}, David M. and {Yusef-Zadeh}, F. and {Palmer}, Patrick and {Goss}, W.~M.},
        title = "{Radio Continuum and Radio Recombination Line Observations of Sagittarius B1 and G0.6-0.0}",
      journal = {\apj},
         year = 1992,
        month = dec,
       volume = {401},
        pages = {168},
          doi = {10.1086/172050},
       adsurl = {https://ui.adsabs.harvard.edu/abs/1992ApJ...401..168M}
}

@ARTICLE{Ho1985,
       author = {{Ho}, P.~T.~P. and {Jackson}, J.~M. and {Barrett}, A.~H. and {Armstrong}, J.~T.},
        title = "{Interactions between the continuum sources in the Galactic Center andtheir immediate molecular environment.}",
      journal = {\apj},
         year = 1985,
        month = jan,
       volume = {288},
        pages = {575-579},
          doi = {10.1086/162823},
       adsurl = {https://ui.adsabs.harvard.edu/abs/1985ApJ...288..575H}
}

@ARTICLE{Ekers1983,
       author = {{Ekers}, R.~D. and {van Gorkom}, J.~H. and {Schwarz}, U.~J. and {Goss}, W.~M.},
        title = "{The radio structure of SGR A.}",
      journal = {\aap},
         year = 1983,
        month = jun,
       volume = {122},
        pages = {143-150},
       adsurl = {https://ui.adsabs.harvard.edu/abs/1983A&A...122..143E}
}

@software{Koch2025radiobeam,
       author = {{Koch}, Eric and {Ginsburg}, Adam and {AKL} and {Rosolowsky}, Erik and {Sip{\H{o}}cz}, Brigitta and {Robitaille}, Thomas and {Thomson}, Alec and {Maret}, S{\'e}bastien and {jdhenshaw} and {O'Brien}, Andrew and {adamginsburg} and {Detiste}, Alexandre and {Privon}, George C. and {Lim}, P.~L.},
        title = "{radio-astro-tools/radio-beam: v0.3.9}",
         year = 2025,
        month = jun,
          eid = {10.5281/zenodo.15677957},
          doi = {10.5281/zenodo.15677957},
      version = {v0.3.9},
    publisher = {Zenodo},
       adsurl = {https://ui.adsabs.harvard.edu/abs/2025zndo..15677957K}
}

@software{Ginsburg2019spectralcube,
       author = {{Ginsburg}, Adam and {Koch}, Eric and {Robitaille}, Thomas and {Beaumont}, Chris and {Adamginsburg} and {Sip{\H{o}}cz}, Brigitta and {ZuHone}, John and {Patra}, Sushobhana and {Jones}, Craig and {Lim}, P.~L. and {Stern}, Kris and {Rosolowsky}, Erik and {Earl}, Nicholas and {De Val-Borro}, Miguel and {Jrobbfed} and {Shuokong} and {Kepley}, Amanda and {Sokolov}, Vlas and {Badger}, The Gitter and {Maret}, S{\'e}bastien and {Garrido}, Juli{\'a}n and {Booker}, Joseph and {Tollerud}, Erik},
        title = "{radio-astro-tools/spectral-cube: Release v0.4.5}",
         year = 2019,
        month = nov,
          eid = {10.5281/zenodo.3558614},
          doi = {10.5281/zenodo.3558614},
      version = {v0.4.5},
    publisher = {Zenodo},
       adsurl = {https://ui.adsabs.harvard.edu/abs/2019zndo...3558614G}
}

@ARTICLE{Lang1997,
       author = {{Lang}, Cornelia C. and {Goss}, W.~M. and {Wood}, O.~S.},
        title = "{VLA H92{\ensuremath{\alpha}} and H115{\ensuremath{\beta}} Recombination Line Observations of the Galactic Center H II Regions: The Sickle (G0.18-0.04) and the Pistol (G0.15-0.05)}",
      journal = {\apj},
         year = 1997,
        month = jan,
       volume = {474},
       number = {1},
        pages = {275-291},
          doi = {10.1086/303452},
       adsurl = {https://ui.adsabs.harvard.edu/abs/1997ApJ...474..275L}
}

@ARTICLE{Walmsley1980,
       author = {{Walmsley}, C.~M. and {Winnewisser}, G. and {Toelle}, F.},
        title = "{Cyanoacetylene and cyanodiacetylene in interstellar clouds}",
      journal = {\aap},
         year = 1980,
        month = jan,
       volume = {81},
       number = {1-2},
        pages = {245-250},
       adsurl = {https://ui.adsabs.harvard.edu/abs/1980A&A....81..245W}
}

@ARTICLE{Kruijssen19,
      author = {{Kruijssen}, J.~M.~D. and {Dale}, J.~E. and {Longmore}, S.~N. and
         {Walker}, D.~L. and {Henshaw}, J.~D. and {Jeffreson}, S.~M.~R. and
         {Petkova}, M.~A. and {Ginsburg}, A. and {Barnes}, A.~T. and
         {Battersby}, C.~D. and {Immer}, K. and {Jackson}, J.~M. and
         {Keto}, E.~R. and {Krieger}, N. and {Mills}, E.~A.~C. and
         {S{\'a}nchez-Monge}, {\'A}. and {Schmiedeke}, A. and {Suri}, S.~T. and
         {Zhang}, Q.},
        title = "{The dynamical evolution of molecular clouds near the Galactic Centre - II. Spatial structure and kinematics of simulated clouds}",
      journal = {\mnras},
         year = 2019,
        month = apr,
      volume = {484},
      number = {4},
        pages = {5734-5754},
          doi = {10.1093/mnras/stz381},
archivePrefix = {arXiv},
      eprint = {1902.01860},
 primaryClass = {astro-ph.GA},
      adsurl = {https://ui.adsabs.harvard.edu/abs/2019MNRAS.484.5734K}
}

@ARTICLE{astropy:2022,
       author = {{Astropy Collaboration} and {Price-Whelan}, Adrian M. and {Lim}, Pey Lian and {Earl}, Nicholas and {Starkman}, Nathaniel and {Bradley}, Larry and {Shupe}, David L. and {Patil}, Aarya A. and {Corrales}, Lia and {Brasseur}, C.~E. and {N{"o}the}, Maximilian and {Donath}, Axel and {Tollerud}, Erik and {Morris}, Brett M. and {Ginsburg}, Adam and {Vaher}, Eero and {Weaver}, Benjamin A. and {Tocknell}, James and {Jamieson}, William and {van Kerkwijk}, Marten H. and {Robitaille}, Thomas P. and {Merry}, Bruce and {Bachetti}, Matteo and {G{"u}nther}, H. Moritz and {Aldcroft}, Thomas L. and {Alvarado-Montes}, Jaime A. and {Archibald}, Anne M. and {B{'o}di}, Attila and {Bapat}, Shreyas and {Barentsen}, Geert and {Baz{'a}n}, Juanjo and {Biswas}, Manish and {Boquien}, M{'e}d{'e}ric and {Burke}, D.~J. and {Cara}, Daria and {Cara}, Mihai and {Conroy}, Kyle E. and {Conseil}, Simon and {Craig}, Matthew W. and {Cross}, Robert M. and {Cruz}, Kelle L. and {D'Eugenio}, Francesco and {Dencheva}, Nadia and {Devillepoix}, Hadrien A.~R. and {Dietrich}, J{"o}rg P. and {Eigenbrot}, Arthur Davis and {Erben}, Thomas and {Ferreira}, Leonardo and {Foreman-Mackey}, Daniel and {Fox}, Ryan and {Freij}, Nabil and {Garg}, Suyog and {Geda}, Robel and {Glattly}, Lauren and {Gondhalekar}, Yash and {Gordon}, Karl D. and {Grant}, David and {Greenfield}, Perry and {Groener}, Austen M. and {Guest}, Steve and {Gurovich}, Sebastian and {Handberg}, Rasmus and {Hart}, Akeem and {Hatfield-Dodds}, Zac and {Homeier}, Derek and {Hosseinzadeh}, Griffin and {Jenness}, Tim and {Jones}, Craig K. and {Joseph}, Prajwel and {Kalmbach}, J. Bryce and {Karamehmetoglu}, Emir and {Ka{l}uszy{'n}ski}, Miko{l}aj and {Kelley}, Michael S.~P. and {Kern}, Nicholas and {Kerzendorf}, Wolfgang E. and {Koch}, Eric W. and {Kulumani}, Shankar and {Lee}, Antony and {Ly}, Chun and {Ma}, Zhiyuan and {MacBride}, Conor and {Maljaars}, Jakob M. and {Muna}, Demitri and {Murphy}, N.~A. and {Norman}, Henrik and {O'Steen}, Richard and {Oman}, Kyle A. and {Pacifici}, Camilla and {Pascual}, Sergio and {Pascual-Granado}, J. and {Patil}, Rohit R. and {Perren}, Gabriel I. and {Pickering}, Timothy E. and {Rastogi}, Tanuj and {Roulston}, Benjamin R. and {Ryan}, Daniel F. and {Rykoff}, Eli S. and {Sabater}, Jose and {Sakurikar}, Parikshit and {Salgado}, Jes{'u}s and {Sanghi}, Aniket and {Saunders}, Nicholas and {Savchenko}, Volodymyr and {Schwardt}, Ludwig and {Seifert-Eckert}, Michael and {Shih}, Albert Y. and {Jain}, Anany Shrey and {Shukla}, Gyanendra and {Sick}, Jonathan and {Simpson}, Chris and {Singanamalla}, Sudheesh and {Singer}, Leo P. and {Singhal}, Jaladh and {Sinha}, Manodeep and {Sip{H{o}}cz}, Brigitta M. and {Spitler}, Lee R. and {Stansby}, David and {Streicher}, Ole and {{{S}}umak}, Jani and {Swinbank}, John D. and {Taranu}, Dan S. and {Tewary}, Nikita and {Tremblay}, Grant R. and {Val-Borro}, Miguel de and {Van Kooten}, Samuel J. and {Vasovi{'c}}, Zlatan and {Verma}, Shresth and {de Miranda Cardoso}, Jos{'e} Vin{'i}cius and {Williams}, Peter K.~G. and {Wilson}, Tom J. and {Winkel}, Benjamin and {Wood-Vasey}, W.~M. and {Xue}, Rui and {Yoachim}, Peter and {Zhang}, Chen and {Zonca}, Andrea and {Astropy Project Contributors}},
        title = "{The Astropy Project: Sustaining and Growing a Community-oriented Open-source Project and the Latest Major Release (v5.0) of the Core Package}",
      journal = {\apj},
         year = 2022,
        month = aug,
       volume = {935},
       number = {2},
          eid = {167},
        pages = {167},
          doi = {10.3847/1538-4357/ac7c74},
archivePrefix = {arXiv},
       eprint = {2206.14220},
 primaryClass = {astro-ph.IM},
       adsurl = {https://ui.adsabs.harvard.edu/abs/2022ApJ...935..167A}
}

@article{astropy:2013,
Adsurl = {http://adsabs.harvard.edu/abs/2013A%26A...558A..33A},
Archiveprefix = {arXiv},
Author = {{Astropy Collaboration} and {Robitaille}, T.~P. and {Tollerud}, E.~J. and {Greenfield}, P. and {Droettboom}, M. and {Bray}, E. and {Aldcroft}, T. and {Davis}, M. and {Ginsburg}, A. and {Price-Whelan}, A.~M. and {Kerzendorf}, W.~E. and {Conley}, A. and {Crighton}, N. and {Barbary}, K. and {Muna}, D. and {Ferguson}, H. and {Grollier}, F. and {Parikh}, M.~M. and {Nair}, P.~H. and {Unther}, H.~M. and {Deil}, C. and {Woillez}, J. and {Conseil}, S. and {Kramer}, R. and {Turner}, J.~E.~H. and {Singer}, L. and {Fox}, R. and {Weaver}, B.~A. and {Zabalza}, V. and {Edwards}, Z.~I. and {Azalee Bostroem}, K. and {Burke}, D.~J. and {Casey}, A.~R. and {Crawford}, S.~M. and {Dencheva}, N. and {Ely}, J. and {Jenness}, T. and {Labrie}, K. and {Lim}, P.~L. and {Pierfederici}, F. and {Pontzen}, A. and {Ptak}, A. and {Refsdal}, B. and {Servillat}, M. and {Streicher}, O.},
Doi = {10.1051/0004-6361/201322068},
Eid = {A33},
Eprint = {1307.6212},
Journal = {\aap},
Month = oct,
Pages = {A33},
Primaryclass = {astro-ph.IM},
Title = {{Astropy: A community Python package for astronomy}},
Volume = 558,
Year = 2013}

@article{astropy:2018,
Adsurl = {https://ui.adsabs.harvard.edu/#abs/2018AJ....156..123T},
Author = {{Price-Whelan}, A.~M. and {Sip{\H{o}}cz}, B.~M. and {G{\"u}nther}, H.~M. and {Lim}, P.~L. and {Crawford}, S.~M. and {Conseil}, S. and {Shupe}, D.~L. and {Craig}, M.~W. and {Dencheva}, N. and {Ginsburg}, A. and {VanderPlas}, J.~T. and {Bradley}, L.~D. and {P{\'e}rez-Su{\'a}rez}, D. and {de Val-Borro}, M. and {Paper Contributors}, (Primary and {Aldcroft}, T.~L. and {Cruz}, K.~L. and {Robitaille}, T.~P. and {Tollerud}, E.~J. and {Coordination Committee}, (Astropy and {Ardelean}, C. and {Babej}, T. and {Bach}, Y.~P. and {Bachetti}, M. and {Bakanov}, A.~V. and {Bamford}, S.~P. and {Barentsen}, G. and {Barmby}, P. and {Baumbach}, A. and {Berry}, K.~L. and {Biscani}, F. and {Boquien}, M. and {Bostroem}, K.~A. and {Bouma}, L.~G. and {Brammer}, G.~B. and {Bray}, E.~M. and {Breytenbach}, H. and {Buddelmeijer}, H. and {Burke}, D.~J. and {Calderone}, G. and {Cano Rodr{\'\i}guez}, J.~L. and {Cara}, M. and {Cardoso}, J.~V.~M. and {Cheedella}, S. and {Copin}, Y. and {Corrales}, L. and {Crichton}, D. and {D{\textquoteright}Avella}, D. and {Deil}, C. and {Depagne}, {\'E}. and {Dietrich}, J.~P. and {Donath}, A. and {Droettboom}, M. and {Earl}, N. and {Erben}, T. and {Fabbro}, S. and {Ferreira}, L.~A. and {Finethy}, T. and {Fox}, R.~T. and {Garrison}, L.~H. and {Gibbons}, S.~L.~J. and {Goldstein}, D.~A. and {Gommers}, R. and {Greco}, J.~P. and {Greenfield}, P. and {Groener}, A.~M. and {Grollier}, F. and {Hagen}, A. and {Hirst}, P. and {Homeier}, D. and {Horton}, A.~J. and {Hosseinzadeh}, G. and {Hu}, L. and {Hunkeler}, J.~S. and {Ivezi{\'c}}, {\v{Z}}. and {Jain}, A. and {Jenness}, T. and {Kanarek}, G. and {Kendrew}, S. and {Kern}, N.~S. and {Kerzendorf}, W.~E. and {Khvalko}, A. and {King}, J. and {Kirkby}, D. and {Kulkarni}, A.~M. and {Kumar}, A. and {Lee}, A. and {Lenz}, D. and {Littlefair}, S.~P. and {Ma}, Z. and {Macleod}, D.~M. and {Mastropietro}, M. and {McCully}, C. and {Montagnac}, S. and {Morris}, B.~M. and {Mueller}, M. and {Mumford}, S.~J. and {Muna}, D. and {Murphy}, N.~A. and {Nelson}, S. and {Nguyen}, G.~H. and {Ninan}, J.~P. and {N{\"o}the}, M. and {Ogaz}, S. and {Oh}, S. and {Parejko}, J.~K. and {Parley}, N. and {Pascual}, S. and {Patil}, R. and {Patil}, A.~A. and {Plunkett}, A.~L. and {Prochaska}, J.~X. and {Rastogi}, T. and {Reddy Janga}, V. and {Sabater}, J. and {Sakurikar}, P. and {Seifert}, M. and {Sherbert}, L.~E. and {Sherwood-Taylor}, H. and {Shih}, A.~Y. and {Sick}, J. and {Silbiger}, M.~T. and {Singanamalla}, S. and {Singer}, L.~P. and {Sladen}, P.~H. and {Sooley}, K.~A. and {Sornarajah}, S. and {Streicher}, O. and {Teuben}, P. and {Thomas}, S.~W. and {Tremblay}, G.~R. and {Turner}, J.~E.~H. and {Terr{\'o}n}, V. and {van Kerkwijk}, M.~H. and {de la Vega}, A. and {Watkins}, L.~L. and {Weaver}, B.~A. and {Whitmore}, J.~B. and {Woillez}, J. and {Zabalza}, V. and {Contributors}, (Astropy},
Doi = {10.3847/1538-3881/aabc4f},
Eid = {123},
Journal = {\aj},
Month = Sep,
Pages = {123},
Primaryclass = {astro-ph.IM},
Title = {{The Astropy Project: Building an Open-science Project and Status of the v2.0 Core Package}},
Volume = {156},
Year = 2018}

@article{Waskom2021,
    doi = {10.21105/joss.03021},
    url = {https://doi.org/10.21105/joss.03021},
    year = {2021},
    publisher = {The Open Journal},
    volume = {6},
    number = {60},
    pages = {3021},
    author = {Michael L. Waskom},
    title = {seaborn: statistical data visualization},
    journal = {Journal of Open Source Software}
}

@software{carta,
       author = {{Comrie}, Angus and {Wang}, Kuo-Song and {Hsu}, Shou-Chieh and {Moraghan}, Anthony and {Harris}, Pamela and {Pang}, Qi and {Pi{\r{A}}ska}, Adrianna and {Chiang}, Cheng-Chin and {Simmonds}, Rob and {Chang}, Tien-Hao and {Jan}, Hengtai and {Lin}, Ming-Yi},
        title = "{CARTA: Cube Analysis and Rendering Tool for Astronomy}",
 howpublished = {Astrophysics Source Code Library, record ascl:2103.031},
         year = 2021,
        month = mar,
          eid = {ascl:2103.031},
       adsurl = {https://ui.adsabs.harvard.edu/abs/2021ascl.soft03031C}
}

@MISC{aplpy,
  author = {{Robitaille}, T. and {Bressert}, E.},
    title = "{APLpy: Astronomical Plotting Library in Python}",
howpublished = {Astrophysics Source Code Library},
     year = 2012,
    month = aug,
archivePrefix = "ascl",
  eprint = {1208.017},
  adsurl = {http://adsabs.harvard.edu/abs/2012ascl.soft08017R}
}

@ARTICLE{krumholz15,
  author = {{Krumholz}, M.~R. and {Kruijssen}, J.~M.~D.},
    title = "{A dynamical model for the formation of gas rings and episodic starbursts near galactic centres}",
  journal = {\mnras},
archivePrefix = "arXiv",
  eprint = {1505.07111},
     year = 2015,
    month = oct,
  volume = 453,
    pages = {739-757},
      doi = {10.1093/mnras/stv1670},
  adsurl = {http://adsabs.harvard.edu/abs/2015MNRAS.453..739K}
}

@ARTICLE{Krumholz16,
  author = {{Krumholz}, M.~R. and {Kruijssen}, J.~M.~D. and {Crocker}, R.~M.
	},
    title = "{A dynamical model for gas flows, star formation and nuclear winds in galactic centres}",
  journal = {\mnras},
archivePrefix = "arXiv",
  eprint = {1605.02850},
     year = 2017,
    month = apr,
  volume = 466,
    pages = {1213-1233},
      doi = {10.1093/mnras/stw3195},
  adsurl = {http://adsabs.harvard.edu/abs/2017MNRAS.466.1213K}
}

\onecolumn

\appendix

\section{Zoom-Ins of \ha line}

The \ha map indicative of ionized gas associated with the H II regions in
the Sgr A mini-spiral \citep[e.g.][]{Zhao2009} and H II region G-0.02-0.07 \citep[e.g.][]{Ekers1983, Mills2011, Uehara2019}, the possible H II region detected in the 20 km s$^{-1}$ cloud \citep{Ho1985,Lu2019b}, the Sgr B1 \citep[e.g.,][]{Mehringer1992,simpson2018a,simpson2021} and Sgr B2 \citep[e.g.][]{Mehringer1993a,DePree2011,Jones2011}, the Arched filaments \citep[e.g.][]{Lang2002}, the pistol and sickle \citep[e.g.][]{Lang1997,Steinke2016}. These sources are significantly detected in the ACES data.

\begin{figure*}
\includegraphics[width=1.0\textwidth]{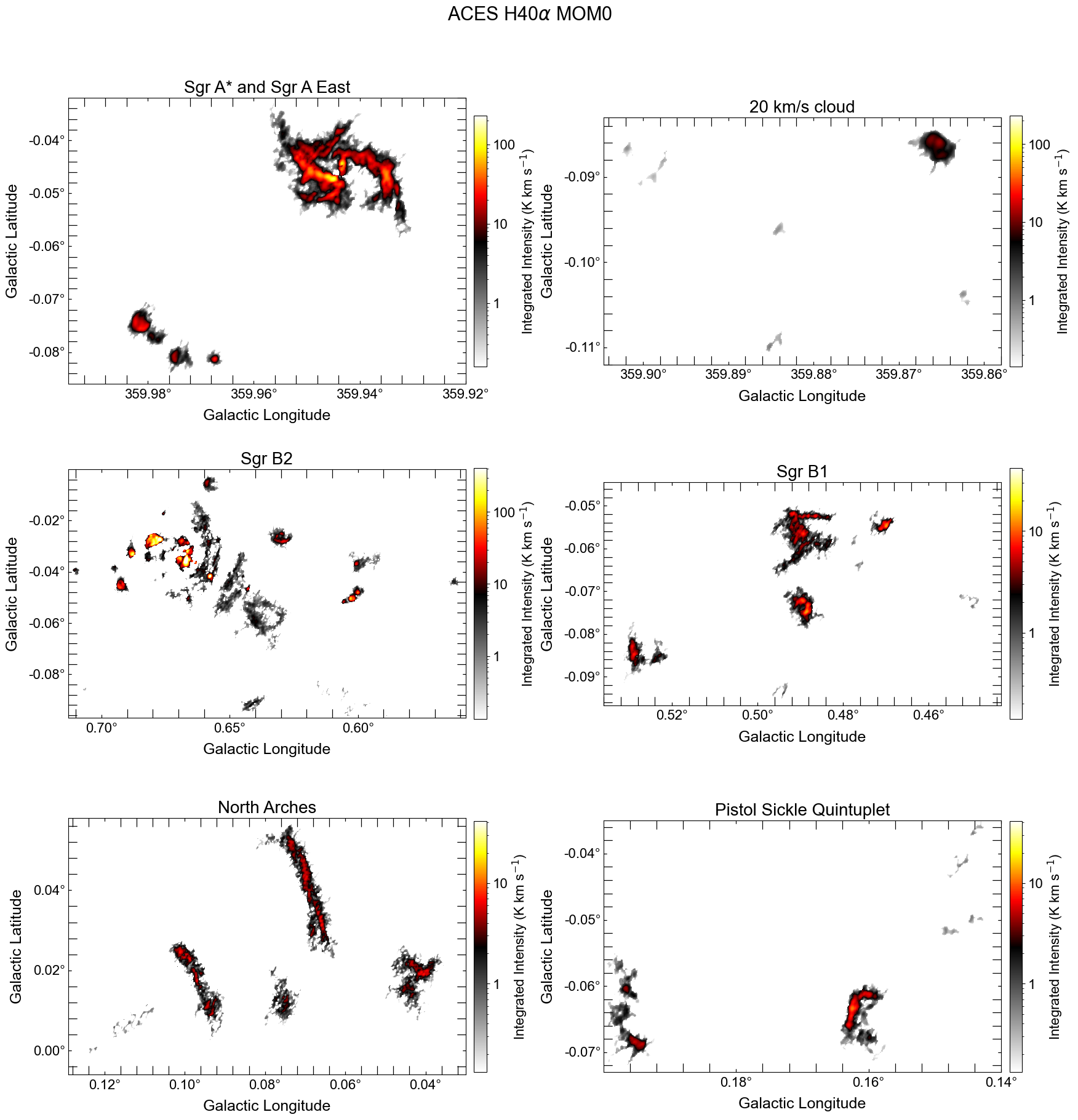}
\caption{The zoom-in views of the ACES \ha\ \texttt{mom0} images show regions with significant detections. For easy comparison, the names of the regions follow the nomenclature adopted in \citet{Ginsburg2025}.}
\label{fig:h40a_sgra}
\end{figure*}

\section{2-D correlation matrix}
In Figure~\ref{fig:2d-matrix}, we present the two-dimensional (2D) Pearson correlation coefficients among the 14 lines and the continuum images. A detailed description of the method is provided in \citetalias{Lu2025}.
The regions used for these calculations are listed in Table~\ref{tab:region}.

We include the correlation matrices to quantify how the spatial distributions of line intensities vary among different molecular tracers across each CMZ field. Each matrix element represents the normalized covariance between the brightness distributions of two spectral lines, measured over all valid pixels.
A high correlation (1) means the lines vary together spatially, tracing similar gas structures or excitation conditions. A low correlation (0) indicates independent spatial variations, while a negative correlation (-1) implies anti-correlated emission.
In general, molecular lines correlate strongly with one another, consistent with their tracing of dense gas structures. However, the overall correlation strength differs across regions: correlations are highest in the TLP, indicating that most molecular tracers vary together spatially, whereas the CND exhibits the lowest correlations. The weaker correlations in the CND suggest that the emission arises from multiple, physically distinct components along the line of sight, leading to less coherent spatial correspondence between tracers. This highlights the increased structural and chemical complexity near Sgr A* compared to regions farther out in the CMZ. We note that the similarity analysis presented in this paper provides a first-order comparison. A more detailed analysis, selecting specific regions or structures, is essential to derive more accurate values.

\begin{table}
\centering
\caption{DS9 regions in Galactic coordinates. Centers are in degrees; lengths are in arcseconds. For \texttt{box}, width/height are full sizes; for \texttt{ellipse}, the values are radii. Position angle (PA) in degrees.}\label{tab:region}
\begin{tabular}{lcccccc}
\hline
Region & Shape & $\ell$ (deg) & $b$ (deg) & Width/Radius$_x$ (arcsec) & Height/Radius$_y$ (arcsec) & PA (deg) \\
\hline
TLP clouds  & box     & 0.096408 & $-0.077577$ & 364 & 129 & 0 \\
CND   & ellipse & 359.944183 & $-0.045176$ & 121 & 164 & 68 \\
Brick & box     & 0.245984 & 0.011100 & 175 & 123 & 0 \\
20 and 50 km s$^{-1}$ cloud & box     & 359.908 & 0.-0.0836 & 675 & 136 & 335 \\
\hline
\end{tabular}
\end{table}


\begin{figure*}
\includegraphics[width=0.9\textwidth]{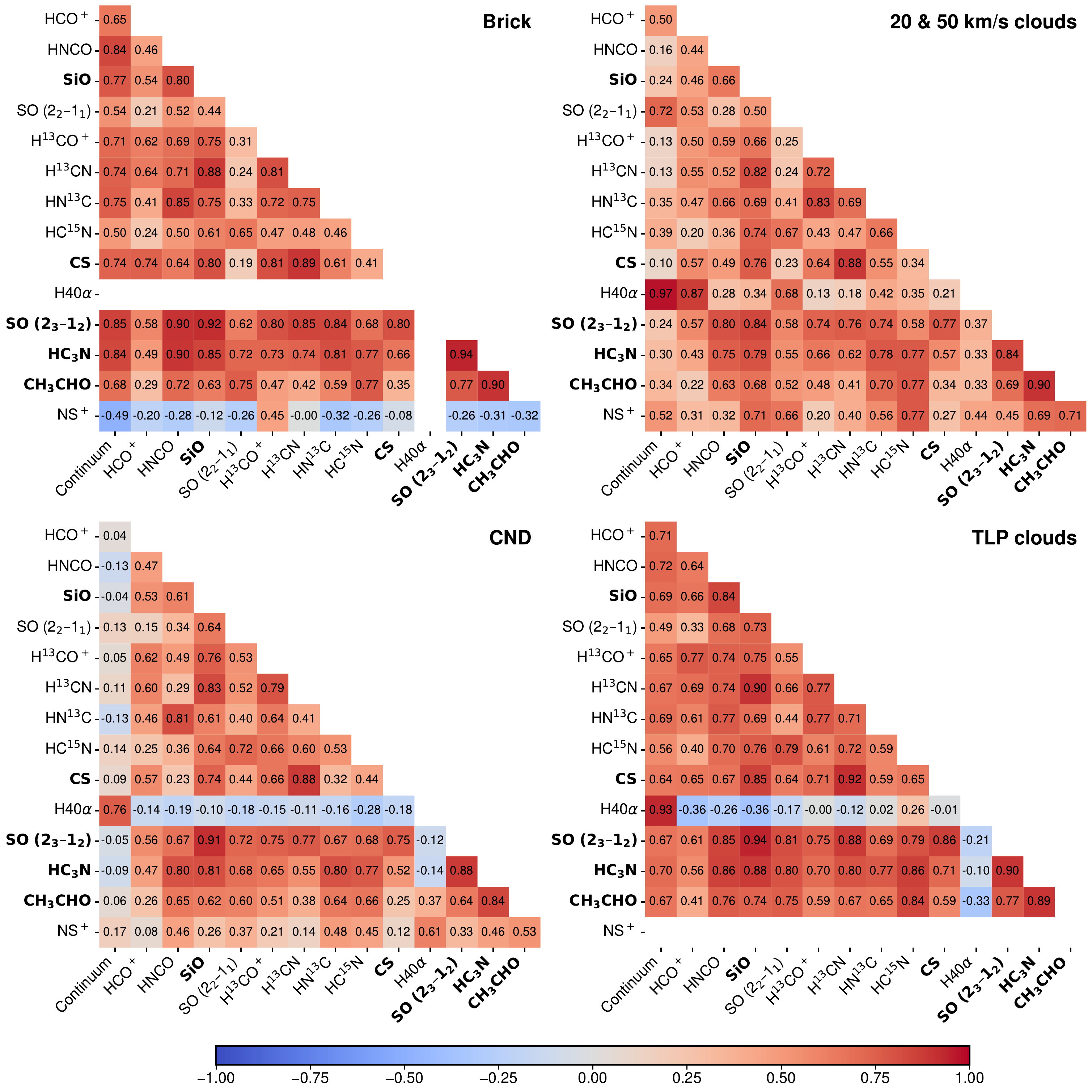}
\caption{
2-D correlation matrix between the continuum and integrated intensities (\texttt{mom 0} images) of the spectral lines presented in this paper, \citetalias{Walker2025}, and \citetalias{Lu2025}. In the Figure 12 of \citetalias{Lu2025}, all the pixels in the ACES mosaic fields are considered. In this paper we present the sub-fields towards the Brick cloud, the 20 and 50 km s$^{-1}$ clouds, and CND, the TLP clouds, the Sgr C and Sgr B regions. Note that the Sgr A* is excluded in the CND region. The Sgr A* mask is centered on $(l,b)=(359.944^{\circ},-0.046^{\circ})$, with a radius of 4.2$''$. Negative values mean anti-correlation, and the blank cells mean the invalid values in the images preventing calculation.
}
\label{fig:2d-matrix}
\end{figure*}

\section{Data release summary}\label{subsec:data_appendix}
\noindent
The data products from the ACES survey follow a uniform naming convention. The filenames for the data products presented in this paper can be constructed using the templates described below, in conjunction with the information provided in Appendix A of \citet{Walker2025}. 

The following subsections are divided according to different product types which occur at the Member ObsUnitSet (MOUS) and Group ObsUnitSet (GOUS) levels\footnote{For a description of ALMA project structures and naming conventions, please refer to Section 2 of the Cycle 9 QA2 guide: \url{https://almascience.eso.org/documents-and-tools/cycle9/alma-qa2-data-products-for-cycle-9}}.
\\
\subsection{Member-level products (12m and 7m)}
We release the data cubes for all SPWs for the 45 ACES regions. These are member-level products, and correspond to the cubes for the 12m and 7m arrays for each field prior to combination. 

We release the cubes for the separate arrays as they are different to those initially delivered by ALMA due to the changes that were made during reprocessing. We do not re-release any of the stand-alone TP products, as we used the same cubes as those available in the ALMA Science Archive (ASA).

In this and subsequent subsections, we describe how to construct the filenames of the released data products by providing filename templates that include placeholder variables. These placeholders can be changed as needed to build the filename of the relevant product.

Following the requirements of the ASA, the filename template is:

\texttt{member.uid\_\_\_A001\_\{\textbf{mous-id}\}.lp\_slongmore.\{\textbf{field\_coords}\}.\{\textbf{array}\}.\{\textbf{frqrange}\}.cube.pbcor.fits}

The placeholder template components should be replaced as follows:
\begin{itemize}
    \item \texttt{\{\textbf{mous-id}\}}: The 12m or 7m MOUS ID for a given field, as listed in Table A1 in \citet{Walker2025}.
    \item \texttt{\{\textbf{field\_coords}\}}: The central coordinates of the field, as listed in Table A1 in \citet{Walker2025} (e.g., \texttt{G000.073+0.184}).
    \item \texttt{\{\textbf{array}\}}: The ALMA array, either \texttt{12m} or \texttt{7m}.
    \item \texttt{\{\textbf{frqrange}\}}: The frequency range, which depends on the SPW:
    \begin{itemize}
        \item SPW 33: \texttt{97.7-99.5GHz}
        \item SPW 35: \texttt{99.6-101.4GHz}
    \end{itemize}
\end{itemize}

For example, to find the 12m-only cube for CS for field \texttt{ao} (this field contains the well-known molecular cloud G0.253+0.016, aka the Brick), the resulting filename would be:

\texttt{member.uid\_\_\_A001\_X15a0\_X190.lp\_slongmore.G000.300+0.063.12m.97.7-99.5GHz.cube.pbcor.fits}

\subsection{Group-level products (Combined cubes per region)}

We also release array-combined products per SPW per region. As these products combine data from multiple MOUSs, they are deemed to be group-level products. The filenames of these products can be constructed as follows:

\texttt{group.uid\_\_\_A001\_X1590\_X30a9.lp\_slongmore.\{\textbf{field\_coords}\}.\{\textbf{arrays}\}.\{\textbf{frqrange}\}.cube.pbcor.fits}

Where the template components are:
\begin{itemize}
    \item \texttt{\{\textbf{field\_coords}\}} and \texttt{\{\textbf{frqrange}\}}: follow the same definition as for the member-level products.
    \item \texttt{\{\textbf{arrays}\}}: \texttt{12m7mTP}
\end{itemize}

As an example, the corresponding CS cube including 12m, 7m, and TP data for field \texttt{ao} would be:

\texttt{group.uid\_\_\_A001\_X1590\_X30a9.lp\_slongmore.G000.300+0.063.12m7mTP.97.7-99.5GHz.cube.pbcor.fits}

\subsection{Group-level products (Full CMZ mosaics)}

Finally, we release the full, contiguous mosaics covering the entire ACES footprint. Again, since these result from the combination of many MOUSs, the full mosaics are at the group level.

In contrast to cubes for the individual fields, we do not release full mosaics for all SPWs, but rather for specific lines. The lines relevant to this paper are given in Table \ref{tab:lines}.
Most of the SPWs are sufficiently broad that they contain significant spectral ranges with no line emission. We therefore opted to not produce full mosaics containing these blank channels in the interest of reducing the file sizes. In addition to the data cubes for each line, we also release a suite of advanced products, including moment maps, noise maps, etc. (see below).

The filename template for the full mosaic cubes and associated products is:

\texttt{group.uid\_\_\_A001\_X1590\_X30a9.lp\_slongmore.cmz\_mosaic.\{\textbf{arrays}\}.\{\textbf{molecule}\}.\{\textbf{suffix}}\}

Where the template components are:
\begin{itemize}
    \item \texttt{\{\textbf{arrays}\}}: \texttt{12m7mTP} 
    \item \texttt{\{\textbf{molecule}\}}: \texttt{HC3N, CS21, H40a, CH3CHO, SO32}
    \item \texttt{\{\textbf{suffix}\}} defines the specific data product. Products released with this paper are:
    \begin{itemize}
        \item \texttt{cube.pbcor.fits}: primary beam corrected image cube.
        \item \texttt{cube.downsampled\_spatially.pbcor.fits}: as previous, spatially smoothed to a 5 arcsec beam and then spatially rebinned by a factor of $9\times9$.
        \item \texttt{integrated\_intensity.fits}: masked integrated intensity map of the full-resolution cube.
        \item \texttt{mad\_std.fits}: noise map of the full-resolution cube.
        \item \texttt{peak\_intensity.fits}: peak intensity map of the full-resolution cube.
        \item \texttt{velocity\_at\_peak\_intensity.fits}: velocity map from the masked full-resolution cube; velocity estimated using the peak intensity of the spectrum for each pixel.
        \item \texttt{PV\_l\_max.fits}: $\ell$–$v$ position–velocity map from the full-resolution cube; maximum intensity taken along Galactic latitude ($b$).
        \item \texttt{PV\_l\_mean.fits}: $\ell$–$v$ position–velocity map from the masked full-resolution cube; mean intensity taken along Galactic latitude ($b$)
        \item \texttt{PV\_b\_max.fits}: $b$–$v$ position–velocity map from the full-resolution cube; maximum intensity taken along Galactic longitude ($\ell$).
        \item \texttt{PV\_b\_mean.fits}: $b$–$v$ position–velocity map from the full-resolution cube; mean intensity taken along Galactic longitude ($\ell$).                
    \end{itemize}
\end{itemize}

\begin{table*}
\centering
\caption{Global statistics for SPW 33 and SPW 35 cubes for all ACES regions, with velocity offset V$_{\textrm{off}}$ added as the second column.}
\label{tab:cubestats_spw33-35}
\begin{tabular}{ccccccccccc}
\hline\hline
Field & V$_{\textrm{off}}$ & SPW$_\textrm{12m}$ & B$_{\textrm{maj}}$ & B$_{\textrm{min}}$ & BPA & Image size & No.\ channels & Pixel size & S$_{peak}$ & $\sigma_{MAD}$ \\
 & (\kms) & & (\arcsec) & (\arcsec) & (deg) & (pix) &  & (\arcsec) & (mJy~beam$^{-1}$) & (mJy~beam$^{-1}$)\\
\hline
a  & 40  & 33 & 1.49 & 1.10 & -67.0 & (2070, 2482) & 3840.0 & 0.20 & 128.3 & 3.9 \\
a  & 40  & 35 & 1.42 & 1.06 & -66.0 & (2070, 2482) & 3840.0 & 0.20 & 112.5 & 3.6 \\
aa & 0   & 33 & 1.63 & 1.44 & -52.0 & (1593, 1456) & 3839.0 & 0.26 & 372.5 & 3.7 \\
aa & 0   & 35 & 1.63 & 1.41 & -52.0 & (1593, 1456) & 3839.0 & 0.26 & 298.2 & 3.4 \\
ab & 0   & 33 & 1.63 & 1.52 & -75.0 & (1465, 1236) & 3836.0 & 0.28 & 199.7 & 3.6 \\
ab & 0   & 35 & 1.60 & 1.48 & -75.0 & (1465, 1236) & 3836.0 & 0.28 & 179.7 & 3.3 \\
ac & 0   & 33 & 1.54 & 1.06 & -71.0 & (1638, 1601) & 3840.0 & 0.19 & 166.4 & 4.3 \\
ac & 0   & 35 & 1.51 & 1.03 & -71.0 & (1638, 1601) & 3840.0 & 0.19 & 145.7 & 4.0 \\
ad & 0   & 33 & 1.79 & 1.78 & -75.0 & (1140, 2527) & 3840.0 & 0.20 & 221.4 & 4.3 \\
ad & 0   & 35 & 1.77 & 1.75 & -75.0 & (1140, 2527) & 3840.0 & 0.20 & 203.2 & 4.0 \\
ae & 0   & 33 & 1.88 & 1.52 &  83.0 & (1313, 1344) & 3840.0 & 0.30 & 253.0 & 3.8 \\
ae & 0   & 35 & 1.87 & 1.51 &  83.0 & (1313, 1344) & 3840.0 & 0.30 & 238.1 & 3.5 \\
af & 0   & 33 & 2.50 & 1.74 &  86.0 & (1417, 1165) & 3840.0 & 0.31 & 509.2 & 3.9 \\
af & 0   & 35 & 2.44 & 1.71 &  86.0 & (1417, 1165) & 3840.0 & 0.31 & 487.0 & 3.7 \\
ag & 0   & 33 & 1.65 & 1.26 & -70.0 & (1698, 1696) & 3840.0 & 0.24 & 149.7 & 4.2 \\
ag & 0   & 35 & 1.62 & 1.24 & -70.0 & (1698, 1696) & 3840.0 & 0.24 & 141.2 & 4.0 \\
ah & 30  & 33 & 1.66 & 1.12 & -61.0 & (1957, 1936) & 3840.0 & 0.20 & 200.0 & 3.5 \\
ah & 30  & 35 & 1.64 & 1.09 & -61.0 & (1957, 1936) & 3840.0 & 0.20 & 188.7 & 3.3 \\
ai & 0   & 33 & 2.11 & 1.52 &  79.0 & (1350, 1601) & 3840.0 & 0.29 & 285.0 & 3.7 \\
ai & 0   & 35 & 2.06 & 1.49 &  79.0 & (1350, 1601) & 3840.0 & 0.29 & 260.8 & 3.4 \\
aj & -30 & 33 & 1.85 & 1.47 &  86.0 & (1414, 1265) & 3836.0 & 0.31 & 492.7 & 3.8 \\
aj & -30 & 35 & 1.83 & 1.45 &  86.0 & (1414, 1265) & 3836.0 & 0.31 & 462.5 & 3.6 \\
ak & 30  & 33 & 1.96 & 1.67 &  80.0 & (736, 988)   & 3836.0 & 0.32 & 256.3 & 3.8 \\
ak & 30  & 35 & 1.92 & 1.64 &  80.0 & (736, 988)   & 3836.0 & 0.32 & 243.7 & 3.6 \\
al & 0   & 33 & 2.37 & 1.79 &  86.0 & (2167, 1422) & 3840.0 & 0.28 & 217.8 & 4.4 \\
al & 0   & 35 & 2.30 & 1.76 &  86.0 & (2167, 1422) & 3840.0 & 0.28 & 205.3 & 4.1 \\
am & 30  & 33 & 2.84 & 1.92 &  89.0 & (2578, 1117) & 3840.0 & 0.30 & 5064.8 & 4.9 \\
am & 30  & 35 & 2.79 & 1.90 &  89.0 & (2578, 1117) & 3840.0 & 0.30 & 4820.0 & 4.5 \\
an & 30  & 33 & 2.31 & 1.67 &  83.0 & (1305, 1336) & 3840.0 & 0.30 & 333.8 & 4.0 \\
an & 30  & 35 & 2.28 & 1.65 &  83.0 & (1305, 1336) & 3840.0 & 0.30 & 312.7 & 3.7 \\
ao & 0   & 33 & 2.28 & 1.58 & -86.0 & (2481, 897)  & 3839.0 & 0.28 & 320.8 & 5.1 \\
ao & 0   & 35 & 2.25 & 1.56 & -86.0 & (2481, 897)  & 3839.0 & 0.28 & 308.6 & 4.7 \\
ap & 0   & 33 & 1.88 & 1.53 &  79.0 & (1314, 1347) & 3836.0 & 0.30 & 469.7 & 3.5 \\
ap & 0   & 35 & 1.86 & 1.51 &  79.0 & (1314, 1347) & 3836.0 & 0.30 & 452.2 & 3.2 \\
aq & 0   & 33 & 1.62 & 1.34 & -78.0 & (1785, 1787) & 3840.0 & 0.22 & 370.2 & 3.8 \\
aq & 0   & 35 & 1.60 & 1.32 & -78.0 & (1785, 1787) & 3840.0 & 0.22 & 349.1 & 3.5 \\
ar & 40  & 33 & 2.13 & 2.11 &  85.0 & (1355, 1401) & 3840.0 & 0.29 & 348.7 & 4.2 \\
ar & 40  & 35 & 2.10 & 2.08 &  85.0 & (1355, 1401) & 3840.0 & 0.29 & 336.9 & 3.9 \\
as & 0   & 33 & 1.77 & 1.23 & -69.0 & (1965, 2020) & 3837.0 & 0.20 & 308.4 & 4.1 \\
as & 0   & 35 & 1.75 & 1.21 & -69.0 & (1965, 2020) & 3837.0 & 0.20 & 296.6 & 3.7 \\
b  & 40  & 33 & 1.57 & 1.55 &  87.0 & (1872, 1925) & 3813.0 & 0.21 & 2100.5 & 3.2 \\
b  & 40  & 35 & 1.53 & 1.52 &  87.0 & (1872, 1925) & 3813.0 & 0.21 & 2043.2 & 2.9 \\
c  & 0   & 33 & 2.53 & 1.74 &  85.0 & (1404, 1436) & 3839.0 & 0.28 & 329.7 & 4.4 \\
c  & 0   & 35 & 2.50 & 1.72 &  85.0 & (1404, 1436) & 3839.0 & 0.28 & 308.9 & 4.1 \\
d  & 30  & 33 & 1.72 & 1.39 & -60.0 & (2022, 1556) & 3839.0 & 0.23 & 156.4 & 4.1 \\
d  & 30  & 35 & 1.69 & 1.37 & -60.0 & (2022, 1556) & 3839.0 & 0.23 & 145.1 & 3.8 \\
e  & -30 & 33 & 2.19 & 1.50 &  83.0 & (1410, 1446) & 3839.0 & 0.28 & 273.6 & 4.6 \\
e  & -30 & 35 & 2.14 & 1.48 &  83.0 & (1410, 1446) & 3839.0 & 0.28 & 260.0 & 4.3 \\
f  & 0   & 33 & 2.20 & 1.32 &  87.0 & (1750, 1442) & 3840.0 & 0.28 & 353.3 & 4.5 \\
f  & 0   & 35 & 2.18 & 1.30 &  87.0 & (1750, 1442) & 3840.0 & 0.28 & 341.0 & 4.2 \\
g  & 30  & 33 & 1.72 & 1.35 & -85.0 & (1572, 1619) & 3840.0 & 0.25 & 198.8 & 4.1 \\
g  & 30  & 35 & 1.68 & 1.32 & -85.0 & (1572, 1619) & 3840.0 & 0.25 & 187.0 & 3.8 \\
h  & 0   & 33 & 1.46 & 1.13 & -78.0 & (1951, 2083) & 3840.0 & 0.21 & 173.2 & 3.9 \\
h  & 0   & 35 & 1.42 & 1.10 & -78.0 & (1951, 2083) & 3840.0 & 0.21 & 160.9 & 3.6 \\
i  & 30  & 33 & 1.47 & 1.27 & -59.0 & (1646, 1613) & 3832.0 & 0.25 & 260.0 & 3.5 \\
i  & 30  & 35 & 1.43 & 1.23 & -59.0 & (1646, 1613) & 3832.0 & 0.25 & 243.5 & 3.2 \\
\hline
\end{tabular}
\end{table*}

\begin{table*}
\centering
\contcaption{}
\begin{tabular}{ccccccccccc}
\hline\hline
Field & V$_{\textrm{off}}$ & SPW$_\textrm{12m}$ & B$_{\textrm{maj}}$ & B$_{\textrm{min}}$ & BPA & Image size & No.\ channels & Pixel size & S$_{peak}$ & $\sigma_{MAD}$ \\
 & (\kms) & & (\arcsec) & (\arcsec) & (deg) & (pix) &  & (\arcsec) & (mJy~beam$^{-1}$) & (mJy~beam$^{-1}$)\\
\hline
j  & 30  & 33 & 2.21 & 1.26 & -87.0 & (1446, 1485) & 3819.0 & 0.27 & 317.1 & 4.6 \\
j  & 30  & 35 & 2.17 & 1.24 & -87.0 & (1446, 1485) & 3819.0 & 0.27 & 301.0 & 4.3 \\
k  & 30  & 33 & 1.63 & 1.01 & -75.0 & (2174, 2167) & 3832.0 & 0.18 & 140.8 & 4.3 \\
k  & 30  & 35 & 1.60 & 0.99 & -75.0 & (2174, 2167) & 3832.0 & 0.18 & 130.7 & 4.0 \\
l  & 0   & 33 & 1.64 & 1.52 & -51.0 & (2383, 1497) & 3840.0 & 0.26 & 851.0 & 4.1 \\
l  & 0   & 35 & 1.61 & 1.50 & -51.0 & (2383, 1497) & 3840.0 & 0.26 & 823.6 & 3.8 \\
m  & 30  & 33 & 2.22 & 1.85 &  81.0 & (1345, 1356) & 3840.0 & 0.29 & 2975.0 & 3.5 \\
m  & 30  & 35 & 2.18 & 1.82 &  81.0 & (1345, 1356) & 3840.0 & 0.29 & 2850.0 & 3.2 \\
n  & 30  & 33 & 2.62 & 1.28 &  75.0 & (1366, 1823) & 3840.0 & 0.19 & 231.5 & 5.3 \\
n  & 30  & 35 & 2.60 & 1.25 &  75.0 & (1366, 1823) & 3840.0 & 0.19 & 218.4 & 4.9 \\
o  & 0   & 33 & 2.04 & 1.56 &  83.0 & (1281, 1230) & 3835.0 & 0.32 & 378.3 & 4.4 \\
o  & 0   & 35 & 2.00 & 1.53 &  83.0 & (1281, 1230) & 3835.0 & 0.32 & 359.2 & 4.1 \\
p  & 0   & 33 & 1.58 & 1.26 &  89.0 & (1856, 1839) & 3836.0 & 0.22 & 270.7 & 3.7 \\
p  & 0   & 35 & 1.54 & 1.23 &  89.0 & (1856, 1839) & 3836.0 & 0.22 & 255.0 & 3.4 \\
q  & 0   & 33 & 2.02 & 1.55 &  88.0 & (1315, 1343) & 3836.0 & 0.30 & 279.5 & 3.3 \\
q  & 0   & 35 & 1.99 & 1.53 &  88.0 & (1315, 1343) & 3836.0 & 0.30 & 262.0 & 3.1 \\
r  & 30  & 33 & 2.34 & 1.36 & -86.0 & (1535, 1395) & 3840.0 & 0.29 & 344.4 & 4.4 \\
r  & 30  & 35 & 2.30 & 1.33 & -86.0 & (1535, 1395) & 3840.0 & 0.29 & 327.0 & 4.1 \\
s  & 0   & 33 & 2.53 & 1.97 &  80.0 & (1310, 1346) & 3839.0 & 0.30 & 618.6 & 3.8 \\
s  & 0   & 35 & 2.48 & 1.94 &  80.0 & (1310, 1346) & 3839.0 & 0.30 & 593.7 & 3.5 \\
t  & 0   & 33 & 2.15 & 1.95 &  65.0 & (1269, 1273) & 3840.0 & 0.31 & 327.7 & 3.9 \\
t  & 0   & 35 & 2.11 & 1.91 &  65.0 & (1269, 1273) & 3840.0 & 0.31 & 310.8 & 3.6 \\
u  & 30  & 33 & 1.71 & 1.18 & -67.0 & (1347, 2051) & 3834.0 & 0.19 & 139.5 & 4.5 \\
u  & 30  & 35 & 1.68 & 1.15 & -67.0 & (1347, 2051) & 3834.0 & 0.19 & 130.2 & 4.2 \\
v  & 30  & 33 & 1.60 & 1.02 & -79.0 & (1362, 1880) & 3819.0 & 0.19 & 236.6 & 4.2 \\
v  & 30  & 35 & 1.58 & 1.00 & -79.0 & (1362, 1880) & 3819.0 & 0.19 & 220.5 & 3.9 \\
w  & 30  & 33 & 2.01 & 1.42 & -79.0 & (1474, 1399) & 3819.0 & 0.28 & 199.0 & 3.4 \\
w  & 30  & 35 & 1.98 & 1.39 & -79.0 & (1474, 1399) & 3819.0 & 0.28 & 189.2 & 3.2 \\
x  & 0   & 33 & 2.20 & 1.49 & -86.0 & (1291, 769)  & 3836.0 & 0.30 & 396.7 & 4.2 \\
x  & 0   & 35 & 2.17 & 1.46 & -86.0 & (1291, 769)  & 3836.0 & 0.30 & 377.1 & 3.9 \\
y  & 0   & 33 & 1.86 & 1.37 & -85.0 & (1428, 1112) & 3835.0 & 0.27 & 209.9 & 4.3 \\
y  & 0   & 35 & 1.83 & 1.34 & -85.0 & (1428, 1112) & 3835.0 & 0.27 & 198.2 & 4.0 \\
z  & 40  & 33 & 1.96 & 1.28 &  79.0 & (1350, 1353) & 3813.0 & 0.29 & 153.1 & 4.3 \\
z  & 40  & 35 & 1.92 & 1.25 &  79.0 & (1350, 1353) & 3813.0 & 0.29 & 142.7 & 4.0 \\
\hline
\end{tabular}
\end{table*}

\clearpage

\section*{Author Affiliations}
\printaffiliation{naoj}{National Astronomical Observatory of Japan, 2-21-1 Osawa, Mitaka, Tokyo 181-8588, Japan}
\printaffiliation{ukarcnode}{UK ALMA Regional Centre Node, Jodrell Bank Centre for Astrophysics, The University of Manchester, Manchester M13 9PL, UK}
\printaffiliation{uflorida}{Department of Astronomy, University of Florida, P.O. Box 112055, Gainesville, FL 32611, USA}
\printaffiliation{eso}{European Southern Observatory (ESO), Karl-Schwarzschild-Stra{\ss}e 2, 85748 Garching, Germany}
\printaffiliation{shao}{Shanghai Astronomical Observatory, Chinese Academy of Sciences, 80 Nandan Road, Shanghai 200030, P.\ R.\ China}
\printaffiliation{naoc_key}{State Key Laboratory of Radio Astronomy and Technology, A20 Datun Road, Chaoyang District, Beijing, 100101, P. R. China}
\printaffiliation{ice_csic}{Institut de Ci\`encies de l'Espai (ICE), CSIC, Campus UAB, Carrer de Can Magrans s/n, E-08193, Bellaterra, Barcelona, Spain}
\printaffiliation{ieec}{Institut d'Estudis Espacials de Catalunya (IEEC), E-08860, Castelldefels, Barcelona, Spain}
\printaffiliation{umd}{University of Maryland, Department of Astronomy, College Park, MD 20742-2421, USA}
\printaffiliation{mpe}{Max-Planck-Institut f\"ur extraterrestrische Physik, Gie\ss enbachstra\ss e 1, 85748 Garching bei M\"unchen, Germany}
\printaffiliation{kansas}{Department of Physics and Astronomy, University of Kansas, 1251 Wescoe Hall Drive, Lawrence, KS 66045, USA}
\printaffiliation{ljmu}{Astrophysics Research Institute, Liverpool John Moores University, 146 Brownlow Hill, Liverpool L3 5RF, The UK}
\printaffiliation{mpia}{{Max Planck Institute for Astronomy, K\"{o}nigstuhl 17, D-69117 Heidelberg, Germany}}
\printaffiliation{ari_heidelberg}{Astronomisches Rechen-Institut, Zentrum f\"{u}r Astronomie der Universit\"{a}t Heidelberg, M\"{o}nchhofstra\ss e 12-14, D-69120 Heidelberg, Germany}
\printaffiliation{COOL}{Cosmic Origins Of Life (COOL) Research DAO, \href{https://coolresearch.io}{https://coolresearch.io}}
\printaffiliation{eso_chile}{European Southern Observatory, Alonso de C\'ordova, 3107, Vitacura, Santiago 763-0355, Chile}
\printaffiliation{jao}{Joint ALMA Observatory, Alonso de C\'ordova, 3107, Vitacura, Santiago 763-0355, Chile}
\printaffiliation{nanjing}{School of Astronomy and Space Science, Nanjing University, 163 Xianlin Avenue, Nanjing 210023, P.R.China}
\printaffiliation{nanjing_key}{Key Laboratory of Modern Astronomy and Astrophysics (Nanjing University), Ministry of Education, Nanjing 210023, P.R.China}
\printaffiliation{cfa}{Center for Astrophysics | Harvard \& Smithsonian, 60 Garden Street, Cambridge, MA 02138, USA}
\printaffiliation{mit}{Haystack Observatory, Massachusetts Institute of Technology, 99 Millstone Road, Westford, MA 01886, USA}
\printaffiliation{colorado}{Center for Astrophysics and Space Astronomy, Department of Astrophysical and Planetary Sciences, University of Colorado, Boulder, CO 80389, USA}
\printaffiliation{uconn}{Department of Physics, University of Connecticut, 196A Auditorium Road, Unit 3046, Storrs, CT 06269, USA}
\printaffiliation{cab_csic}{Centro de Astrobiolog{\'i}a (CAB), CSIC-INTA, Carretera de Ajalvir km 4, 28850 Torrej{\'o}n de Ardoz, Madrid, Spain}
\printaffiliation{iaa_taipei}{Academia Sinica Institute of Astronomy and Astrophysics, Astronomy-Mathematics Building, AS/NTU No.1, Sec. 4, Roosevelt Rd, Taipei 10617, Taiwan}
\printaffiliation{chalmers}{Space, Earth and Environment Department, Chalmers University of Technology, SE-412 96 Gothenburg, Sweden}
\printaffiliation{clap}{{Como Lake centre for AstroPhysics (CLAP), DiSAT, Universit{\`a} dell’Insubria, via Valleggio 11, 22100 Como, Italy}}
\printaffiliation{iop_epfl}{Institute of Physics, Laboratory for Galaxy Evolution and Spectral Modelling, EPFL, Observatoire de Sauverny, Chemin Pegasi 51, 1290 Versoix, Switzerland}
\printaffiliation{oaq}{Observatorio Astron\'omico de Quito, Observatorio Astron\'omico Nacional, Escuela Polit\'ecnica Nacional, 170403, Quito, Ecuador}
\printaffiliation{inaf_arcetri}{INAF Arcetri Astrophysical Observatory, Largo Enrico Fermi 5, Firenze, 50125, Italy}
\printaffiliation{jbca}{Jodrell Bank Centre for Astrophysics, The University of Manchester, Manchester M13 9PL, UK}
\printaffiliation{nrao}{National Radio Astronomy Observatory, 520 Edgemont Road, Charlottesville, VA 22903, USA}
\printaffiliation{ita_heidelberg}{Universit\"{a}t Heidelberg, Zentrum f\"{u}r Astronomie, Institut f\"{u}r Theoretische Astrophysik, Albert-Ueberle-Str 2, D-69120 Heidelberg, Germany}
\printaffiliation{anu}{Research School of Astronomy and Astrophysics, Australian National University, Canberra, ACT 2611, Australia}
\printaffiliation{iaa_csic}{Instituto de Astrof\'{i}sica de Andaluc\'{i}a, CSIC, Glorieta de la Astronomía s/n, 18008 Granada, Spain}
\printaffiliation{ucn}{Instituto de Astronom\'ia, Universidad Cat\'olica del Norte, Av. Angamos 0610, Antofagasta, Chile}
\printaffiliation{cassaca}{Chinese Academy of Sciences South America Center for Astronomy, National Astronomical Observatories, CAS, Beijing 100101, China}
\printaffiliation{ias}{Institute for Advanced Study, 1 Einstein Drive, Princeton, NJ 08540, USA}
\printaffiliation{ucl}{Department of Physics and Astronomy, University College London, Gower Street, London WC1E 6BT, UK}
\printaffiliation{izw_heidelberg}{Universit\"{a}t Heidelberg, Interdisziplin\"{a}res Zentrum f\"{u}r Wissenschaftliches Rechnen, Im Neuenheimer Feld 225, 69120 Heidelberg, Germany}
\printaffiliation{radcliffe}{Elizabeth S. and Richard M. Cashin Fellow at the Radcliffe Institute for Advanced Studies at Harvard University, 10 Garden Street, Cambridge, MA 02138, U.S.A.}
\printaffiliation{ipm}{Institute for Research in Fundamental Sciences (IPM), School of Astronomy, Tehran, Iran}
\printaffiliation{villanova}{Department of Physics, Villanova University, 800 E. Lancaster Ave., Villanova, PA 19085, USA}
\printaffiliation{ulaserena}{Departamento de Astronom\'ia, Universidad de La Serena, Ra\'ul Bitr\'an 1305, La Serena, Chile}
\printaffiliation{iff_csic}{Instituto de Física Fundamental (CSIC), Calle Serrano 121-123, 28006, Madrid, Spain}
\printaffiliation{gbo}{Green Bank Observatory, P.O. Box 2, Green Bank, WV 24944, USA}
\printaffiliation{utokyo}{Institute of Astronomy, The University of Tokyo, Mitaka, Tokyo 181-0015, Japan}
\printaffiliation{umass}{Department of Astronomy, University of Massachusetts, Amherst, MA 01003, USA}
\printaffiliation{aberystwyth}{Department of Physics, Aberystwyth University, Ceredigion, Cymru, SY23 3BZ, UK}
\printaffiliation{kiaa_pku}{Kavli Institute for Astronomy and Astrophysics, Peking University, Beijing 100871, People's Republic of China}
\printaffiliation{pku_astro}{Department of Astronomy, School of Physics, Peking University, Beijing, 100871, People's Republic of China}

\bsp
\label{lastpage}
\end{document}